\documentclass[10pt]{aastex62}
\usepackage{amsmath}
\usepackage[mathscr]{euscript}
\usepackage{stmaryrd}
\usepackage[T1]{fontenc}
\usepackage[utf8]{inputenc}
      
\DeclareFontFamily{U}{BOONDOX-calo}{\skewchar\font=45 }
\DeclareFontShape{U}{BOONDOX-calo}{m}{n}{ <-> s*[1.05] BOONDOX-r-calo}{}
\DeclareFontShape{U}{BOONDOX-calo}{b}{n}{ <-> s*[1.05] BOONDOX-b-calo}{}
\DeclareMathAlphabet{\mathcalboondox}{U}{BOONDOX-calo}{m}{n}

\DeclareFontFamily{OT1}{pzc}{}
\DeclareFontShape{OT1}{pzc}{m}{it}{<-> s * [1.10] pzcmi7t}{}
\DeclareMathAlphabet{\mathpzc}{OT1}{pzc}{m}{it}
\DeclareMathOperator{\UnitStep}{U}

\newcommand{\Metric}{g}
\newcommand{\Continuity}{\ensuremath{\mathcal{C}}}
\newcommand{\ContinuityHodgeMin}{\ensuremath{\Continuity^2}}
\newcommand{\ContinuouslyDifferentiable}{\ensuremath{\Continuity^1}}
\newcommand{\ContinuityStokes}{\ensuremath{\Continuity^1}}
\newcommand{\ContinuityHairy}{\ensuremath{\Continuity^2}}
\newcommand{\ContinuityCurrentMin}{\ensuremath{\Continuity^3}}
\newcommand{\ContinuityMin}{\ContinuityCurrentMin}

\newcommand{\Rcyl}{\ensuremath{R}}
\newcommand{\Lcyl}{\ensuremath{L}}

\newcommand{\Surf}{\ensuremath{\Manifold}}
\newcommand{\surf}{\mathscr{S}}
\newcommand{\Manifold}{\ensuremath{\mathcal{S}}}
\newcommand{\genus}{\ensuremath{\mathfrak{g}}}
\newcommand{\Energy}{\ensuremath{\mathcal{E}}}
\newcommand{\hell}{\ensuremath{\mathcalboondox{h}}}
\newcommand{\Hell}{\ensuremath{\mathcal{H}}}
\newcommand{\Jfirst}{\ensuremath{\mathcal{J}}}
\newcommand{\Harmonic}{\ensuremath{\boldsymbol{\Omega}}}

\newcommand{\rhoP}{\ensuremath{\varrho_\mathrm{P}}}
\newcommand{\HarmonicP}{\ensuremath{\boldsymbol{\Harmonic}_\mathrm{P}}}

\newcommand{\Em}{\ensuremath{\mathrm{Em}}}
\newcommand{\Sh}{\ensuremath{\mathrm{Sh}}}
\newcommand{\NI}{\ensuremath{\mathrm{NI}}}

\newcommand{\ZETA}{\ensuremath{\zeta}}
\newcommand{\ZETAE}{\ensuremath{\ZETA_\Em}}
\newcommand{\ZETAER}{\ensuremath{{\ZETA_R}}}
\newcommand{\ZETAT}{\ensuremath{\ZETA_\Sh}}
\newcommand{\ZETAPot}{\ensuremath{\ZETA_\mathrm{P}}}
\newcommand{\ZETANI}{\ensuremath{\ZETA_\NI}}
\newcommand{\ZETAPotNI}{\ensuremath{{\ZETA}_{\mathrm{P}\NI}}}

\newcommand{\Volume}{\ensuremath{\mathcal{V}}}
\newcommand{\Contour}{\ensuremath{\ell}}
\newcommand{\Kernel}{\ensuremath{\boldsymbol{\mathbb{K}}}}
\newcommand{\GreenNP}{\ensuremath{\mathcal{G}_\mathrm{N}}}
\newcommand{\GreenF}{\ensuremath{\mathcal{G}_\mathrm{F}}}
\newcommand{\A}{\ensuremath{\boldsymbol{A}}}
\newcommand{\tang}{\ensuremath{\mathcal{S}}}
\newcommand{\BthF}{\ensuremath{d}}
\newcommand{\Inner}{\ensuremath{0}}
\newcommand{\Outer}{\ensuremath{1}}

\newcommand{\phiPot}{\ensuremath{\phi_\mathrm{P}}}
\newcommand{\PHI}{\ensuremath{\chi}}
\newcommand{\PhiPot}{\ensuremath{\chi_\mathrm{P}}}

\newcommand{\dHelicity}{\ensuremath{\dot\Hell}}
\newcommand{\Helicity}{\ensuremath{\Hell}}
\newcommand{\dHelicityV}{\ensuremath{{\dot\Helicity_\Vol}}}
\newcommand{\dHelicityS}{\ensuremath{{\dot\Helicity_\Surf}}}
\newcommand{\HelicityS}{\ensuremath{\Helicity_\Surf}}

\newcommand{\ENI}{\ensuremath{\E_\NI}}
\newcommand{\In}{\ensuremath{B_n}}

\newcommand{\helicity}{\ensuremath{\hell}}
\newcommand{\HelicityBF}{\ensuremath{\Hell}}
\newcommand{\dHelicityBF}{\ensuremath{\dot\HelicityBF}}
\newcommand{\dHelicityBFS}{\ensuremath{{\dot\HelicityBF_\Surf}}}
\newcommand{\dHelicityBFV}{\ensuremath{{\dot\HelicityBF_\Vol}}}
\newcommand{\HelicityBFS}{\ensuremath{\HelicityBF_\Surf}}

\newcommand{\XI}{\ensuremath{\xi}}

\newcommand{\nhat}{\ensuremath{\widehat{\boldsymbol{n}}}}

\newcommand{\uhat}{\ensuremath{\widehat{\boldsymbol{u}}}}

\newcommand{\rhat}{\ensuremath{\widehat{\boldsymbol{r}}}}
\newcommand{\um}{\ensuremath{\widehat{\boldsymbol{m}}}}
\newcommand{\ut}{\ensuremath{\widehat{\boldsymbol{t}}}}
\newcommand{\rhohat}{\ensuremath{\widehat{\boldsymbol{\rho}}}}
\newcommand{\thetahat}{\ensuremath{\widehat{\boldsymbol{\theta}}}}

\newcommand{\xhat}{\ensuremath{\widehat{\boldsymbol{x}}}}
\newcommand{\yhat}{\ensuremath{\widehat{\boldsymbol{y}}}}
\newcommand{\zhat}{\ensuremath{\widehat{\boldsymbol{z}}}}
\newcommand{\J}{\ensuremath{\boldsymbol{J}}}
\newcommand{\ARef}{\ensuremath{\boldsymbol{A}_\mathrm{R}}}
\newcommand{\ERef}{\ensuremath{\boldsymbol{E}_\mathrm{R}}}
\newcommand{\psiRef}{\ensuremath{\psi_\mathrm{R}}}
\newcommand{\APot}{\ensuremath{\boldsymbol{A}_\mathrm{P}}}

\newcommand{\bpot}{\ensuremath{P}}

\newcommand{\BPot}{\ensuremath{\boldsymbol{\bpot}}}

\newcommand{\APotC}{\ensuremath{\boldsymbol{A}_\mathrm{PC}}}
\newcommand{\EPot}{\ensuremath{\boldsymbol{E}_\mathrm{P}}}
\newcommand{\epot}{\ensuremath{\boldsymbol{\Sigma}_\mathrm{P}}}
\newcommand{\Tepot}{\ensuremath{\boldsymbol{\Sigma}_{\mathrm{P}\NI}}}
\newcommand{\psiPot}{\ensuremath{\psi_\mathrm{P}}}
\newcommand{\eLL}{\ensuremath{l}}
\newcommand{\ELL}{\ensuremath{\boldsymbol}{\eLL}}
\newcommand{\gsign}{}
\newcommand{\gauge}{\ensuremath{\Lambda}}
\newcommand{\GaugeP}{\gauge_\mathrm{P}}
\newcommand{\GaugePp}{\gauge_\mathrm{P}'}

\newcommand{\Link}{\ensuremath{\mathcal{L}}}
\newcommand{\Flux}{\ensuremath{\Psi}}
\newcommand{\Vol}{\ensuremath{\Volume}}
\newcommand{\Cont}{\ensuremath{\Contour}}
\newcommand{\B}{\ensuremath{\boldsymbol{B}}}
\newcommand{\BRef}{\ensuremath{\boldsymbol{B}_\mathrm{R}}}
\newcommand{\Bcl}{\ensuremath{\boldsymbol{B}_\mathrm{cl}}}
\newcommand{\E}{\ensuremath{\boldsymbol{E}}}

\newcommand{\f}{\ensuremath{\boldsymbol{f}}}

\newcommand{\F}{\ensuremath{\boldsymbol{F}}}
\newcommand{\g}{\ensuremath{\boldsymbol{g}}}

\newcommand{\h}{\ensuremath{\boldsymbol{h}}}
\newcommand{\grad}{\ensuremath{\boldsymbol{\nabla}}}
\newcommand{\gradp}{\ensuremath{\boldsymbol{\nabla}_\Manifold}}
\newcommand{\Lapp}{\ensuremath{{\nabla}_\Manifold^2}}

\newcommand{\x}{\ensuremath{\boldsymbol{x}}}

\renewcommand{\v}{\ensuremath{\boldsymbol{v}}}
\newcommand{\cross}{\ensuremath{\boldsymbol{\times}}}

\newtheorem{theorem}{Theorem}
\newtheorem{corollary}{Corollary}

\makeatletter
\def\input@path{{./picts/}}
\makeatother

\graphicspath{{./picts/}}

\received{2019 April 4}
\revised{2019 June 5}
\accepted{2019 June 10}
\submitjournal{ApJ}

\shorttitle{On Solar Helicity and Energy Transport}
\shortauthors{Schuck and Antiochos}

\begin{document}

\title{Determining the Transport of Magnetic Helicity and Free Energy
   in the Sun's Atmosphere}

\correspondingauthor{Peter W. Schuck}
\email{peter.schuck@nasa.gov}

\author[0000-0003-1522-4632]{Peter W. Schuck}
\author[0000-0003-0176-4312]{Spiro K. Antiochos}
\address{Heliophysics Science Division, NASA Goddard Space Flight Center 
8800 Greenbelt Rd., Greenbelt, MD 20771, USA}



\begin{abstract}
  The most important factors determining solar coronal activity are
  believed to be the availability of magnetic free energy and the
  constraint of magnetic helicity conservation. Direct measurements of
  the helicity and magnetic free energy in the coronal volume are
  difficult, but their values may be estimated from measurements of
  the helicity and free energy transport rates through the
  photosphere. We examine these transport rates for a topologically
  open system such as the corona, in which the magnetic fields have a
  nonzero normal component at the boundaries, and derive a new formula
  for the helicity transport rate at the boundaries. In addition, we
  derive new expressions for helicity transport due to flux
  emergence/submergence versus photospheric horizontal motions. The
  key feature of our formulas is that they are manifestly gauge
  invariant. Our results are somewhat counterintuitive in that only
  the lamellar electric field produced by the surface potential
  transports helicity across boundaries, and the solenoidal electric
  field produced by a surface stream function does not contribute to
  the helicity transport. We discuss the physical interpretation of
  this result. Furthermore, we derive an expression for the free
  energy transport rate and show that a necessary condition for free
  energy transport across a boundary is the presence of a closed
  magnetic field at the surface, indicating that there are current
  systems within the volume. We discuss the implications of these
  results for using photospheric vector magnetic and velocity field
  measurements to derive the solar coronal helicity and magnetic free
  energy, which can then be used to constrain and drive models for
  coronal activity.

\end{abstract}



\makeatletter
\def\smallunderbrace#1{\mathop{\vtop{\m@th\ialign{##\crcr
   $\hfil\displaystyle{#1}\hfil$\crcr
   \noalign{\kern3\p@\nointerlineskip}%
   \tiny\upbracefill\crcr\noalign{\kern3\p@}}}}\limits}
\makeatother
\section{Introduction} \label{sec:intro}
The Sun's atmosphere, the chromosphere--corona, is characterized by
ubiquitous bursts of energy release ranging from sporadic giant
coronal mass ejections (CMEs) and eruptive flares with energies up to
$10^{33}$ erg \cite[e.g.,][]{Forbes2000,Aulanier2013} to ever-present
nanoflares with energies of the order of $10^{24}$ erg or less
\cite[]{Parker1988,Klimchuk2006}.  All these forms of solar activity
share a common underlying scenario. First, magnetic energy is injected
into the corona either directly by the convective motions of the
high-$\beta$ photosphere acting on the low-$\beta$ coronal magnetic
field, or by the emergence of pre-stressed magnetic fields through the
photosphere and into the corona. Second, the energy builds up until
some fast process such as an ideal or resistive instability releases
the magnetic energy, transferring it to the plasma in the form of
heating, mass motion, and/or energetic particle acceleration. The
amount of energy released (the size of the coronal event) depends on
the dynamical constraints of the system. For the corona, there are two
general dynamical constraints due primarily to the time scales for
coronal evolution. Because the Alfv\'{e}n speed $V_\mathrm{A}$ is three
orders of magnitude or more faster in the corona than in the photosphere,
the magnetic flux through the photosphere is expected to be
approximately constant during an energy release event. Accordingly,
the minimum energy state for the coronal field is the unique potential
(current-free) field with that photospheric normal flux
distribution. This result implies that the maximum available energy
for release is only the ``free'' energy, defined as the difference
between the energy of the coronal field and the energy of this
potential state. \par
However, even this free energy overestimates the available energy for a
coronal event, because the system is further constrained by the magnetic
topology.  In the corona, the time scale for resistive dissipation
$\tau_\mathrm{d}\sim{L}^2/\eta$ is very large compared to observed
evolutionary time scales $\tau_\mathrm{t}=V_\mathrm{A}/L$ where $L$ is the
scale of the system and $\eta$ is the magnetic diffusivity.  In other words,
the corona has a very large Lundquist number
$\tau_\mathrm{d}/\tau_\mathrm{t}=L\,V_\mathrm{A}/\eta\simeq10^9$ which means
that magnetic disturbances can cross the system long before they can be
diffused.  In this situation, the coronal evolution must be primarily ideal,
which imposes additional constraints on the final state.  If the evolution
were perfectly ideal, then every field line would be a constant of the motion,
leading to an infinite number of constraints, and the final state along with
the energy available for release through an ideal evolution would be highly
constrained by the initial magnetic topology.  Fortunately, this is not the
case; because the corona does have finite resistivity, magnetic
reconnection can occur wherever current sheets form. For this reason,
reconnection is widely believed to be the dominant energy release mechanism in
most forms of solar activity, including CMEs, flares
\cite[e.g.,][]{Karpen2012}, jets \cite[e.g.][]{Wyper2017}, and coronal heating
\cite[]{Klimchuk2006}. Reconnection, however, preserves the topological
constraint of total magnetic helicity, which is described in detail directly
below. The implication is that the only relevant constraints for determining
the energy available for explosive coronal activity are the magnetic free
energy and the magnetic helicity. Accurate measurement of these quantities
would greatly enable both the understanding and prediction of solar activity, but
this would require measuring the full vector magnetic field in the coronal
volume, which is not experimentally feasible at present. Because the free energy
and helicity are injected into the corona through the photosphere, a more
observationally feasible approach to constraining coronal energy release would
be to measure the time-integrated transport rate of free energy and helicity
through the photosphere. The goal of this paper is to derive rigorous
expressions that are straightforward to understand physically and can be used
effectively for measuring the free energy and helicity transport with
photospheric observations. \par
\subsection{Magnetic Helicity}
While energy is undoubtedly the best known and most widely used quantity 
throughout physics, magnetic helicity is not generally well known
or understood; therefore, we start with a
discussion of its salient properties. Magnetic helicity is simply the
quantification of magnetic flux linkages. The topological concept of
linking numbers \cite[]{GaussWerkeV,Calugareanu1959,White1969} and
their connections with helicity introduced by Lord Kelvin \cite[then
Sir William][]{Thomson1868}, plays a fundamental role in science from
astrophysics \cite[]{Woltjer1958} to fluid dynamics
\cite[]{Moffatt1969}, and biochemistry
\cite[]{Fuller1971,Crick1976}. The Gauss\footnote{See note dated 1833
  January 22 in \cite{GaussWerkeV} as discussed extensively in
  \cite{Ricca2011}.} linking number $\Link_{ij}$ represents the number
of times a closed curve $i$ encircles a second closed curve $j$ in
space $\left(i\neq{j}\right)$. The
\citeauthor{Calugareanu1959}-\citeauthor{White1969} linking number
$\Link_i$ represents the number of crossings of the edges of a ribbon
curve, which may be further decomposed,
$\Link_{i}\equiv\mathrm{Tw}_i+\mathrm{Wr}_i$ where $\mathrm{Tw}_i$ is
the twist and $\mathrm{Wr}_i$ is the writhe \cite[]{Calugareanu1959,Pohl1968,White1969}.
Twist is a measure of how much a ribbon is twisted about its own
axis and writhe is a measure of the nonplanarity of the ribbon axis
itself \cite[]{Dennis2005}.
Correspondingly, helicity measures the total linkage of magnetic
field lines in a plasma or vortex lines in a fluid, and is given by
the simple expression \cite[see review
  by][]{Moffatt2014}
\begin{subequations}
\begin{equation}
H=\int_\Vol{d^3x}\,\A\cdot\grad\cross\A,\label{eqn:Helicity}
\end{equation}
where $\Vol$ is a connected volume, $\A$ is the vector potential, and
$\B=\grad\cross\A$ relates the vector potential to the observable
magnetic field $\B$. For fluid flow, $\A$ is replaced by the fluid
velocity $\v$ and $\boldsymbol{\omega}=\grad\cross\v$ is the
vorticity. Note that in contrast with energy, $H$ is not positive
definite. Helicity is a pseudo-scalar that changes sign under a
transformation from a right-handed to a left-handed frame of reference,
i.e., $H\neq\reflectbox{$H$}=-H$, where $H$ and \reflectbox{$H$}
represent the helicity in the right-handed and left-handed coordinate
systems. \par
For a set of $N$ closed flux tubes or vortices, the
expression for helicity,~(\ref{eqn:Helicity}) can be shown to be formally
related to the linking numbers
\cite[]{Moffatt1969,Berger1984a,Moffatt1992,Ricca2002}
\begin{equation}
H=\sum_{i}^N\Link_i\,\Flux_i+2\,\sum_{i\neq{j}}^{N,N}\Link_{ij}\,\Flux_i\,\Flux_j,\label{eqn:Helicity:Link}
\end{equation}
\end{subequations}
where $\Flux_i$ is the toroidal flux of each elemental flux tube or
the circulation of each vortex. The first term is identified as the
``self-helicity'' and the second term as the ``mutual helicity.''
Despite the broad relevance of helicity, the concept is fraught with
practical challenges in application to astrophysical plasmas. The
magnetic helicity is only defined in terms of the integral over a
closed volume $\Vol$ bounded by the surface $\Surf$. Thus, a complete
knowledge of $\A$ and $\B$ in $\Vol$ is required to compute
$H$. Furthermore, the field must satisfy strong constraints in order
for $H$, as defined by~(\ref{eqn:Helicity}), to be physically meaningful.
Eichinvarianz or gauge invariance, a fundamental principle of modern physics,
is a manifestation of the unobservability of the electric and magnetic 
potentials $\left(\psi/c,\A\right)$ \cite[]{Weyl1919,Jackson2001}. Gauge-transforming~(\ref{eqn:Helicity}) with $\A\rightarrow\A+\grad\gauge$ and
using~(\ref{eqn:Div_psi_f}) and the Gauss\---Ostrogradsky
theorem,~(\ref{eqn:Gauss}) produces
\begin{equation}
H\rightarrow{H}-\oint_\Surf{dS}\,\nhat\cdot\left(\gauge\,\B\right),
\end{equation}
where $\nhat$ is the inwardly\footnote{This is simply so that
  $\nhat$ points in the radial direction in the photosphere and into
  the coronal volume, which is conceptually convenient for practical
  solar calculations.} directed normal to the volume $\Vol$.
Therefore, $H$ is uniquely defined only for a magnetically isolated
plasma, where the normal component of $\B$ vanishes on all bounding
surfaces $\left.\B\cdot\nhat\right|_\Surf=0$. (Hereafter, the term
``isolated'' is equivalent to ``magnetically isolated.'')  \par
For ideal plasma motions in an isolated system, the quadratic
invariant $H$ is conserved, notwithstanding complex dynamical
evolution of the system.  Just over 60 yr ago, \cite{Woltjer1958}
demonstrated that helicity is preserved for the evolution of an
isolated ideal low-$\beta$ plasma and that linear force-free fields
are the minimum energy state of this system with a prescribed value of
$H$. If the evolution were truly ideal, there would be an infinite
number of constraints, in which case the helicity would not be very
useful. \cite{Taylor1974}, however, conjectured that even for a weakly
dissipative isolated plasma, the magnetic helicity $H$ is a robust
invariant, meaning that $H$ is approximately conserved even in the
presence of dissipation and is insensitive to the details of the
nonequilibrium mechanisms involved in the relaxation to a lower energy
state.  If correct, \citeauthor{Taylor1974}'s (\citeyear{Taylor1974})
bold conjecture constrains the final state of the field and plasma
even under large-scale reconfiguration, due to magnetic reconnection.
\citeauthor{Taylor1974}'s hypothesis has been largely verified in
laboratory devices, such as reverse field pinches, where conducting
walls with $\left.\B\cdot\nhat\right|_\Surf=0$ isolate the plasma
\cite[]{Butt1976}. Furthermore, helicity conservation has been
observed in numerous MHD simulations with high Lundquist number
\cite[e.g.,][]{MacNeice2004,Knizhnik2017}. The magnetic helicity $H$
is also a robust invariant in periodic geometries in the absence of a
mean magnetic field. However, the presence of a mean field in a
periodic system destroys the topological concept of linkage and $H$ is
no longer a robust invariant \cite[]{Berger1997,Watson2001}. \par
\subsection{Application to the Corona}
Astrophysical plasmas, and in particular the solar corona, can rarely
be considered isolated or periodic. To address this situation,
\cite{Berger1984a} and \cite{Finn1985} contemporaneously introduced
the concept of relative helicity $\Helicity$, for systems with
magnetic fields having $\left.\B\cdot\nhat\right|_\Surf\neq0$ at one
or more system boundaries.  \cite{Berger1984a} and \cite{Finn1985}
defined relative helicity $\Helicity$ in terms of an arbitrary
reference magnetic field $\BRef$ that satisfies
$\left.\BRef\cdot\nhat\right|_\Surf=\left.\B\cdot\nhat\right|_\Surf$
on all surfaces, and with no constraint on the vector
potentials. The two definitions are equivalent.\label{ref:RefField} \par
The magnetic field lines that thread the boundaries are defined as ``open''
fields. For the corona, the natural boundary is the photosphere, in
which case the normal flux certainly does not vanish there; on the
contrary, every field line in the corona is expected to connect to the
photosphere. Note that the term ``open'' is used very
  differently here than in the usual solar physics terminology where
  ``open'' denotes coronal flux that connects to the photosphere at
  only one end, as in a coronal hole. In the terminology of this paper
  and of the standard helicity literature, all the coronal
    flux is open, because it intersects the boundary, at least,
  once. \par
An important point that must be emphasized is that both the helicity
and the helicity transport rate are defined rigorously only for a
finite closed volume. This implies that measuring the magnetic field
at a single surface, the photosphere, is not sufficient to determine
the coronal helicity or its evolution; however, it is not possible with
present instrumentation to measure the field accurately in the hot
corona. As a result, the approximation that is generally made is to
consider a coronal volume that is finite but sufficiently large to
contain an active region, for example, and then to assume that the
helicity within this volume is due only to injection or loss through
the photosphere.  If a CME/eruptive flare occurs in the active region,
the assumption clearly breaks down, but even without eruptions,
reconnection may transport helicity into or out of a particular
coronal volume \cite[]{Antiochos2013}. On the other hand, given the
available solar observations, the assumption to ignore all
boundaries but the photosphere is virtually unavoidable.  \par
With the results of \cite{Berger1984a}, \cite{Berger1984b} extended
the \citeauthor{Taylor1974} conjecture to a coronal system and argued
that relative helicity $\HelicityBF$ is a robust invariant for a flare
or other form of fast solar activity. Of course, solar eruptions do
change the relative helicity in the lower corona by ejecting
twisted field out into the heliosphere. The relative helicity is also
changed by the transport of twisted field across the photospheric
surface $\Surf$ and the twisting and tangling of footpoints by motion
in the surface which can link coronal field lines. \cite{Berger1984a}
derived a simple expression for the relative helicity transport rate for a
closed volume $\Vol$ bounded by a surface $\Surf$
\begin{equation}
\dot\HelicityBF\equiv\frac{\partial\HelicityBF}{\partial
  t}=-\overbrace{2\,c\,\int_{\Vol}d^3x\,\E\cdot\B}^{\dHelicityBFV}-\overbrace{2\,c\,\oint_{\Surf}dS\,\nhat\cdot\left(\APotC\cross\E+\frac{1}{c}\,\frac{\partial\phiPot}{\partial
  t}\,\APotC\right)}^{\dHelicityBFS},\label{eqn:BergerField1984}  
\end{equation}  
where $\E$ is the electric field, $\phiPot$ is
the magnetic scalar potential,
\begin{equation}
\BPot=-\grad\phiPot,\label{eqn:Potential}
\end{equation}
determined from
\begin{subequations}
\begin{align}
\nabla^2\phiPot&=0\qquad\x\in\Vol,\label{eqn:Laplace}\\
\left.\B\cdot\nhat\right|_\Surf&=\left.\BPot\cdot\nhat\right|_\Surf=-\left.\frac{\partial\phiPot}{\partial n}\right|_\Surf,\label{eqn:Laplacer}
\end{align}  
\end{subequations}
and $\APotC$ is the unique vector potential of the potential magnetic
field determined by
\begin{subequations}
\begin{equation}
\BPot=\grad\cross\APotC,\label{eqn:BP:APC}  
\end{equation}  
in the Coulomb gauge
\begin{equation}
\grad\cdot\APotC=0,\label{eqn:Coulomb}
\end{equation}  
with the boundary condition that the normal component of the vector
potential vanishes at the surface $\Surf$
\begin{equation}
\left.\nhat\cdot\APotC\right|_\Surf=0.\label{eqn:APC:Neumann}
\end{equation}
\end{subequations}
This boundary condition can always be satisfied by adding the
appropriate gradient of a harmonic function to any $\APotC$
\cite[]{Cantarella2002}. Relationships~(\ref{eqn:Coulomb})
and~(\ref{eqn:APC:Neumann}) define an intrinsically solenoidal vector
$\APotC=\grad\cross\f$ according to Helmholtz's theorem (see
Appendix~\ref{sec:Helmholtz} and \cite{Kemmer1977}).  \par
The first integral $\dHelicityBFV$ in
  (\ref{eqn:BergerField1984}) represents the generation and
  dissipation of helicity in the volume $\Vol$ and the second integral
  $\dHelicityBFS$ represents the helicity transport rate across
  the boundary $\Surf$. Generally, the electric field $\E$ may
  comprise both ideal and nonideal terms through a generalized Ohm's
  law. Indeed, $\E$ may be nonideal in highly localized regions of the
  volume permitting reconnection but ideal throughout most of the
  volume.  For this scenario, the generation and dissipation of
  helicity in the volume $\dHelicityBFV$ can be ignored, and
  $\dHelicityBF\simeq\dHelicityBFS$
  \cite[]{Taylor1974,Berger1984b}. Nonetheless, if $\E$ can be
  measured directly on the surface, then $\dHelicityBFS$ is the
  surface helicity transport rate regardless of the nonideal
  properties of the evolution on the surface or in the volume. However,
  if the evolution in the volume is substantially nonideal, then
  $\dHelicityBF\neq\dHelicityBFS$.
Using the ideal magnetohydrodynamics (MHD) Ohm's law,
\begin{equation}
  \E=-\frac{1}{c}\,\v\cross\B,\label{eqn:Ohm}
\end{equation}
where $\v$ is the single fluid bulk plasma velocity, the surface helicity
transport rate simplifies to
\begin{equation}
\boxed{\frac{\partial\HelicityBFS}{\partial t}=2\,\oint_{\Surf}dS\,\underbrace{\left(\APotC\cdot\B\right)\,v_n}_{\mathrm{emergence}}-2\,\oint_{\Surf}dS\,
\underbrace{\left(\APotC\cdot\v\right)\,B_n}_{\mathrm{shearing}}}\label{eqn:Berger1984}
\end{equation}
where the subscript ``$n$'' indicates the normal component of a
vector. \cite{Berger1984b,Berger1999}, and
\cite{Berger2000} have identified these two terms as the emergence and
shearing terms \cite[see
  also][]{Kusano2002a,Kusano2003,Yamamoto2009}. The emergence term is,
in principle, due to the emergence through the photosphere of twisted
field from the solar interior, and the shearing term is presumed to
represent the helicity generated by the shearing and twisting of field
lines by tangential motions on the photospheric surface.  There is a
long history of estimating each term independently
\cite[]{Chae2001b,Nindos2003a,Pevtsov2003,Pariat2005,Demoulin2007a,Liu2014}. However,
there are several concerns with the interpretation and use of this
equation:
\begin{enumerate}
\item The terms in~(\ref{eqn:Berger1984}) are neither in part, nor in
  whole, manifestly gauge invariant. Indeed,~(\ref{eqn:Berger1984}) is
  expressed explicitly in the Coulomb
  gauge,~(\ref{eqn:BP:APC})\--(\ref{eqn:APC:Neumann}). While the
  combination of the emergence and shearing terms with other terms
  implicitly contained in~(\ref{eqn:Berger1984}) is gauge invariant
  (see~(\ref{eqn:BergerField}) in \S~\ref{sec:Relative:Helicity}),
  neither term is independently gauge invariant, and their physical
  interpretation is ambiguous. If the separation of helicity transport
  into emergence and shearing is physically valid, there should be a
  gauge-invariant expression with that interpretation. Consequently,
  without further justification, the two surface integrals generally
  cannot be considered independently as observables.
\item Central to the interpretation of relative helicity is the role
  of the flux
  threading the bounding surface. If the system is isolated,
  $\left.\B\cdot\nhat\right|_\Surf=0$, then $\Helicity$ should be a robust
  invariant. In other words, helicity transport across the surface $\Surf$
  must require a finite 
  $\left.\B\cdot\nhat\right|_\Surf$; however,~(\ref{eqn:Berger1984}) is
  only indirectly related to $\left.\B\cdot\nhat\right|_\Surf$ through
  the tangential components of $\APotC$.
\item \cite{Prior2014} state at the top of p. 2, that when ``the
  boundary conditions $\left.\B\cdot\nhat\right|_\Surf$ are changing
  in time, ..... the evolution of the relative helicity will mix up
  both real topological changes in $\B$ and those simply due to the
  change of $\BPot$.''
\item While the combination of the emergence and shearing terms is
  independent of the flow parallel to the magnetic field because they
  result from~(\ref{eqn:Ohm}), each term, by itself, is dependent on
  the value of the parallel flow $v_\parallel$. Thus, to have any hope
  of estimating each term independently, one has to formally subtract
  the flow parallel to the magnetic field
  $\v_\perp\rightarrow\v-\left(\v\cdot\B\right)\,\B/\left(\B\cdot\B\right)$. However,
  if one subtracts this flow then, for example, $v_{\perp{n}}$
  explicitly depends on the values of the flow tangent to the surface,
  complicating the interpretation of~(\ref{eqn:Berger1984})!
\end{enumerate}  
We conclude from this discussion that while~(\ref{eqn:Berger1984}) is
a fully accurate expression for the helicity transport rate through a
closed boundary, it does not afford a clear physical
interpretation. Consequently, we undertake below to derive new
expressions for both the helicity and free energy transport into the
corona that allow for physically intuitive interpretation and
straightforward calculation from data. The organization of the paper
is as follows. In \S~\ref{sec:Relative:Helicity}, we first discuss key
features of the \cite{Finn1985} and \cite{Berger1984a} relative
helicity formulas that are needed for our derivations. In
\S~\ref{sec:Rate}, we revisit the expression for the relative helicity
transport rate derived by \cite{Berger1984a} and propose an
alternative expression that is manifestly gauge invariant on the
surface. \S~\ref{sec:Implications} discusses the implications of the
new expression: its gauge invariance, its equivalence with the
\cite{Berger1984a} results, and its relationship with the lamellar
electric field on the boundary. \S~\ref{sec:Emerge:Shear} develops the
gauge-invariant emerging and shearing terms and briefly discusses
nonideal effects.  In \S~\ref{sec:Energy}, we turn to the free energy
and derive an expression for the free energy transport rate, which
shows that the transport of free energy across the surface $\Surf$
requires electric currents in the enclosed volume\----in other words, the
coronal field must be nonpotential.  Appendix~\ref{sec:EPot:Revisited}
derives a general expression for the reference electric field.  The
other Appendices provide some vector relationships in a volume
(\S~\ref{sec:Vector}), a brief introduction to vector calculus on a
surface (\S~\ref{sec:Surface}), some integral
relationships~(\S~\ref{sec:Integral}), and the Helmholtz theorem in
volumes and the Helmholtz\---Hodge theorem on surfaces
(\S~\ref{sec:Helmholtz})\----all of which are necessary for this
paper. Given the detailed nature of some of the expressions derived in
this paper, we tried to include all the material required for the
reader to understand the derivations without having to consult outside
sources.  \par
\section{The Relative Helicity\label{sec:Relative:Helicity}}
As discussed in \S~\ref{sec:intro}, astrophysical plasmas can rarely be
considered isolated. To address the field lines threading bounding
surfaces of a plasma, \cite{Finn1985} introduced the relative
helicity formula (equivalent to \cite{Berger1984a})
\begin{subequations}
\begin{equation}
\Helicity\equiv\int_\Vol{d^3x}\,\helicity,\label{eqn:Finn}
\end{equation}
where
\begin{equation}
\helicity\equiv\left(\A+\ARef\right)\cdot\left(\B-\BRef\right),\label{eqn:Finn-Antonsen}
\end{equation}
\end{subequations}
is the helicity integrand, and $\BRef$ and $\ARef$ are arbitrary
reference fields. While using the term ``density'' for the helicity integrand $\helicity$ is
tempting, doing so would be misleading. Unlike a true
density, $\helicity$ is not a physically meaningful quantity, because helicity
is a topological property that cannot be localized. The integrand $\helicity$
is not an observable; its value is at the discretion of the
observer through the gauge freedom.
Any observable with respect to $\helicity$
requires an integral  over a spatial volume $\Vol$ with appropriate boundary
conditions on the vector potentials and magnetic fields.  \par
Magnetic fields are assumed to be solenoidal which implies
that 
\begin{equation}
\oint_\Surf{dS}\,\nhat\cdot\B=0,\label{eqn:Solenoidal}
\end{equation}
 for every and any
  closed surface, or equivalently, that there are no magnetic monopoles
anywhere,
\begin{subequations}
\begin{equation}
 \grad\cdot\B=\grad\cdot\BRef=0\qquad\mbox{(everywhere).}\label{eqn:Monopoles}
\end{equation}
The absence of monopoles permits the magnetic fields $\B$ and $\BRef$ to be
expressed in terms of vector potentials
\begin{equation}
\B=\grad\cross\A,\qquad\mbox{and}\qquad\BRef=\grad\cross\ARef.\label{eqn:BA}
\end{equation}
\end{subequations}\par
The gauge invariance of the relative helicity is proven by gauge-transforming~(\ref{eqn:Finn})\--(\ref{eqn:Finn-Antonsen}) with $\A\rightarrow\A+\grad\gauge'$ and
$\ARef\rightarrow\ARef+\grad\gauge''$.  Defining $\gauge=\gauge'+\gauge''$
\begin{equation}
\Helicity'=\int_\Vol{d^3x}\,\left(\A+\ARef+\grad\gauge\right)\cdot\left(\B-\BRef\right),\label{eqn:Helicity:Explicit}
\end{equation}
and using the original definition,~(\ref{eqn:Finn})\--(\ref{eqn:Finn-Antonsen}), this becomes
\begin{equation}
\Helicity'=\Helicity+\int_\Vol{d^3x}\,\grad\gauge\cdot\left(\B-\BRef\right),
\end{equation}
Applying~(\ref{eqn:Div_psi_f}) produces
\begin{equation}
\Helicity'=\Helicity+\int_\Vol{d^3x}\,\grad\cdot\left[\gauge\,\left(\B-\BRef\right)\right]-\int_\Vol{d^3x}\,\gauge\,\grad\cdot\left(\B-\BRef\right).
\end{equation}
The second term can be converted to an integral of the surface bounding
$\Vol$ with~(\ref{eqn:Gauss})
\begin{equation}
  \Helicity'=\Helicity-\oint_{\Surf}dS\,\nhat\cdot\left[\gauge\,\left(\B-\BRef\right)\right]-\int_\Vol{d^3x}\,\gauge\,\grad\cdot\left(\B-\BRef\right).\label{eqn:BD:Bn}
\end{equation}
If there are no monopoles,~(\ref{eqn:Monopoles}) and the normal component of
$\B$ and $\BRef$ match on the boundary,
\begin{subequations}
\begin{equation}
\left.\nhat\cdot\left(\B-\BRef\right)\right|_\Surf=\left.\nhat\cdot\grad\cross\left(\A-\ARef\right)\right|_\Surf=0,\label{eqn:Boundary:Bn}
\end{equation}
then~(\ref{eqn:Finn})\--(\ref{eqn:Finn-Antonsen}) is gauge invariant
$\Helicity'=\Helicity$, with respect to
independent gauge transformations of $\A$ and $\ARef$. Note,
however, that~(\ref{eqn:Boundary:Bn}) does imply some restrictions on
the relative values of~$\A$ and~$\ARef$ at the boundary. 
The null space of $\nhat\cdot\grad\cross\f$ is $\f=\gradp\gauge+\tau\,\nhat$
on $\Surf$ (see
Appendix~\ref{sec:Helmholtz:Surface} Equations~(\ref{eqn:null_space_i})\--(\ref{eqn:null_space_f})). Here,
the subscript ``$\Surf$'' indicates that the gradient (or vector) tangent to
the surface $\Surf$ (see Appendix~\ref{sec:Surface} and in
particular~(\ref{eqn:fS})), and $\gradp$ is the surface gradient operator
defined in~(\ref{eqn:GradS_psi}).  Consequently, the boundary
condition~(\ref{eqn:Boundary:Bn}) implies that the tangential components
of~$\A$ and~$\ARef$ must be equivalent to within the gradient of a scalar on
the surface~$\Surf$,
\begin{equation}
 \A-\ARef=\tau\,\nhat+\gradp\gauge\qquad\in\Surf.\label{eqn:Boundary:A1}
\end{equation}
\end{subequations}
\par
Note that gauge invariance for the \citeauthor{Finn1985}
formula~(\ref{eqn:Finn})\--(\ref{eqn:Finn-Antonsen}) is a direct result of the
fact that there is no relative flux $\B-\BRef$ threading the bounding
surface~(\ref{eqn:Boundary:Bn}). Thus, we again see that the normal component
of the magnetic flux at the bounding surface plays a crucial role for defining
a rigorous helicity observable. \citeauthor{Finn1985}'s insight into defining
relative helicity in terms of $\B-\BRef$ is fundamentally related to the
Helmholtz theorem~(see Appendix~\ref{sec:Helmholtz:Volume}) and an isolated
system. The vector field $\B$, while solenoidal
$\left(\grad\cdot\B=0\right)$ is not intrinsically solenoidal for an
arbitrary volume $\Vol$ bounded by $\Surf$, because $\B$ generally admits a
mixed description $\B=\grad\cross\f-\grad\xi$ in $\Vol$ according to the
Helmholtz theorem. However, the ``closed field''
\cite[]{Kusano1995,Berger1999}, $\Bcl=\B-\BRef=\grad\cross\f$ in $\Vol$ with
$\left.\nhat\cdot\Bcl\right|_\Surf=0$, is intrinsically solenoidal by
construction and, therefore, magnetically isolated.\par
\section{The Relative Helicity Transport Rate\label{sec:Rate}}
Because the \cite{Finn1985} formula is fully rigorous and consistent
with \cite{Berger1984a}, its time dependence can be explored to
determine the helicity transport
rate. Expanding~(\ref{eqn:Finn-Antonsen}) and using~(\ref{eqn:BA})
followed by~(\ref{eqn:Div_f_x_g}), the integrand $\helicity$ becomes
\begin{equation}
\helicity=\A\cdot\B-\ARef\cdot\BRef+\grad\cdot\left(\A\cross\ARef\right).\label{eqn:dh:symmetric}
\end{equation}
Taking the time
derivative,
\begin{equation}
\frac{\partial\helicity}{\partial t}=\frac{\partial}{\partial
t}\left(\A\cdot\B-\ARef\cdot\BRef\right)+\grad\cdot\frac{\partial}{\partial t}\left(\A\cross\ARef\right),
\end{equation}
and using Faraday's Law of induction,
\begin{equation}
\frac{\partial \B}{\partial t}=-c\,\grad\cross\E,\label{eqn:Faraday}
\end{equation}
and the relationship between the electric field, the magnetic vector potential
$\A$ and electric scalar potential $\psi$
\begin{equation}
\E\equiv-\frac{1}{c}\frac{\partial\A}{\partial t}-\grad\psi,\label{eqn:EAPsi}
\end{equation}
yields
\begin{align}
\frac{\partial\helicity}{\partial t}=&-c\,\left(\E\cdot\B-\ERef\cdot\BRef\right)-c\,\left(\B\cdot\grad\psi-\BRef\cdot\grad\psiRef\right)-c\,\A\cdot\grad\cross\E+c\,\ARef\cdot\grad\cross\ERef\nonumber\\
&+c\,\grad\cdot\left(\ARef\cross\E-\A\cross\ERef\right)-c\,\grad\cdot\left(\A\cross\grad\psiRef-\ARef\cross\grad\psi\right),\label{eqn:twentyone}
\end{align}
where the reference electric field $\ERef$ satisfies Faraday's
law~(\ref{eqn:Faraday}) for $\BRef$ and the corresponding
relationship~(\ref{eqn:EAPsi}) between the reference electric field
$\ERef$, the magnetic vector potential $\ARef$, and the electric scalar
potential $\psiRef$.  Applying~(\ref{eqn:Div_f_x_g}) to the third,
fourth, and fifth groups of terms, using~(\ref{eqn:BA}), and the absence
of monopoles~(\ref{eqn:Monopoles}), and regrouping the results, the
helicity rate of change becomes
\begin{equation}
\frac{\partial\helicity}{\partial t}=-2\,c\,\left(\E\cdot\B-\ERef\cdot\BRef\right)+c\,\grad\cdot\left[\left(\A+\ARef\right)\cross\left(\E-\ERef\right)\right]-c\,\grad\cdot\left[\left(\B-\BRef\right)\,\left(\psi+\psiRef\right)\right].\label{eqn:h:symmetric}
\end{equation}
Using~(\ref{eqn:Div_f_x_g}) and~(\ref{eqn:Faraday}), \cite{Berger1984a}
express the volumetric rate of change of the self-helicity of the reference field
as
\begin{subequations}
\begin{align}
\ERef\cdot\BRef=\ERef\cdot\grad\cross\ARef=&\grad\cdot\left(\ARef\cross\ERef\right)+\ARef\cdot\grad\cross\ERef,\nonumber\\
=&\grad\cdot\left(\ARef\cross\ERef\right)-\frac{1}{c}\,\ARef\cdot\frac{\partial\BRef}{\partial
t}.\label{eqn:Reference_Helicity}
\end{align}
The choice of a potential reference field~(\ref{eqn:Potential}) permits
further simplification,
\begin{align}
\EPot\cdot\BPot=&\grad\cdot\left(\APot\cross\EPot\right)+\frac{1}{c}\,\APot\cdot\frac{\partial\grad\phiPot}{\partial
t},\nonumber\\
=&\grad\cdot\left(\APot\cross\EPot\right)+\grad\cdot\left(\frac{1}{c}\,\frac{\partial\phiPot}{\partial
t}\,\APot\right)-\frac{1}{c}\,\frac{\partial\phiPot}{\partial t}\,\grad\cdot\APot,\label{eqn:Potential:Helicity}
\end{align}
\end{subequations}
producing
\begin{equation}
\frac{\partial\helicity}{\partial
t}=-2\,c\,\E\cdot\B-2\,\frac{\partial\phiPot}{\partial
t}\,\grad\cdot\APot+2\,\grad\cdot\left(\frac{\partial\phiPot}{\partial
t}\,\APot\right)+c\,\grad\cdot\left[\left(\A+\APot\right)\cross\left(\E-\EPot\right)+2\,\APot\cross\EPot\right]-c\,\grad\cdot\left[\left(\B-\BPot\right)\,\left(\psi+\psiPot\right)\right].
\end{equation}
Integrating over the volume $\Vol$ and
applying~(\ref{eqn:Gauss}) produces
\begin{align}
\frac{\partial\Helicity}{\partial
t}=&-2\,c\,\int_{\Vol}d^3x\left(\E\cdot\B+\frac{1}{c}\,\frac{\partial\phiPot}{\partial
t}\,\grad\cdot\APot\right)-2\,\oint_{\Surf}dS\,\nhat\cdot\left(\frac{\partial\phiPot}{\partial
t}\,\APot\right)\nonumber\\
&-c\,\oint_{\Surf}dS\,\nhat\cdot\left[\left(\A+\APot\right)\cross\left(\E-\EPot\right)+2\,\APot\cross\EPot\right]+c\,\oint_{\Surf}dS\,\nhat\cdot\left[\left(\B-\BPot\right)\,\left(\psi+\psiPot\right)\right].\label{eqn:canceled}
\end{align}
Expressing the volumetric rate of change of self-helicity of the reference
field according to~(\ref{eqn:Potential:Helicity}) is a critical choice in the
derivation of the helicity transport equation. The left-hand side
of~(\ref{eqn:Potential:Helicity}) is, in principle, composed of electric and
magnetic field observables, albeit of potential reference fields. In contrast,
the right-hand side involves the vector potential and consequently, as we
shall see below, constraints on a gauge-dependent quantity.
Applying the magnetic field boundary conditions~(\ref{eqn:Boundary:Bn}) and
constraints on $\A$ implied by boundary conditions~(\ref{eqn:Boundary:A1}),
results in
\begin{equation}
\frac{\partial\Helicity}{\partial
t}=-2\,c\,\int_{\Vol}d^3x\left(\E\cdot\B+\frac{1}{c}\,\frac{\partial\phiPot}{\partial
t}\,\grad\cdot\APot\right)-2\,c\,\oint_{\Surf}dS\,\nhat\cdot\left(\APot\cross\E+\frac{1}{c}\,\frac{\partial\phiPot}{\partial
t}\,\APot\right).
\end{equation}
The normal component of $\APot$ can always be eliminated on any
surface,
\begin{subequations}
\begin{equation}
\left.\APot\cdot\nhat\right|_\Surf=0,\label{eqn:Boundary:APn}
\end{equation}
through a gauge transformation with Neumann boundary conditions \cite[see also
  \S~III.B in][]{Clegg2000b},
\begin{equation}
\APot'\rightarrow\APot+\grad\gauge,\qquad\nabla^2\gauge=0,\qquad\left.\nhat\cdot\grad\gauge\right|_\Surf=-\left.\nhat\cdot\APot\right|_\Surf,\quad\Longrightarrow\quad\left.\nhat\cdot\APot'\right|_\Surf=0.\label{eqn:AP:Neumann}  
\end{equation} 
\end{subequations}
Using the ideal MHD relationship~(\ref{eqn:Ohm}) for the surface electric
field produces
\begin{equation}
\frac{\partial\Helicity}{\partial
  t}=-2\,c\,\int_{\Vol}d^3x\left(\smash{\overbrace{\E\cdot\B}^{\left(\mathrm{\ref{eqn:Ohm}}\right)}+\frac{1}{c}\,\frac{\partial\phiPot}{\partial
      t}\,\overbrace{\grad\cdot\APot}^{\left(\mathrm{\ref{eqn:Coulomb}}\right)}}\vphantom{\frac{\partial\phiPot}{\partial
      t}}\right)-2\,\oint_{\Surf}dS\,\frac{\partial\phiPot}{\partial
  t}\,\overbrace{\nhat\cdot\APot}^{\left(\mathrm{\ref{eqn:Boundary:APn}}\right)}
+2\,\oint_{\Surf}dS\,\underbrace{\left(\APot\cdot\B\right)\,v_n}_{\mathrm{emergence}}-2\,\oint_{\Surf}dS\,\underbrace{\left(\APot\cdot\v\right)\,B_n}_{\mathrm{shearing}},
\label{eqn:BergerField}
\end{equation}
where because of the referenced equations in the overbraces the first
three terms vanish, resulting finally in the expression
in~(\ref{eqn:Berger1984}). The simplicity of~(\ref{eqn:Berger1984}) is
a consequence of the choice of gauge, which eliminates several terms
from the more general expression~(\ref{eqn:BergerField}) and the ideal
Ohm's law.  Note that setting the volume term $\grad\cdot\APot=0$ and
the surface term $\left.\nhat\cdot\APot\right|=0$, leaves the surface
term $2\,\APot\cross\EPot$ originating from the volumetric rate of
change $\EPot\cdot\BPot$ in~(\ref{eqn:Potential:Helicity}). This term
is used to cancel another surface term of the same form originating
from $\left(\A+\APot\right)\cross\left(\E-\EPot\right)$
in~(\ref{eqn:h:symmetric}).  Thus, the volumetric rate of
  change of the self-helicity of the potential reference field is not
  guaranteed to be zero in the Coulomb gauge unless the surface
  integral of $\APot\cross\EPot$ is zero! We address this issue
directly in \S~\ref{sec:equiv}. \par
As noted by \cite{Prior2014}, the remaining emergence and shearing
surface terms can entangle the transport of the self-helicity of
the potential reference field across the boundary with changes in the
self-helicity of the potential reference field in the volume $\Vol$.
Additionally, while~(\ref{eqn:BergerField}) is gauge
invariant,~(\ref{eqn:Berger1984}) is not, because it is written
explicitly in a particular gauge. Without further justification, the
emergence and shearing terms cannot be considered observables as they
are not independently gauge invariant. On the other hand, emergence
and shearing are clearly physical processes that, in principle, are
observable as long as all components of the velocity and magnetic
fields can be measured at the boundary; consequently, there must exist
manifestly gauge-invariant expressions for these quantities. \par
 Given the concerns with~(\ref{eqn:BergerField}) and by
 extension~(\ref{eqn:Berger1984}), we endeavor to derive an alternate
 expression for the helicity transport across a surface.  Restarting
 from~(\ref{eqn:h:symmetric}) but with a potential reference
 field~(\ref{eqn:Potential}), the rate of change of the helicity integrand becomes
\begin{equation}
\frac{\partial\helicity}{\partial
  t}=-2\,c\,\left(\E\cdot\B-\EPot\cdot\BPot\right)+c\,\grad\cdot\left[\left(\A+\APot\right)\cross\left(\E-\EPot\right)\right]-c\,\grad\cdot\left[\left(\B-\BPot\right)\,\left(\psi+\psiPot\right)\right].
\end{equation}
Integrating over the volume and applying the
Gauss\---Ostrogradsky theorem~(\ref{eqn:Gauss}) produces 
\begin{equation}
\frac{\partial\Helicity}{\partial
  t}=-2\,c\,\int_\Vol{d^3x}\,\left(\E\cdot\B-\EPot\cdot\BPot\right)-c\,\oint_\Surf{dS}\,\nhat\cdot\left[\left(\A+\APot\right)\cross\left(\E-\EPot\right)\right]+c\,\oint_\Surf{dS}\,\nhat\cdot\left[\left(\B-\BPot\right)\,\left(\psi+\psiPot\right)\right].\label{eqn:New}
\end{equation}
The second term in the volume integral represents the generation of helicity
in the volume caused by changes in the potential reference field. Instead of
employing~(\ref{eqn:Reference_Helicity})\--(\ref{eqn:Potential:Helicity}),
we require that the evolution of the potential reference
  field does not generate any relative helicity in the volume:
\begin{equation}
\int_{\Vol}d^3x\,\EPot\cdot\BPot\equiv0.\label{eqn:Reference}
\end{equation}
This critically important ansatz merits some discussion
because it elevates the potential magnetic field to a special status,
which in fact, the potential field $\BPot$ does possess. For a given
distribution of normal flux at a closed boundary, the potential field
is physically the unique ground state of the system. It is the only
field that has no sources i.e., electric currents, anywhere in the
volume. Note that because MHD ignores the displacement current, then
all currents are due to actual material sources. Consequently, the
potential field $\BPot$ is not simply some convenient reference field,
but a well-defined state for a physical system. Any deviation from the
potential field anywhere in the volume increases the energy of the
system. Furthermore, if a system evolves through a sequence of
potential states, then no free energy can be generated by this
evolution. These unique properties of the potential field also apply
to the relative helicity $\Helicity$ of the system.  As long as the
system evolves through potential states, at least in a quasi-static
sense, then no relative helicity can be generated.  By construction,
in the \cite{Berger1984a} or \cite{Finn1985} formalisms, there is no
self-helicity of the reference field $\BRef$ and in our formulation, none
can be generated purely by changes in the potential reference field
$\BRef=\BPot$. This choice directly addresses the concern raised by
\cite{Prior2014} in the introduction.\par
The reference electric field $\EPot$ that satisfies Faraday's law for
the changing reference magnetic field $\BPot$ and the
ansatz~(\ref{eqn:Reference}) above can now be determined. In
general, $\EPot$ can be decomposed into an solenoidal (inductive)
piece $\epot=\grad\cross\f$ and an irrotational (electrostatic)
component $\grad\GaugeP$:
\begin{subequations}
\begin{equation}
\EPot=\epot+\grad\GaugeP,\label{eqn:EP:Decomposition}
\end{equation}
which must satisfy the following conditions for compatibility with
the ansatz~(\ref{eqn:Reference}):
\begin{align}
\grad\cdot\epot=&0\qquad\in\Vol,\label{eqn:eP:Div}\\
\left.\nhat\cdot\epot\right|_\Surf=&0,\label{eqn:eP:Neumann}\\
\nabla^2\GaugeP=&\rhoP\qquad\in\Vol,\label{eqn:LambdaP:Poisson}\\
\left.\GaugeP\right|_\Surf=&\mbox{constant},\label{eqn:LambdaP:Dirichlet}
\end{align}
\end{subequations}
and $\grad\cdot\EPot=\rhoP$ is arbitrary. These conditions guarantee
that the $\EPot$ does not generate any change in the relative helicity
and, as noted in~(\ref{eqn:Dirichlet}) of
Appendix~\ref{sec:Helmholtz}, the boundary conditions ensure
orthogonality between~$\epot$ and $\grad\GaugeP$. In particular,
conditions~(\ref{eqn:eP:Div}) and~(\ref{eqn:eP:Neumann}) ensure that
$\epot$ is intrinsically solenoidal and does not change the
self-helicity of the potential reference field,
\begin{subequations}
\begin{equation}
\int_{\Vol}d^3x\,\epot\cdot\BPot=-\int_{\Vol}d^3x\,\epot\cdot\grad\phiPot=\int_{\Vol}d^3x\,\phiPot\,\left(\vphantom{\grad\cdot\epot}\smash{\overbrace{\grad\cdot\epot}^{(\ref{eqn:eP:Div})}}\right)+\oint_{\Surf}dS\,\smash{\overbrace{\nhat\cdot\left(\phiPot\,\epot\right)}^{(\ref{eqn:eP:Neumann})}}\equiv0,\label{eqn:eP}
\end{equation}
and Laplace's equation,~(\ref{eqn:Laplace}) and
condition~(\ref{eqn:LambdaP:Dirichlet}) combined with the solenoidal property
of $\B$ in~(\ref{eqn:Solenoidal}) ensure that $\GaugeP$ does not change the
self-helicity of the potential reference field,
\begin{equation}
\int_{\Vol}d^3x\,\grad\GaugeP\cdot\BPot=-\int_{\Vol}d^3x\,\grad\GaugeP\cdot\grad\phiPot=\int_{\Vol}d^3x\,\GaugeP\,\smash{\overbrace{\nabla^2\phiPot}^{(\ref{eqn:Laplace})}}+\oint_{\Surf}dS\,\smash{\overbrace{\GaugeP\,\nhat\cdot\grad\phiPot}^{(\ref{eqn:Solenoidal})\,(\ref{eqn:LambdaP:Dirichlet})}}\equiv0.\label{eqn:EP:GaugeP}
\end{equation}
\end{subequations}
This surface integral on the right is zero by the solenoidal property
of magnetic fields when $\GaugeP$ is a constant on $\Surf$ and the
volume integral on the right is zero by~(\ref{eqn:Laplace}). Note that
 $\gradp\GaugeP=0$ and $\Lapp\GaugeP=0$ on $\Surf$,
and $\grad\GaugeP$ does not contribute to $\grad\cross\EPot$ in $\Vol$
or on $\Surf$ because it is a complete gradient. \par
Additionally, $\EPot$ must satisfy Faraday's law~(\ref{eqn:Faraday})
for $\partial\BPot/\partial t$ which specifies $\grad\cross\EPot$ and
by extension $\grad\cross\epot$. Temporarily deferring the calculation
of $\epot$ to \S~\ref{sec:EPot} and Appendix~\ref{sec:EPot:Revisited},
we note that the existence and uniqueness of $\epot$ by~(\ref{eqn:Faraday}),
(\ref{eqn:eP:Div}), and (\ref{eqn:eP:Neumann}) have been firmly
established by numerous authors \cite[see][and references
  therein]{Girault1986,Jiang1994,Amrouche1998,Amrouche2013,Cheng2017}. Thus,
while $\EPot$ exhibits some arbitrariness through $\grad\GaugeP$ and
$\rhoP$, the existence and uniqueness of $\epot$ throughout $\Vol$
permit the expression of~(\ref{eqn:New}) with boundary
condition~(\ref{eqn:Boundary:Bn}) and
ansatz~(\ref{eqn:Reference}) as
\begin{equation}
\frac{\partial\Helicity}{\partial
  t}=-2\,c\,\int_{\Vol}d^3x\,\left[\E\cdot\B-\phiPot\,\smash{\overbrace{\grad\cdot\epot}^\mathrm{(\ref{eqn:eP:Div})}}-\GaugeP\,\smash{\overbrace{\nabla^2\phiPot}^{(\ref{eqn:Laplace})}}\right]+c\,\oint_{\Surf}dS\,\nhat\cdot\left[\smash{2\,\overbrace{\phiPot\,\epot}^\mathrm{(\ref{eqn:eP:Neumann})}}+2\,\smash{\overbrace{\GaugeP\,\grad\phiPot}^{(\ref{eqn:Solenoidal})\,(\ref{eqn:LambdaP:Dirichlet})}}-\left(\A+\APot\right)\cross\left(\E-\epot-\smash{\overbrace{\grad\GaugeP}^{(\ref{eqn:LambdaP:Dirichlet})}}\right)\right].\label{eqn:RH}
\end{equation}
The terms with overbraces are zero because of the referenced
equations resulting in
\begin{equation}
\boxed{\dHelicity\equiv\frac{\partial\Helicity}{\partial
  t}=-\overbrace{2\,c\,\int_{\Vol}d^3x\,\E\cdot\B}^{\dHelicityV}-\overbrace{c\,\oint_{\Surf}dS\,\nhat\cdot\left[\left(\A+\APot\right)\cross\left(\E-\EPot\right)\right]}^\dHelicityS.\label{eqn:RH_Simple}}
\end{equation}
As expected, $\GaugeP$, the irrotational constituent of $\EPot$, plays
no role in helicity transport. The final surface integral involves
only the tangential components of $\EPot$, which are unique on
$\Surf$, since $\nhat\cross\EPot\equiv\nhat\cross\epot$, i.e.,
${\EPot}_\Surf=\epot$.  We show in the next section that the
representation above for the surface helicity transport $\dHelicityS$
has major advantages over the \citeauthor{Berger1984a}
formula~(\ref{eqn:Berger1984}).\par
\section{Implications}\label{sec:Implications}
\subsection{Gauge Invariance\label{sec:gauge_invariance}}
The most important advantage of Equation~(\ref{eqn:RH_Simple}) for
determining the helicity transport across the boundary is that the
surface helicity transport rate $\dHelicityS$ is independently and
manifestly gauge invariant. Because $\dHelicityV$ is comprised entirely
of observables $\E$ and $\B$, the gauge invariance of $\dHelicity$
depends on the gauge invariance of $\dHelicityS$.  Gauge-transforming
$\dHelicityS$ with $\A\rightarrow\A+\grad\gauge'$ and
$\APot\rightarrow\APot+\grad\gauge''$, where $\gauge=\gauge'+\gauge''$
produces
\begin{equation}
\frac{\partial\HelicityS'}{\partial t}=-c\,\oint_{\Surf}dS\,\nhat\cdot\left(\A+\APot\right)\cross\left(\E-\EPot\right)-c\,\oint_{\Surf}dS\,\nhat\cdot\grad\gauge\cross\left(\E-\EPot\right),
\end{equation}
this can be rewritten with~(\ref{eqn:Curl_psi_f}) and~(\ref{eqn:Faraday})
\begin{equation}
\frac{\partial\HelicityS'}{\partial t}=\frac{\partial \HelicityS}{\partial t}
  -2\,c\,\oint_{\Surf}dS\,\nhat\cdot\grad\cross\left[\gauge\,\left(\E-\EPot\right)\right]-2\,\oint_{\Surf}dS\,\gauge\,\nhat\cdot\frac{\partial}{\partial
    t}\left(\B-\BPot\right),\label{eqn:RH_Simply_gauge}
\end{equation}
which proves ${\partial \HelicityS'}/{\partial t}\equiv{\partial
  \HelicityS}/{\partial t}$ and by extension the gauge invariance of
${\partial \Helicity}/{\partial t}$, because the second term is zero
by Stokes' theorem on a closed surface~(\ref{eqn:StokesClosed}) and
the third term is zero by the boundary
condition~(\ref{eqn:Boundary:Bn}). This new formula directly addresses
our primary concern in using the \cite{Berger1984a} formula for
helicity transport\----namely, the lack of manifest gauge invariance
of its surface helicity transport rate. This demonstrates that the
volumentric rate of change of helicity $\dHelicityV$ and the
surface helicity transport $\dHelicityS$ are independently
observables inasmuch as the surface $\Surf$ is closed.\par
\subsection{Equivalence with \cite{Berger1984a}\label{sec:equiv}}
The \cite{Berger1984a} result~(\ref{eqn:BergerField1984}) and the new
expression~(\ref{eqn:RH_Simple}) are equivalent for a closed
surface. First, we note that by using the boundary
relation~(\ref{eqn:Boundary:A1}) which relates the tangential
components of the vector potential, our surface helicity transport
expression~$\dHelicityS$ above can be written entirely in terms of
$\A$ or $\APot$ as
\begin{equation}
\frac{\partial\HelicityS}{\partial
  t}=-2\,c\,\oint_{\Surf}dS\,\nhat\cdot\left[\A\cross\left(\E-\EPot\right)\right]
=-2\,c\,\oint_{\Surf}dS\,\nhat\cdot\left[\APot\cross\left(\E-\EPot\right)\right].\label{eqn:BFequivalence}
\end{equation}
Comparing the expression on the right with the \cite{Berger1984a}
formula, which uses the Coulomb gauge, $\APotC$, we see 
that they are equivalent if and only if
\begin{equation}
 \oint_{\Surf}dS\,\nhat\cdot\left(\APotC\cross\EPot\right)=0.\label{eqn:BFproof}
\end{equation}
This result follows directly from our
ansatz~(\ref{eqn:Reference}), which can be written with~(\ref{eqn:Potential:Helicity}) as:
\begin{equation}
  \int_{\Vol}d^3x\,\EPot\cdot\BPot=-\oint_{\Surf}dS\,\nhat\cdot\left(\APot\cross\EPot\right)-\oint_{\Surf}dS\,\nhat
  \cdot\left(\frac{1}{c}\,\frac{\partial\phiPot}{\partial
      t}\,\APot\right)-\int_{\Vol}d^3x\,\frac{1}{c}\,\frac{\partial
    \phiPot}{\partial t}\,\grad\cdot\APot\equiv0.\label{eqn:RHS}
\end{equation}
Substituting $\APotC$ in the expression
above, we note that the last two terms on the right vanish due to the
properties~(\ref{eqn:Coulomb}) and ~(\ref{eqn:APC:Neumann}) that
define the Coulomb gauge vector
potential, and with the properties of $\EPot$ given by~(\ref{eqn:EP:Decomposition}\---\ref{eqn:EP:GaugeP}), this becomes
\begin{equation}
\int_{\Vol}d^3x\,\EPot\cdot\BPot=-\oint_{\Surf}dS\,\nhat\cdot\left(\APotC\cross\EPot\right)=-\oint_{\Surf}dS\,\nhat\cdot\left(\APotC\cross\epot\right)\equiv0,\label{eqn:Equivalence}
\end{equation}
and the \cite{Berger1984a} result is, indeed, equivalent to the new expression
when integrated over a closed surface; consequently, one is free to
use the old result if it is computationally more advantageous.\par
\subsection{Electric Field Determining Helicity Transport \label{sec:lamellar}}
Another important implication of our expression is that the transport across
the boundary is due solely to the lamellar part of the electric field on the
boundary $\E_\Surf\sim\gradp\ZETA$ \cite[]{Scharstein1991}. To prove this, we
must determine the solenoidal and lamellar components of $\EPot$ and $\E$ on
the bounding surface $\Surf$. The Helmholtz\---Hodge theorem (see
Appendix~\ref{sec:Helmholtz:Surface}) provides the mathematical framework for
this decomposition given the surface components of $\EPot$ and $\E$. The
electric field $\E$ is a specified field in the helicity transport problem,
but ${\EPot}_\Surf=\epot$ is related to changes in the normal component of the
surface magnetic field.  The existence and uniqueness of solutions of the
div-curl system represented by $\epot$, e.g.,~(\ref{eqn:Faraday}),
(\ref{eqn:eP:Div}), and (\ref{eqn:eP:Neumann}), are well established
\cite[]{Girault1986,Jiang1994,Amrouche1998,Amrouche2013,Cheng2017}. Thus,
$\epot$ can always be computed from changes in the normal component of the
magnetic field $\partial B_n/\partial t$. However, the details of determining
$\epot$ in general geometries adds unnecessary complexity, given the simple
geometries that are typically of interest to solar physicists (Cartesian,
which is flat; spherical with $\nhat=\rhat$, which has the same curvature
along the orthogonal principal directions, etc.). In these simple geometries,
the surface components of ${\EPot}_\Surf=\epot$ can be determined from
$\partial B_n/\partial t$ and directly measured using a $\dot\B$ loop embedded
in the surface.  Thus, in \S~\ref{sec:EPot}\---\ref{sec:Helicity} below, we
assume that $\epot$ has a priori been determined and develop the general
formalism for computing helicity transport. We defer the discussion of
determining $\epot$ from $\partial B_n/\partial t$ in general curvilinear
geometries to Appendix~\ref{sec:EPot:Revisited}.  To emphasize the essential
results for the reader, we highlight below in boxed equations the general
results and the results that apply only to simple geometries most relevant to
the Sun.\par
Up to this point, the detailed properties of the surface $\Surf$ have largely
been ignored. However, moving forward, the geometry of the differentiable
surface $\Surf$ bounding the volume $\Vol$ becomes intertwined with the
definition of the differential operators (see Appendix~\ref{sec:Surface}).
For the remainder of \S~\ref{sec:Implications}, we assume that the bounding
surface $\Surf$ is at least a $\ContinuityMin$, closed (compact and without
boundaries), and simply connected hypersurface, where $\Continuity^k$
indicates continuity up to the $k$th derivative. This includes the
$\Continuity^\infty$ spherical surface, which is the most relevent for solar
physics. \cite{Reusken2019} notes that the $\ContinuityHodgeMin$ continuity of
$\Manifold$ is sufficient but not necessary for the definition of the surface
differential operators in the Helmholtz\---Hodge decomposition on
$\Manifold$. Indeed, Chapter 7 of \cite{Morrey1966} extends the
Helmholtz\---Hodge decomposition to $\Continuity^1$ surfaces with and without
boundaries using differential forms. However, $\ContinuityCurrentMin$ is
sufficient for the existence of
$\grad\cross\grad\cross\E\propto\partial\J/\partial t$ on the surface $\Surf$,
which is necessary for the analysis of relative helicity transport. Our
results can be extended mutatis mutandis to multiply connected
domains\footnote{The simply connected assumption may be relaxed by noting that
  any multiply connected volume can be transformed into a simply connected
  volume by $\genus$ cuts, where $\genus$ is the genus of the bounding surface
  $\Surf$ which is equivalent to the number of holes (see
  Appendix~\ref{sec:Helmholtz:Volume}).\label{foot:Simply}} and surfaces that
are $\ContinuityMin$ almost everywhere in a set of measure zero sense
\cite[]{Halmos1974}.  \par
\subsubsection{The Reference Electric Field $\EPot$ on the Bounding Surface $\Surf$\label{sec:EPot}}
As shown in \S~\ref{sec:Rate}, the reference electric field $\EPot$ that
changes the potential field $\BPot$ can be decomposed into an instrinsically
solenoidal part $\epot$ and irrotational part $\grad\GaugeP$. The
intrinsically solenoidal constituent $\epot$ is unique, whereas the
irrotational constituent $\grad\GaugeP$ is arbitrary, subject to Dirichlet
boundary conditions~(\ref{eqn:LambdaP:Dirichlet}) on $\GaugeP$ such that
${\EPot}_\Surf=\epot$.  The irrotational part plays no role in helicity
generation and transport, and thus helicity transport through the boundary
only requires determining the unique tangential components of $\epot$.  On the
surface $\Manifold$, the intrinsically solenoidal component $\epot$ may be
decomposed with the Helmholtz\---Hodge theorem outlined in
Appendix~\ref{sec:Helmholtz:Surface} \cite[see
  also][]{Hodge1952,Scharstein1991,Bladel1993a,Bhatia2013,ONeil2018,Reusken2019}
in the Dupin surface coordinate system $\left(u_1,u_2,n\right)$ discussed in
Appendix~\ref{sec:Surface} \cite[see
  also][]{Weatherburn1955,Tai1992,Bladel2007}
\begin{equation}
\epot=\tau_\mathrm{P}\,\nhat-\frac{1}{c}\,\nhat\cross\gradp\frac{\partial\PhiPot}{\partial
  t}-\gradp\ZETAPot.\label{eqn:Helmholtz:Surface:EPot}
\end{equation}
Here, the variable $\epot$ is understood to be differentiable on $\Manifold$
and in some neighborhood of $\Manifold$. In other words, $\EPot$ exists not
just on the surface $\Manifold$ but also in the volume $\Vol$ and consequently
is a function of the parameterization of the surface by $u_1$ and $u_2$ in the
Dupin surface coordinate system, as well as $n$, the coordinate normal to the
surface $\Manifold$. The terms in~(\ref{eqn:Helmholtz:Surface:EPot}) are
mutually orthogonal when integrated over a closed surface $\Manifold$. In the case of a multiply-connected volume, a mutually
orthogonal harmonic term $\HarmonicP$ can be added to the
decomposition on the surface or the volume can be converted to a
simply connected one with the appropriate cuts and auxiliary surfaces
as mentioned in footnote~\ref{foot:Simply}. However, for a simply
connected volume bounded by a $\Continuity^k$ surface with $k\ge2$, such
as a sphere with $k=\infty$, the dimension of the harmonic space is zero,
$\Harmonic=0$, and the harmonic field can thus be ignored in the
decomposition \cite[see Lemma 4.3 in][]{Reusken2019}. \par
The second term in~(\ref{eqn:Helmholtz:Surface:EPot}) is purely
solenoidal as
$\grad\cdot\left[\nhat\cross\gradp\left({\partial\PhiPot}/{\partial
  t}\right)\right]=0$, and the third term is irrotational with respect to the
normal component of the curl $\nhat\cdot\grad\cross\gradp\ZETAPot=0$,
but it is not perfectly irrotational. In particular, $\gradp\ZETAPot$ is
not irrotational with respect to the three-dimensional curl operator
because it is an incomplete gradient of a scalar (see
discussion at the end of Appendix~\ref{sec:Surface}).  Thus,
$\gradp\ZETAPot$ contains a solenoidal component in general
three-dimensional curvilinear coordinates. Nonetheless, this term is
commonly referred to as the ``lamellar term'' in the Helmholtz\---Hodge
decomposition \cite[]{Scharstein1991} because its so-called ``surface
curl'' $\nhat\cdot\gradp\cross\gradp\ZETAPot$ is identically
zero. \par
The lamellar and solenoidal components of  $\EPot$ on $\Surf$ are determined by
\begin{subequations}
\begin{equation}
\Lapp\ZETAPot = -\gradp\cdot\left[\left(\nhat\cross\EPot\right)\cross\nhat\right]=-\gradp\cdot\epot \qquad\in\Manifold,\label{eqn:Helmholtz:Surface:DivS:EPot}
\end{equation}
\begin{equation}
\nhat\cdot\gradp\cross\left(\frac{1}{c}\,\nhat\cross\gradp\frac{\partial\PhiPot}{\partial
  t}\right)= -\nhat\cdot\gradp\cross\epot=\frac{1}{c}\,\frac{\partial\bpot_n}{\partial t}\Longrightarrow
\Lapp\PhiPot=\bpot_n\qquad\in\Manifold,\label{eqn:Helmholtz:Surface:PhiPot}
\end{equation}
where~(\ref{eqn:n_Curl_n_x_GradS_psi}) has been used and
\begin{equation}
\left.\tau_\mathrm{P}\right|_\Manifold=0,\label{eqn:tauP}
\end{equation}
to satisfy~(\ref{eqn:eP:Neumann}). Here,~(\ref{eqn:eP:Div})
with~(\ref{eqn:Helmholtz:Surface:PhiPot}),~(\ref{eqn:tauP}),
and~(\ref{eqn:DivS_f}) implies
\begin{equation}
\Lapp\ZETAPot=\frac{\partial\tau_\mathrm{P}}{\partial n}.\label{eqn:Reviewer}
\end{equation}
\end{subequations}
Note that the only finite homogeneous solution
for~(\ref{eqn:Helmholtz:Surface:DivS:EPot})
or~(\ref{eqn:Helmholtz:Surface:PhiPot}), ($\Lapp\ZETAPot\mbox{ and
}\Lapp\PhiPot=0$, on $\Manifold$) are $\PhiPot=\mbox{constant}$ and
$\ZETAPot=\mbox{constant}$, i.e., there is no nonzero harmonic vector on
$\Manifold$ which is at least $\ContinuityHairy$ everywhere.\footnote{This is
  related to the hairy ball theorem, which can be stated as a continuous
  tangent vector field on a sphere must have at least one zero
  \cite[]{Brouwer1912}. Thus, a constant tangent vector (other than zero) is
  forbidden on a sphere and shapes continuously deformable to a sphere. See
  Chapter~II, \S~y22 in \cite{Shubin2001} and the Laplace equation on a sphere
  discussed by \cite{Esparza-Lopez2016}). } Thus, $\PhiPot$ and $\ZETAPot$ are
unique on a closed, simply connected, $\Continuity^{k\ge2}$ surface
$\Manifold$ to within a constant.  For simple, smooth, unbounded geometries
such as a plane or a sphere,~(\ref{eqn:Helmholtz:Surface:PhiPot})
and~(\ref{eqn:tauP}) are sufficient to uniquely determine $\epot$ on $\Surf$
because $\ZETAPot$ is $\mbox{constant}$ on $\Surf$ as discussed below.\par
However,~(\ref{eqn:Helmholtz:Surface:DivS:EPot}) implies that in general
curvilinear coordinates $\PhiPot$ and $\ZETAPot$ are coupled.  Substituting
$\EPot$ into Faraday's Law~(\ref{eqn:Faraday}) for $\BPot$ and
using~(\ref{eqn:Curl_psi_f}) and~(\ref{eqn:Tangent:Der}) on the first term and
rearranging,
\begin{equation}
\frac{\partial \BPot}{\partial t}=-c\,\grad\cross\EPot=
\grad\cross\left(\nhat\cross\gradp\frac{\partial\PhiPot}{\partial
  t}\right)-c\,\left(\tau_\mathrm{P}+\frac{\partial\ZETAPot}{\partial n}\right)\grad\cross\nhat+c\,\nhat\cross\gradp\left(\tau_\mathrm{P}+\frac{\partial\ZETAPot}{\partial n}\right)\qquad\in\Manifold.\label{eqn:Curl_EPot}
\end{equation}  
Here, $\ZETAPot$ only appears in the form $\partial\ZETAPot/\partial
n$, a consequence of its manifestation as an incomplete gradient.  Taking
the curl of (\ref{eqn:Curl_EPot}), and
using~(\ref{eqn:Curl_n}),~(\ref{eqn:Curl_psi_f}),
and~(\ref{eqn:Curl_Laplacian}),
\begin{equation}
\frac{\partial \grad\cross\BPot}{\partial t}=
-\grad\cross\left\lbrace\grad\cross\left[\grad\cross\left(\frac{\partial\PhiPot}{\partial
  t}\,\nhat\right)\right]+c\,\grad\cross\left[\left(\tau_\mathrm{P}+\frac{\partial\ZETAPot}{\partial n}\right)\,\nhat\right]\right\rbrace\qquad\in\Manifold.\label{eqn:Curl_Curl_EPot}
\end{equation}  
By definition, the potential magnetic field cannot have any associated
currents in the volume, thus the normal component provides
a constraint on ${\partial\ZETAPot}/{\partial n}$ with
$\tau_\mathrm{P}=0$ on $\Manifold$:
\begin{equation}
\Lapp\left(\tau_\mathrm{P}+\frac{\partial\ZETAPot}{\partial n}\right)=\nhat\cdot\grad\cross\left\lbrace\grad\cross\left[\left(\tau_\mathrm{P}+\frac{\partial\ZETAPot}{\partial n}\right)\,\nhat\right]\right\rbrace=-\frac{1}{c}\,\nhat\cdot\grad\cross\left[\nabla^2\left(\frac{\partial\PhiPot}{\partial
  t}\,\nhat\right)\right]\qquad\in\Manifold.\label{eqn:n_Curl_Curl_EPot}
\end{equation}  
Generally, $\gradp\ZETAPot$ produces a solenoidal component when the
total curl is applied and this solenoidal component is dependent on
the missing part of the gradient: ${\partial\ZETAPot}/{\partial
  n}$. This constraint requires the existence of
$\Lapp\left({\partial\ZETAPot}/{\partial n}\right)$, which our
assumption of $\ContinuityCurrentMin$ of $\Surf$ ensures. For several
geometries, such as planes and spheres, the right-hand side
of~(\ref{eqn:n_Curl_Curl_EPot}) vanishes and
$\Lapp\left({\partial\ZETAPot}/{\partial n}\right)=0$,
$\ZETAPot=\mbox{constant}$ on $\Surf$, and $\ZETAPot$ can essentially
be ignored in the subsequent analysis. \par
\subsubsection{The Electric Field $\E$}
The electric field $\E$ may also be decomposed with the Helmholtz\---Hodge
theorem
\begin{equation}
\E=\tau\,\nhat-\frac{1}{c}\,\nhat\cross\gradp\frac{\partial\PHI}{\partial t}-\gradp\ZETA\qquad\in\Manifold.\label{eqn:Helmholtz:Surface:E}
\end{equation}
Again, the solenoidal and lamellar components are determined by
\begin{subequations}
\begin{equation}
\Lapp\ZETA=-\grad\cdot\left[\left(\nhat\cross\E\right)\cross\nhat\right]=-\gradp\cdot\E_\Surf\qquad\in\Manifold,\label{eqn:Helmholtz:Surface:DivS:E}
\end{equation}
and
\begin{equation}
\nhat\cdot\gradp\cross\left(\frac{1}{c}\,\nhat\cross\gradp\frac{\partial\PHI}{\partial
  t}\right) = -\nhat\cdot\gradp\cross\E=\frac{1}{c}\,\frac{\partial\bpot_n}{\partial t}\Longrightarrow
\Lapp\PHI=\bpot_n\qquad\in\Manifold,\label{eqn:Helmholtz:Surface:PHI}
\end{equation}
where~(\ref{eqn:n_Curl_n_x_GradS_psi}) was used.
\end{subequations}
The electric field $\E$ must satisfy the same boundary conditions on the
normal component of the curl as $\epot$,\footnote{Practically, for a numerical
  simulation, the instantaneous electric field data corresponding to the
  left-hand side of~(\ref{eqn:Helmholtz:Surface:PHI}) may not be equal to the
  magnetic data corresponding to the right hand-hand side
  of~(\ref{eqn:Helmholtz:Surface:PHI}) because of discretization, dissipation,
  etc. Indeed, deviations from equality for ${\partial B_n}/{\partial
    t}=-c\,\nhat\cdot\grad\cross\E$ is a indication of how well the induction
  equation is satisfied by the numerical data from the simulation.} and thus,
\begin{equation}
\frac{\partial\PHI}{\partial
  t}\equiv\frac{\partial\PhiPot}{\partial
  t}\qquad\in\Manifold,\label{eqn:PHI}
\end{equation}
to within a constant. For any geometry, ${\partial\PHI}/{\partial
  t}$ and ${\partial\PhiPot}/{\partial t}$ can be determined directly 
from the change in the normal component of the magnetic field at the
surface.
\subsubsection{The Surface Helicity Transport $\dHelicityS$\label{sec:Helicity}}
Substituting
(\ref{eqn:Helmholtz:Surface:EPot}) and (\ref{eqn:Helmholtz:Surface:E})
into~(\ref{eqn:RH_Simple}) produces the surface helicity transport rate
\begin{equation}
\frac{\partial\HelicityS}{\partial
  t}=-c\,\oint_{\Manifold}dS\,\nhat\cdot\left[\left(\A+\APot\right)\cross\left(\E-\EPot\right)\right]=-c\,\oint_{\Manifold}dS\,\nhat\cdot\left[\grad\left(\ZETA-\ZETAPot\right)\cross\left(\A+\APot\right)\right].\label{eqn:RH_Simpler}
\end{equation}
Using~(\ref{eqn:Curl_psi_f}), the general surface helicity transport
rate takes the simple form
\begin{equation}
\boxed{\frac{\partial\HelicityS}{\partial
    t}=-c\,\oint_{\Manifold}dS\,\nhat\cdot\grad\cross\left[\left(\ZETA-\ZETAPot\right)\,\left(\A+\APot\right)\right]+c\,\oint_{\Manifold}dS\,\left(\ZETA-\ZETAPot\right)\,\nhat\cdot\grad\cross\,\left(\A+\APot\right)\equiv2\,c\,\oint_{\Manifold}dS\,\left(\ZETA-\ZETAPot\right)\,B_n,}\label{eqn:RH_Simpliest}
\end{equation}
where the first surface integral vanishes
by~(\ref{eqn:StokesClosed}). \par
This simple and somewhat surprising expression~(\ref{eqn:RH_Simpliest})
provides key insight into helicity transport. First, it shows that only the
lamellar ($\sim$irrotational) part of $\E$ is involved in the helicity
transport across $\Surf$.  In simple geometries, this part, $\gradp\ZETA$,
would be considered the electrostatic component, whereas the twisting and
tangling of field lines might be expected to be due to an inductive electric
field. Resolution to this apparent paradox can be achieved by noting that for
ideal motions,
\begin{equation}
\grad\cdot\E\propto\grad\cdot\left(\v\cross\B\right)=\B\cdot\grad\cross\v -
\v \cdot\grad\cross\B,  \label{eqn:DivE}
\end{equation} 
which shows that the irrotational part of $\E$ is related to the transport
of vorticity and electric current, as expected for helicity
injection. Second, the difference $\ZETA-\ZETAPot$
can be interpreted as that part of $\E$ not involved in changing
the normal component of $\B$ at the boundary, implying 
that  the instantaneous change of the potential
  reference magnetic field at the boundary does not contribute to
  helicity transport.  \par
Although the mean value of $\ZETA-\ZETAPot\equiv\ZETA_0$ is subject to
the choice of the observer,~(\ref{eqn:RH_Simpliest}) is manifestly
gauge invariant.  The solenoidal property of the magnetic field
$\oint_\Manifold{dS}\,\B_n=0$ ensure that ${\partial\HelicityS}/{\partial
  t}$ is independent of this constant $\ZETA_0$. Nonetheless, the
integrand is only determined to within a constant $\ZETA_0$ times the
normal component of $\B$, i.e., $\ZETA_0\,B_n$. Consequently the value
of the ``helicity flux density'' can be adjusted at an arbitrary location to be
positive, negative, or zero by fiat through the choice of $\ZETA_0$
without changing the observable ${\partial\HelicityS}/{\partial
  t}$. Given that the sign of 
$\left(\ZETA-\ZETAPot\right)\,B_n$ cannot be determined uniquely at
any location, helicity transport \textit{cannot} be assigned a unique
\textit{local} interpretation; thereby, demonstrating again that the
concept of ``helicity flux density'' is not meaningful. This point is
proven explicitly in the example discussed in \S~\ref{sec:Linked} below.  \par
As noted above, in curvilinear coordinates, $\PhiPot$ and $\ZETAPot$ are
generally coupled.  However, in several geometries (Cartesian which is flat,
spherical with $\nhat=\rhat$ which has the same curvature along the orthogonal
principal directions, etc) ${\partial\ZETAPot}/{\partial n}$ and
${\partial\PhiPot}/{\partial t}$ completely decouple \--- the right-hand side
of~(\ref{eqn:Curl_Curl_EPot}\label{ref:projection}) has no projection along
the normal component and $\Lapp{\partial\ZETAPot}/{\partial n}=0$ on
$\Manifold$. Indeed, for spherical boundaries the intrinsically solenoidal
reference electric field for the potential magnetic field may be represented
\textit{everywhere} in $\Vol$ by
$\epot=-{c}^{-1}\,\rhat\cross\gradp\left({\partial\PhiPot}/{\partial
  t}\right)$ with $\rhat\cdot\epot=0$ and
$\nabla^2\left({\partial\PhiPot}/{\partial t}\right)=0$
\cite[]{Backus1986}. Thus for the volume between two spherical boundaries,
there is no radial component of $\epot$ anywhere! In this case
$\gradp\cdot\epot=\grad\cdot\epot-{\partial\rhat\cdot\epot}/{\partial
  r}\equiv0$ on every radial surface in $\Vol$ and in particular
$\ZETAPot=\mbox{constant}$ on the boundaries $\Manifold$.\footnote{This
  special geometry also leads to the analogous representation for $\APotC$ for
  the potential magnetic field leading to $\gradp\cdot\APotC=0$ on $\Manifold$
  \cite[e.g.,][]{Berger2000}.} We emphasize that this is not the case in
general curvilinear coordinates. However, for simple, smooth, unbounded,
geometries such as a plane or sphere, the helicity transport equation admits a
particularly simple form
\begin{equation}
\boxed{\frac{\partial\HelicityS}{\partial
  t}\equiv2\,c\,\oint_{\Manifold}dS\,\ZETA\,B_n.}\label{eqn:RH_Spherical}
\end{equation} 
This form has been noted by \cite{Berger1999} for application to
laboratory plasmas with flux conserving boundary conditions such that
$\B\cdot\nhat|_\Surf=0$ and $\E\equiv-\gradp\psi$.\label{ref:flux} We emphasize that
we have made no such boundary assumptions in
deriving~(\ref{eqn:RH_Spherical}). Indeed we argue below in the
conclusions that the expression~(\ref{eqn:RH_Spherical}) above is, in
fact, the form that should generally be used for measuring the
helicity transport through the photosphere and into the corona. \par
\section{Surface Helicity Transport: Emerging, Shearing, and Nonideal Effects\label{sec:Emerge:Shear}}
For ideal boundary motions, the surface helicity transport $\partial
\HelicityS/\partial t$ can be changed by two MHD processes: (1)
emergence\---the transport of linked flux across the surface $\Surf$
and (2) shearing\----the twisting and tangling of footpoints by
motions in the surface. Using the helicity transport
expression~(\ref{eqn:RH_Simpliest}), we can now derive the helicity
transport terms due to emergence and shearing. For comparison with
previous work, recall that for ideal surface motions
\cite{Berger1984b} decomposed the integrand of the surface helicity
transport~(\ref{eqn:Berger1984}) into two terms: one involving the
motion of magnetic flux through $\Surf$ (emergence) and a second which
transports helicity through the surface by motions tangent to 
$\Surf$ (shearing)
\begin{equation}
\frac{\partial\HelicityBFS}{\partial
  t}=-2\,\oint_{\Surf}dS\,\nhat\cdot\APotC\cross\left(\v\cross\B\right)=2\,\oint_{\Surf}dS\,\underbrace{\left(\APotC\cdot\B\right)\,v_n}_{\mathrm{Emergence}}-2\,\oint_{\Surf}dS\,
\underbrace{\left(\APotC\cdot\v\right)\,B_n}_{\mathrm{Shearing}}.
\end{equation}
The first expression involves $\v\cross\B$, which ensures that
velocity parallel $v_\parallel$ to the magnetic field plays no role in
helicity transport through $\Surf$. However, once this integrand is
decomposed into emerging and shearing constituents, in the second
expression, each term can individually depend on $v_\parallel$. To
clarify this point, the decomposition~(\ref{eqn:fS}) adapted to the
magnetic field is
\begin{subequations}
\begin{align}
\v_\parallel=&\left(\v\cdot\B\right)\,\B/B^2,\\
\v_\perp=&\left(\B\cross\v\right)\cross\B/B^2=\v-v_\parallel\,\B/B,\label{eqn:vperp}
\end{align}
\end{subequations}
which separates plasma flows parallel to the magnetic field
$\left(\v_\parallel\right)$ and perpendicular to the magnetic field
$\left(\v_\perp\right)$ by construction. The total velocity in terms
of the magnetic field vector is
\begin{equation}
\v=\v_\perp+v_\parallel\,\B/B,
\end{equation}  
producing the emerging and shearing helicity transport terms
\begin{subequations}
\begin{align}
\left(\frac{\partial\HelicityBFS}{\partial
  t}\right)_{\Em}=&2\,\oint_{\Surf}dS\,\left(\APotC\cdot\B_\Surf\right)\,\left(\v_{\perp{n}}+v_\parallel\,B_n/B\right),\\
\left(\frac{\partial\HelicityBFS}{\partial
  t}\right)_{\Sh}=&-2\,\oint_{\Surf}dS\,\left[\APotC\cdot\left(\v_{\perp\Surf}+v_\parallel\,\B_\Surf/B\right)\right]\,B_n.
\end{align}
\end{subequations}
The $\v_\parallel$ terms locally cancel when combined
in~(\ref{eqn:Berger1984}) and thus lead to no net pointwise helicity
transport.  However, these parallel velocities, which play no role in
the magnetic evolution, bias the individual emergence and shearing
terms. To correct this bias, only the components of the perpendicular
velocity~(\ref{eqn:vperp}) are usually used to compute the emerging
and shearing helicity transport:
\begin{subequations}
\begin{align}
\left(\frac{\partial\HelicityBFS}{\partial
  t}\right)_{\Em}=&2\,\oint_{\Surf}dS\,\left(\APotC\cdot\B_\Surf\right)\,\v_{\perp{n}},\\
\left(\frac{\partial\HelicityBFS}{\partial
  t}\right)_{\Sh}=&-2\,\oint_{\Surf}dS\,\left(\APotC\cdot\v_{\perp\Surf}\right)\,B_n,
\end{align}
\end{subequations}
but these expressions are still explicitly gauge dependent.  \par
A more rigorous procedure is to split the contributions to the
$\v\cross\B$ electric field into emergence and shearing terms by
constructing electric fields corresponding to the two MHD surface
processes:
\begin{subequations}
\begin{alignat}{2}
\mbox{Emerging}\quad&&\E_\Em&=-v_{\perp{n}}\nhat\cross\B_\Surf/c,\label{eqn:E:Emerging}\\
\mbox{Shearing}\quad&&\E_\Sh&=-\left(\v_{\perp\Surf}\cross{B}_n\,\nhat+\v_{\perp\Surf}\cross\B_\Surf\right)/c,\label{eqn:E:Shearing}
\end{alignat}
\end{subequations}
each of which individually satisfies $\E_\Em\cdot\B=\E_\Sh\cdot\B=0$.
Note that the shearing term supports an electric field in the normal
direction except when $\B_\Surf=0$. For ideal
motions,~(\ref{eqn:Helmholtz:Surface:DivS:E}) separates into two terms
for emerging and shearing:
\begin{subequations}
\begin{align}
\Lapp\ZETA_\Em=&\frac{1}{c}\,\gradp\cdot\left(\v_{\perp{n}}\,\nhat\cross\B_\Surf\right)\quad\in\Manifold.\label{eqn:ZETAEM:Ideal}\\
\Lapp\ZETA_\Sh=&\frac{1}{c}\,\gradp\cdot\left(\v_{\perp\Surf}\cross\nhat\,{B_n}\right)\quad\in\Manifold,\label{eqn:ZETASH:Ideal}
\end{align} 
\end{subequations}
where by superposition, $\zeta=\zeta_\Em+\zeta_\Sh$. Note that the
$\left(\nhat\cross\E_\Sh\right)\cross\nhat$
in~(\ref{eqn:Helmholtz:Surface:DivS:E}) under the divergence kills the normal
electric field $\v_\Surf\cross\B_\Surf$. In many unbounded geometries of
interest (spheres with $\nhat=\rhat$, planes, etc.),
${\ZETAPot}_{i}=\mbox{constant}$, and ${\ZETAPot}_{i}$ has no effect on the
emergence or shearing helicity transport and can simply be ignored.  The
surface helicity transport equation is then
\begin{equation}
\boxed{\frac{\partial \HelicityS}{\partial t}=2\,c\,\sum_{i}\oint_{\Surf}{dS}\,\ZETA_i\,B_n,}\label{eqn:Emergence:Shearing}
\end{equation}
where, here and below, the sum over $i\in\left(\Em,\Sh\right)$ represents the
emerging and shearing terms. Each term is individually gauge
invariant in this formalism by the solenoidal property of $\B$ and the
uniqueness, to within arbitrary constants, of the $\ZETA_i$'s. Consequently,
each term in the sum can be considered independently as an observable.\par
In general curvilinear coordinates, the helicity transport equation
involves ${\ZETAPot}_{i}$
\begin{equation}
\boxed{\frac{\partial \HelicityS}{\partial t}=2\,c\,\sum_{i}\oint_{\Surf}{dS}\,\left(\ZETA_i-{\ZETAPot}_{i}\right)\,B_n,}\label{eqn:Emergence:Shearing:General}
\end{equation}
and the individual contributions to $\epot$ from emergence and
shearing must be determined. Because $\epot$ is determined directly from
the changes in the normal component of the magnetic field
at the boundary, then as with the electric field above, we can
maintain gauge invariance by splitting these changes into the
contributions from emerging and shearing velocities:
\begin{subequations}
\begin{align}
\left(\frac{\partial \bpot_n}{\partial t}\right)_\Em=\Lapp\frac{\partial{\PhiPot}_{\Em}}{\partial
  t}=&\nhat\cdot\gradp\cross\left(v_{\perp{n}}\,\nhat\cross\B_\Surf\right)\qquad\in\Manifold,\label{eqn:Chi:Em}\\
\left(\frac{\partial \bpot_n}{\partial t}\right)_\Sh=\Lapp\frac{\partial{\PhiPot}_{\Sh}}{\partial
  t}=&\nhat\cdot\gradp\cross\left(\v_{\perp\Surf}\cross\nhat\,B_n\right)\qquad\in\Manifold.\label{eqn:Chi:Sh}
\end{align}
\end{subequations}
Each of these processes individually can produce a change in $\BPot$ at the
surface $\Surf$ and the $\partial {\PhiPot}_{i}/\partial t$'s can be established
directly from these changes. Again, the normal electric field
$\nhat\cdot\grad\cross\left(\v_\Surf\cross\B_\Surf\right)=0$ does not affect
the value $\partial{\PhiPot}_\Em/\partial t$ or $\partial
{\PhiPot}_{\Sh}/\partial t$ and by superposition, $\partial{\PhiPot}/\partial
t=\partial{\PhiPot}_{\Em}/\partial t+\partial {\PhiPot}_{\Sh}/\partial t$. The
emergence and shearing terms for $\epot$ can now be expressed in terms of the
relevant constituents of $\partial B_n/\partial t$ using~(\ref{eqn:EP:Unique:Surface}):
\begin{subequations}
\label{eqn:Missing:Green}
\begin{align}
{\epot}_{\Em}\left(\x,t\right)=&c^{-1}\,\int_\Vol{d^3x'}\,\Kernel_\Vol\left(\x,\x'\right)\cdot\grad'
\oint_\Surf dS''\,\GreenNP\left(\x',\x''\right)\,\left[\frac{B_n\left(\x'',t\right)}{\partial t}\right]_\Em,\\
{\epot}_{\Sh}\left(\x,t\right)=&c^{-1}\,\int_\Vol{d^3x'}\,\Kernel_\Vol\left(\x,\x'\right)\cdot\grad'
\oint_\Surf dS''\,\GreenNP\left(\x',\x''\right)\,\left[\frac{B_n\left(\x'',t\right)}{\partial t}\right]_\Sh.
\end{align}
\end{subequations}
Taking the surface divergence of the potential reference electric
fields produces expressions for the emerging and shearing lamellar
potentials in terms of the motions corresponding to the two ideal MHD
surface processes:
\begin{subequations}
\begin{align}
\Lapp{\ZETAPot}_\Em\left(\x,t\right)=&-c^{-1}\,\gradp\cdot\int_\Vol{d^3x'}\,\Kernel_\Vol\left(\x,\x'\right)\cdot\grad'
\oint_\Surf dS''\,\GreenNP\left(\x',\x''\right)\,\nhat''\cdot\gradp''\cross\left[v_{\perp{n}}\left(\x'',t\right)\,\nhat''\cross\B_\Surf\left(\x'',t\right)\right],\label{eqn:Zeta:Em}\\
\Lapp{\ZETAPot}_\Sh\left(\x,t\right)=&-c^{-1}\,\gradp\cdot\int_\Vol{d^3x'}\,\Kernel_\Vol\left(\x,\x'\right)\cdot\grad'
\oint_\Surf dS''\,\GreenNP\left(\x',\x''\right)\,\nhat''\cdot\gradp''\cross\left[\v_{\perp\Surf}\left(\x'',t\right)\cross\nhat''\,B_n\left(\x'',t\right)\right],\label{eqn:Zeta:Sh}
\end{align}
\end{subequations}
where again, by superposition,
${\ZETAPot}={\ZETAPot}_\Em+{\ZETAPot}_\Sh$. Note that
(\ref{eqn:Chi:Em})\--(\ref{eqn:Chi:Sh}) and
(\ref{eqn:Zeta:Em})\--(\ref{eqn:Zeta:Sh}) directly couple the
${\PhiPot}_i$ and ${\ZETAPot}_i$. The ${\ZETAPot}_i$ are unique to
within an arbitary constant and dependent only on observables $\v$ and
$\B$.  \par
For application to observations, it is useful to re-express the ideal
plasma velocity in terms of the constituents that produce the changes
in the normal component of the magnetic field $\v_{\In}$ and transport
helicity $\v_{\Helicity}$ across the boundary.  For an ideal electric
field, the Helmholtz\---Hodge decomposition on $\Surf$ has
\begin{equation}
\tau=\frac{1}{B_n}\,\B_\Surf\,\cdot\left[\frac{1}{c}\,\nhat\cross\gradp\frac{\partial\PHI}{\partial t}+\gradp\ZETA\right],
\end{equation}
in~(\ref{eqn:Helmholtz:Surface:E}) to enforce $\E\cdot\B=0$.  The
ideal plasma velocity can be expressed on $\Surf$ as 
\begin{subequations}
\begin{equation}
\v=\v_\parallel\,\B/B+\sum_{i}\left(\v_{\In,{i}}+\v_{\Helicity,{i}}\right).
\end{equation}
where
\begin{align}
\v_{\In,{i}}=&\frac{c}{B^2}\,\left\lbrace {\epot}_i 
-\frac{\nhat}{B_n}\,\B_\Surf\cdot{\epot}_i\right\rbrace\cross\B,\nonumber\\
=&\frac{c}{B^2}\,\B\cross\left\lbrace\nhat\cross\gradp\left(\frac{1}{c}\,\frac{\partial{\PhiPot}_i}{\partial t}\right)
+\gradp{\ZETAPot}_i-\frac{\nhat}{B_n}\,\B_\Surf\cdot\left[\nhat\cross\gradp\left(\frac{1}{c}\,\frac{\partial{\PhiPot}_i}{\partial t}\right)+\gradp{\ZETAPot}_i\right]\right\rbrace,\label{eqn:vB} \\ 
\v_{\Helicity,{i}}=&\frac{c}{B^2}\,\left\lbrace \left(\E_i-{\epot}_i\right)_\Surf 
-\frac{\nhat}{B_n}\,\B_\Surf\cdot\left(\E_i-{\epot}_i\right)\right\rbrace\cross\B,\nonumber\\
=&\frac{c}{B^2}\,\B\cross\left[\gradp\left({\ZETA}_i-{\ZETAPot}_i\right)-\frac{\nhat}{B_n}\,\B_\Surf\cdot\gradp\left({\ZETA}_i-{\ZETAPot}_i\right)
\right],\label{eqn:vH}
\end{align} 
\end{subequations}
are the ideal MHD constituent velocities that change the normal
component of the magnetic field and that transport helicity,
respectively. Note that all three components of the ideal plasma
velocity can be reconstructed from just the surface components of the
ideal electric field consituents produced by the emerging $\E_\Em$ and
shearing $\E_\Sh$ processes, due to the ideal MHD constraints
$\v_{\In,i}\cdot\B=0$ and $\v_{\Helicity,i}\cdot\B=0$. The total ideal
electric field is reconstructed exactly by the sum of the
$\v\cross\B$ produced by the constituent velocities
$\sum_i{\E_i}=-\sum_i{\left(\v_{\Helicity,{i}}+\v_{\In,{i}}\right)\cross\B/c}$. If
we interpret $\v_{\In}$ as reconstructing the reference electric field
that produces changes in $\BPot$ on $\Surf$, then
\begin{subequations}
\begin{align}
{\EPot}_i=&-\v_{\In,i}\cross\B/c+\grad{\GaugePp}_i,\nonumber\\
=&-\nhat\cross\gradp\left(\frac{1}{c}\,\frac{\partial{\PhiPot}_i}{\partial t}\right)
-\gradp{\ZETAPot}_i+\frac{\nhat}{B_n}\,\B_\Surf\cdot\left[\nhat\cross\gradp\left(\frac{1}{c}\,\frac{\partial{\PhiPot}_i}{\partial t}\right)+\gradp{\ZETAPot}_i\right]+\grad{\GaugePp}_i,\\
=&{\epot}_i+\grad{\GaugeP}_i\qquad\in\Surf.\nonumber
\end{align}
In other words, $\v_{\In,{i}}$ is the ideal MHD plasma velocity
consistent with the change in the normal component of the magnetic
field.  Here, $\v_{\In,{i}}$ reconstructs the surface components of
$\EPot$ exactly, and the freedom in $\GaugeP$ can account for the
normal component of $\EPot$ produced by $\v\cross\B$
\begin{equation}
\frac{\partial{\GaugeP}_i}{\partial n}=\frac{\partial{\GaugePp}_i}{\partial n}+\frac{1}{B_n}\,\B_\Surf\cdot\left[\nhat\cross\gradp\left(\frac{1}{c}\,\frac{\partial{\PhiPot}_i}{\partial t}\right)+\gradp{\ZETAPot}_i\right]\qquad\in\Surf,
\end{equation}
which, with (\ref{eqn:DivS_GradS_psi}), requires that 
\begin{equation}
\grad\cdot{\EPot}_i\equiv\rhoP=\overbrace{\Lapp{\GaugeP}_i}^{(\ref{eqn:LambdaP:Dirichlet})}-\Jfirst\,\frac{\partial{\GaugeP}_i}{\partial
  n}+\frac{\partial^2{\GaugeP}_i}{\partial n^2}\qquad\in\Surf,
\end{equation}
where the term with the overbrace is zero by the referenced equation.
 As an example of the considerable freedom in $\GaugeP$ relevant to solar
physics, consider the space between two spherical shells at $r_1$ and
$r_2$. Writing
\begin{equation}
\GaugeP=\frac{\left(r-r_1\right)\,\left(r-r_2\right)^2\,E_{n1}\left(\vartheta,\varphi\right)+\left(r-r_1\right)^2\,\left(r-r_2\right)\,E_{n2}\left(\vartheta,\varphi\right)}{\left(r_2-r_1\right)^2},
\end{equation}
\end{subequations}
which satisfies $\GaugeP=0$ at both $r=r_1$ and $r=r_2$ while leaving
$\partial\GaugeP/\partial n$ arbitrary at those two surfaces,
demonstrates that $\EPot$ can accomodate the normal component of $\E$
produced by $\v\cross\B$.  \par
While ${\ZETAPot}_\Em$ and ${\ZETAPot}_\Sh$ are not observables,
$\v_{\Helicity,\Em}$ and $\v_{\Helicity,\Sh}$ depend on the gradients
of the corresponding lamellar potentials and thus can be used to
determine where the two MHD surface processes are operating. Note,
however, that a finite $\v_{\Helicity,\Em}$ or $\v_{\Helicity,\Sh}$
does not necessarily imply a finite helicity transport due to the
nonlocal nature of helicity. This is demonstrated directly below in \S~\ref{sec:Linked}.\par
\subsection{Nonideal Effects\label{sec:Nonideal}}
Although we have focused on ideal transport effects, we emphasize
that~(\ref{eqn:RH_Simple}) and~(\ref{eqn:RH_Simpliest}) do not
inherently require ideal evolution to estimate
$\partial\HelicityS/\partial t$, which is itself a gauge-invariant
observable. If we have independent measurements of $\E$, $\v$, and
$\B$, then the nonideal electric field $\ENI$ can be estimated from
\begin{subequations}
\begin{equation}
\ENI=\E+\frac{1}{c}\,\v\cross\B,
\end{equation}
which produces a nonideal change in the normal component of $\B$ given by
\begin{equation}
\left(\frac{\partial B_n}{\partial t}\right)_\mathrm{NI}=-c\,\nhat\cdot\grad\cross\ENI.
\end{equation}
\end{subequations}
Once $\ENI$ and $\left({\partial B_n}/{\partial t}\right)_\mathrm{NI}$
are established, then the corresponding
$\partial{\PHI_\NI}/\partial t$ and $\ZETANI$ can be
computed directly from the Helmholtz\---Hodge decomposition of
$\ENI$. Similarly, the nonideal ${\Tepot}$ can in principle be
determined as in Appendix~{\ref{sec:EPot:Revisited} and then
  $\ZETAPotNI$ can be computed from the
  Helmholtz\---Hodge decomposition of ${\Tepot}$. The nonideal
  surface helicity transport takes the form
\begin{equation}
\boxed{\left(\frac{\partial \HelicityS}{\partial t}\right)_\mathrm{NI}=2\,c\,\sum_{i}\oint_{\Surf}{dS}\,\left(\ZETANI-\ZETAPotNI\right)\,B_n.}
\end{equation}
Of course, generally, a nonideal electric field will produce a
volumetric contribution in addition to the surface term above, leading
to a nonideal term in $\dHelicityV$ and an overall nonideal
contribution to the helicity transport rate $\dHelicity$.
\subsection{The Emergence of Disconnected Linked Flux\label{sec:Linked}}
\begin{figure}
\centerline{\includegraphics[width=4in]{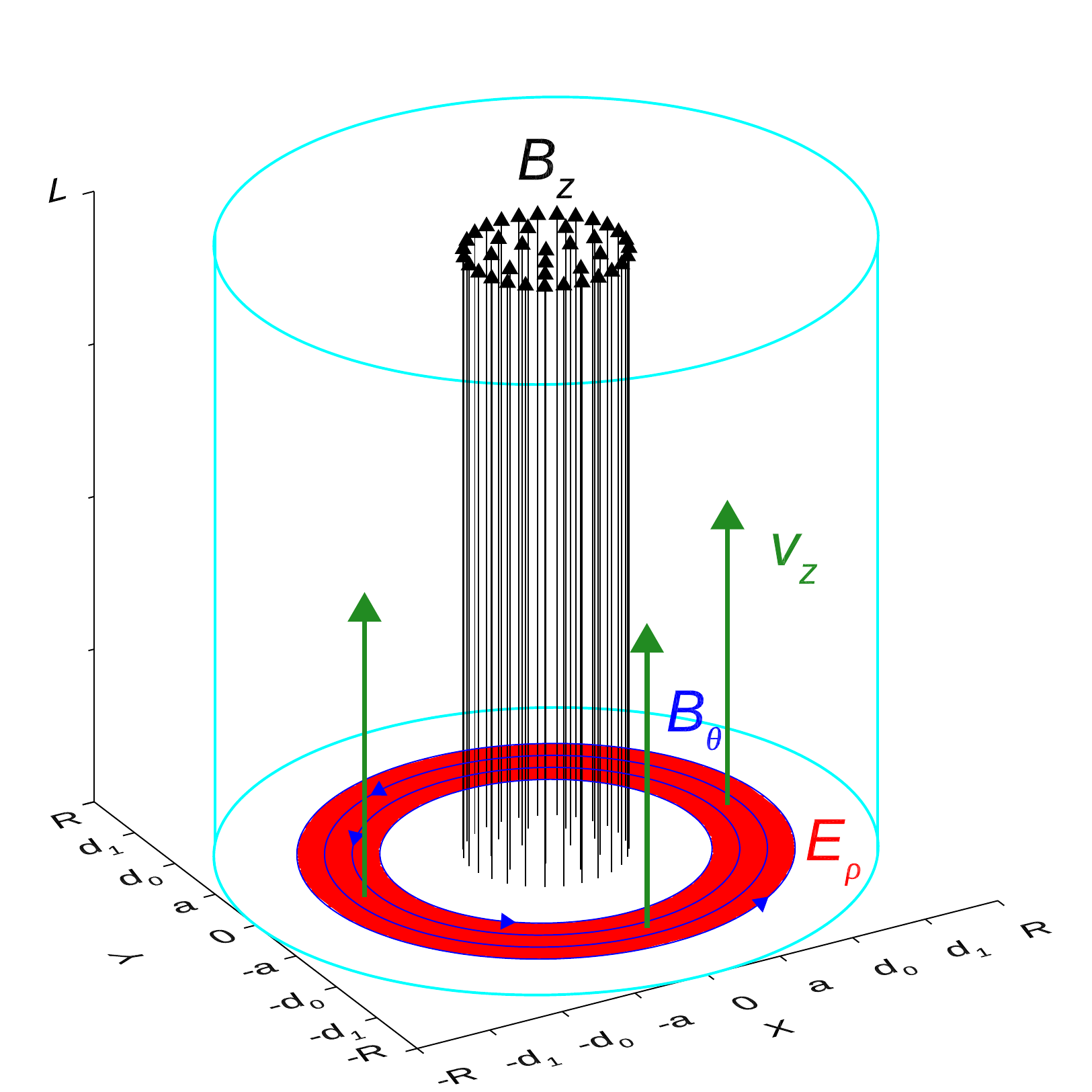}}
\caption{The emergence of disconnected linked field. The bounding
  cylindrical surface (cyan) contains a uniform column of vertical
  field $B_z$ for $\rho\le{a}$ (black arrows). Closed rings of
  azimuthal field $B_\theta$ are injected at the bottom boundary
  $\BthF_\Inner\le{\rho}\le\BthF_\Outer$ (blue arrows) by a uniform vertical
  flow $v_z$ (green arrows), which produces a ring of radial electric
  field $E_\rho=v_z\,B_\theta/c$ (red).\label{fig:linked_flux}}
\end{figure}
To clarify the nonlocal nature of our expressions for helicity
transport, consider the simple case of the emergence of a horizontal ring of
closed field (zero self-helicity) that links with a vertical column of
untwisted field, Figure~\ref{fig:linked_flux}. In this case all the helicity
generated is due to the linkage of the two systems. The key point is that this
case demonstrates clearly  that the concept of a ``helicity flux
density'' is not valid. Let the domain be the cylinder
$\rho\in\left[0,\Rcyl\right]$ and $z\in\left[0,\Lcyl\right]$ shown in
Figure~\ref{fig:linked_flux}. The initial magnetic field is given by
\begin{subequations}
\begin{align}
\B\left(\rho,z\right)=&B_1\,\UnitStep\left(-z\right)\,\left[\UnitStep\left(\rho-\BthF_\Inner\right)-\UnitStep\left(\rho-\BthF_\Outer\right)\right]\,\thetahat+B_0\,\left[1-\UnitStep\left(\rho-a\right)\right]\zhat,
\end{align}  
where $\Rcyl\gg \BthF_\Outer>\BthF_\Inner>a$, and the velocity is simply a
constant vertical flow,
\begin{equation}
\v\left(\rho,z\right)=v_{0}\,\zhat,  
\end{equation}  
\end{subequations}
with the unit 
step function is defined as
\begin{equation}
  \UnitStep\left(x\right)=\left\lbrace\begin{aligned}0\quad&{x<0},\\
  1\quad&{x\ge0}.
  \end{aligned}\right.
\end{equation}  
Neither cap of the cylinder is flux balanced, but the entire surface
satisfies $\oint_\Surf{dS}\,B_n=0$.  The potential due to emergence is
\begin{subequations}
\begin{equation}
\ZETAE\left(\rho,0\right)=\ZETA_0-\frac{v_{0}\,B_1}{c}\,\left[\left(\rho-\BthF_\Inner\right)\,\UnitStep\left(\rho-\BthF_\Inner\right)-\left(\rho-\BthF_\Outer\right)\,\UnitStep\left(\rho-\BthF_\Outer\right)\right],
\end{equation}
where we redefine the arbitrary constant $\ZETA_0$ in terms of the value of
$\ZETAE$, as $\rho\rightarrow{\Rcyl}$ namely $\ZETAER$,
\begin{equation}
\ZETA_0=\ZETAER+\frac{v_0\,B_1}{c}\,\left(\BthF_\Outer-\BthF_\Inner\right), 
\end{equation}
then by continuity,
\begin{equation}
\ZETAE\left(\rho\ge{\BthF_\Outer},0\right)=\ZETAE\left(\Rcyl,z\right)=\ZETAE\left(\rho,\Lcyl\right)=\ZETAER
\end{equation}
\end{subequations}
Because no magnetic field penetrates the sides for $\rho=\Rcyl$ and
$z\in\left[0,\Lcyl\right]$ or the caps for $\rho\ge{a}$, the helicity transport rate is then
\begin{align}
\frac{d\Helicity}{dt}=&\overbrace{2\,c\,B_0\,\pi\,a^2\,\ZETAER+2\,B_0\,\pi\,a^2\,B_1\,\left(\BthF_\Outer-\BthF_\Inner\right)\,v_0}^{\mbox{bottom surface}}-\overbrace{2\,c\,B_0\,\pi\,a^2\,\ZETAER}^{\mbox{top surface}}\nonumber\\
=&2\,B_0\,\pi\,a^2\,B_1\,\left(\BthF_\Outer-\BthF_\Inner\right)\,v_0.\label{eqn:seventyone}
\end{align}
The choice of constant $\ZETAER$ has no influence on the net helicity
transport rate.\par
Intuitively, we would say that the helicity transport in this situation
is ``emerging'' at the bottom boundary between
$\rho=\BthF_\Inner$ and
$\rho=\BthF_\Outer$, where closed field is being transported across the
boundary. However, the choice of constants influences this
interpretation.  First, we choose
$\ZETAE\left(0,0\right)=0$ and
$\ZETAER=\left(v_0\,B_1/c\right)\,\left(\BthF_\Outer-\BthF_\Inner\right)$. This
implies counterintuitively that there is no helicity transport
through the bottom boundary, and the helicity enters through top
boundary. Alternatively, if we choose
$\ZETAER=0$, then there is no helicity transport through the top
boundary and the helicity appears to enter through the bottom
boundary. Note that in both cases, the integrand is only nonzero at
boundary locations where
$B_n\neq0$, i.e., where normal field penetrates the bounding surface,
but there are no footpoint motions there, because
$\v\cross\B=0$! Even knowing the field line connectivity for this
situation does not resolve the nonlocality of where helicity is
injected on the surface $\Surf$. We conclude, therefore, that helicity
injection, just like helicity itself, is inherently
nonlocal. Associating a local interpretation to the
helicity transport is incorrect; consequently, observational
quantities such as ``helicity flux maps'' are misleading
\cite[]{Pariat2005,Pariat2007,Chandra2010,Romano2011,Vemareddy2012,Vemareddy2015}.
\par
\section{Free Energy Flux}\label{sec:Energy}
As with relative helicity, free energy is defined only with respect to a
reference potential field. To determine its transport, let us consider
a closed volume $\Vol$ with boundary $\Surf$ and magnetic field
$\B$. There is a unique potential field $\BPot$
that satisfies $\grad\cross\BPot=0$ in $\Vol$ and
$\left.\BPot\cdot\nhat\right|_\Surf=\left.\B\cdot\nhat\right|_\Surf$. Following
\cite{Berger1999}, the closed field in the volume may be defined~as
\begin{equation}
\Bcl\equiv\B-\BPot\qquad\mbox{with}\qquad\left.\Bcl\cdot\nhat\right|_\Surf=0.\label{eqn:closed}
\end{equation}
The closed field carries all the current in the volume. Note, however,
that $\Bcl$ may be potential with $\grad\cross\Bcl=0$ in some parts of
the volume, or even on some or all of the surface.  At the surface
$\Surf$, $\Bcl$ represents the tangential field produced by current
systems in the volume or normal to the surface.\par
Because the corona is low $\beta$, the energy transfer into the corona is
almost completely via the magnetic field. In this case, the rate of
energy transfer through the surface $\Surf$ into $\Vol$ is given by
the Poynting flux:
\begin{equation}
  \frac{d\Energy}{dt}=\frac{c}{4\,\pi}\,\oint_\Surf{dS}\,\nhat\cdot\left(\E\cross\B\right).\label{eqn:energy:flux}
\end{equation}
The total potential magnetic energy in the volume $\Vol$ is
\begin{equation}
\Energy_\mathrm{P}=\frac{1}{8\,\pi}\,\int_\Vol{d^3x}\left(\BPot\cdot\BPot\right).\label{eqn:potential:energy}
\end{equation}
The goal is to derive the equivalent equation for the rate of potential
energy transfer across the surface $\Surf$
and subtract this from~(\ref{eqn:energy:flux}) to obtain the rate of free
energy transfer across the surface $\Surf$
\begin{equation}
\frac{d\Energy_\mathrm{F}}{dt}=\frac{d\Energy}{dt}-\frac{d\Energy_\mathrm{P}}{dt}.\label{eqn:Free}
\end{equation}
Because $\BPot$ is the potential field, it may be determined
from~(\ref{eqn:Potential}) in the volume $\Vol$,
and the last term in~(\ref{eqn:Free}) above may be recast as 
\begin{align}
\frac{d\Energy_\mathrm{P}}{dt}=\frac{1}{4\,\pi}\,\int_\Vol{d^3x}
  \BPot\cdot\frac{\partial \BPot}{\partial t}&=
-\frac{1}{4\,\pi}\,\int_\Vol{d^3x}\grad\phiPot\cdot\frac{\partial
                                               \BPot}{\partial t}\nonumber\\
&=-\frac{1}{4\,\pi}\,\int_\Vol{d^3x}\left[\grad\cdot\left(\phiPot\,\frac{\partial
                                               \BPot}{\partial t}\right)
  -\phiPot\,\frac{\partial}{\partial t}\left(\grad\cdot\BPot\right)\right],\label{eqn:Energy:Phi}\\
&=\gsign\frac{1}{4\,\pi}\,\oint_\Surf dS\,\phiPot\frac{\partial
  \bpot_n}{\partial t}.\nonumber
\end{align}
(Recall that we are using the convention that $\nhat$ on the
surface $\Surf$ points into the coronal volume $\Vol$).\par
Using Faraday's law of induction~(\ref{eqn:Faraday}),\footnote{Note that
  $\EPot$ from the previous discussion can be substituted for $\E$ here.}
\begin{equation}
\frac{d\Energy_\mathrm{P}}{dt}=-\gsign\frac{c}{4\,\pi}\,\oint_\Surf
dS\,\nhat\cdot\left(\phiPot\,\grad\cross\E\right).
\end{equation}
with~(\ref{eqn:Curl_psi_f}) and~(\ref{eqn:StokesClosed}), this becomes
\begin{equation}
\frac{d\Energy_\mathrm{P}}{dt}=\gsign\frac{c}{4\,\pi}\,\oint_\Surf
dS\,\nhat\cdot\left[\E\cross\BPot-\grad\cross\left(\phiPot\,\E\right)\right]=\gsign\frac{c}{4\,\pi}\,\oint_\Surf
dS\,\nhat\cdot\left(\E\cross\BPot\right).\label{eqn:energy:flux:potential}
\end{equation}
Substituting~(\ref{eqn:energy:flux}) and~(\ref{eqn:energy:flux:potential})
into~(\ref{eqn:Free}), the rate of change in the free energy is
\begin{equation}
\boxed{\frac{d\Energy_\mathrm{F}}{dt}=\frac{c}{4\,\pi}\,\oint_\Surf
dS\,\nhat\cdot\left[\E\cross\left(\B-\BPot\right)\right]
=\frac{c}{4\,\pi}\,\oint_\Surf
dS\,\nhat\cdot\left(\E\cross\Bcl\right).}\label{eqn:EFREE}
\end{equation}
The result that the free energy transport depends on the presence of a finite
$\Bcl$ at the boundary may seen somewhat obscure, but in fact, it can be
understood from straightforward force arguments. If the field is purely
potential at the boundary, then it exerts no stress there, and any
instantaneous dynamics of the boundary do no work on the field. The presence
of a $\Bcl$ is required for the boundary to add/subtract energy to/from the
field. Furthermore, while the Poynting vector $\E\cross\B$ defines a
physically valid flux density, the free energy transport depends on the
details of the boundary, which determines the potential field
$\BPot$. Consequently, unlike energy transport \cite[see Vol. II,
  p. 27-6\---27-8][]{Feynman1989}, free energy (and helicity) transport has
physical significance only in the context of a specified volume. \par
We also note that free energy transport across the boundary requires currents in the volume
$\Vol$ or at the surface $\Surf$. This can be seen by
substituting~(\ref{eqn:Helmholtz:Surface:E})
with~(\ref{eqn:Helmholtz:Surface:DivS:E}\---\ref{eqn:Helmholtz:Surface:PHI}) for $\E$ producing,
\begin{equation}
\frac{d\Energy_\mathrm{F}}{dt}=-\oint_\Manifold
dS\,\nhat\cdot\left\lbrace\frac{1}{4\,\pi}\,\left[\nhat\cross\gradp\left(\frac{\partial\PhiPot}{\partial t}\right)\right]\cross\Bcl+\frac{c}{4\,\pi}\,\gradp\ZETA\,\cross\Bcl\right\rbrace.\label{eqn:eighty}
\end{equation}
Using~(\ref{eqn:fxgxh}) on the first term with $\Bcl\cdot\nhat$ followed
by~(\ref{eqn:DivS_psi_f}) and using~(\ref{eqn:CurlS_psi_f}) on the second term
produces
\begin{equation}
\frac{d\Energy_\mathrm{F}}{dt}=\oint_\Manifold
dS\,\nhat\cdot\left\lbrace\frac{\nhat}{4\,\pi}\,\left[\gradp\cdot\left(\frac{\partial\PhiPot}{\partial t}\,\Bcl\right)-\left(\frac{\partial\PhiPot}{\partial t}\right)\,\gradp\cdot\Bcl\right]-\frac{c}{4\,\pi}\,\gradp\cross\left(\ZETA\,\Bcl\right)+\frac{c}{4\,\pi}\,\ZETA\,\gradp\cross\Bcl\right\rbrace.\label{eqn:eightyone}
\end{equation}
Applying~(\ref{eqn:StokesClosed}) and~(\ref{eqn:SDT:closed}) and noting
that any current though the surface must be produced by the curl of
$\Bcl$,
\begin{equation}
\nhat\cdot\grad\cross\Bcl=\nhat\cdot\grad\cross\B=\frac{4\,\pi}{c}\,J_n,\label{eqn:Amperes}
\end{equation}
leads to
\begin{equation}
\boxed{\frac{d\Energy_\mathrm{F}}{dt}=\oint_\Manifold
dS\,\left[\ZETA\,J_n-\frac{1}{4\,\pi}\,\frac{\partial\PhiPot}{\partial t}\gradp\cdot\Bcl\right].}\label{eqn:Freedom}
\end{equation}
The first term involves local currents normal to the surface and the
lameller electric field, whereas the second term describes purely
inductive changes in the surface fields due to currents in the volume
and involves the solenoidal (inductive) electric field.  We conclude
that as with helicity, a net free energy transport requires the
presence of electric currents. Furthermore, the free energy and free
energy transport are nonlocal quantities that depend on the surface
$\Surf$ bounding the volume $\Vol$ that determines the potential
field. Whereas the Poynting flux $\E\cross\B$ can be calculated
locally and is invariant to changes in the shape of the volume away
from the local point of interest, the local free energy transport
$\E\cross\Bcl$ can change in response to nonlocal modifications in the
shape of the volume.\par
\section{Discussion and Conclusions\label{sec:Conclusions}}
The most important conclusion from the results above is that there
does, indeed, exist a gauge-invariant expression for the surface
helicity transport, namely,~(\ref{eqn:RH_Simple})
which we repeat here for completeness:
\begin{align*}
\frac{\partial\HelicityS}{\partial
  t}=&-c\,\oint_{\Surf}dS\,\nhat\cdot\left[\left(\A+\APot\right)\cross\left(\E-\EPot\right)\right].\\
\end{align*}
Although this expression may not always have computational advantages over the
\cite{Berger1984a} expression for the helicity transport, it has major
theoretical advantages. It resolves the long-standing concern that, while the
volumetric helicity itself could be readily expressed in a fully
gauge-invariant form using the \cite{Berger1984a} or \cite{Finn1985} formulas,
the surface helicity transport apparently could not. The \cite{Berger1984a}
expression explicitly requires the Coulomb gauge. Our expression above is
valid in any gauge for either $\A$ or $\APot$ and with arbitrary
$\grad\cdot\EPot=\rhoP$. \par
The key physical insight that is used to derive this gauge-invariant
expression is the requirement that no self-helicity is generated in the
potential field by its evolution\----in other words,
ansatz~(\ref{eqn:Reference}).  In fact, this ansatz is
physically no different than the standard assumption used in every discussion
of helicity --- that a potential magnetic field has zero helicity. Given this
assumption, then~(\ref{eqn:Reference}) inevitably follows. Mathematically, this
ansatz leads to the decomposition of the reference electric field,
\begin{equation}
\EPot=\epot+\grad\GaugeP=-\frac{1}{c}\,\frac{\partial\APot}{\partial t}-\grad\psiPot,\label{eqn:EPot:APC}
\end{equation} 
into a unique intrinsically solenoidal reference electric field
$\epot$ and the gradient of an arbitrary function $\GaugeP$ which
satisfies Dirichlet boundary conditions on $\Surf$.  For a closed
domain, the potential magnetic field is determined uniquely by its
boundary values, and its temporal evolution is also uniquely specified
by the evolution of these boundary values; consequently, there should
exist a unique intrinsically solenoidal electric field $\epot$
corresponding to the changing $\BPot$.  The conditions on $\epot$ and
$\GaugeP$ are summarized by Faraday's law~(\ref{eqn:Faraday}) which
relates $\grad\cross\epot$ and ${\partial\BPot}/{\partial t}$ and
conditions~(\ref{eqn:eP:Div})\--(\ref{eqn:LambdaP:Dirichlet}).  Note,
however, that $\EPot$ itself is not unique because its
divergence is arbitrary, and thus, the vector potential
in~(\ref{eqn:EPot:APC}) admits a gauge transformation.\par
From the general surface helicity transport expression above, it is
possible to derive a somewhat simpler, but still fully general,
expression~(\ref{eqn:RH_Simpliest}), included here for
convenience:
\begin{displaymath}
  \frac{\partial\HelicityS}{\partial
  t}\equiv2\,c\,\oint_{\Manifold}dS\,\left(\ZETA - \ZETAPot\right)\,B_n.
  \end{displaymath}
\begin{enumerate}
\item This expression is manifestly gauge invariant as it is only
  dependent on the potentials $\ZETA$ and $\ZETAPot$ and the
  observable $B_n$. While $\ZETA$ and $\ZETAPot$
  are still arbitrary to within a constant, e.g., 
  $\ZETA\rightarrow\ZETA+\ZETA_0$, this constant $\ZETA_0$ has no
  effect on the rate of change of relative helicity by virtue of the solenoidal
  property of the magnetic field $\left(\oint_{\Manifold}dS\,\ZETA_0\,B_n=0\right)$.
\item This expression is explicitly dependent on the flux
  threading the surface. 
\begin{itemize}
\item The helicity $\Helicity$ can only be changed by boundary motions if
  there is flux threading the bounding surface.  Isolation means
  $\left.\B\cdot\nhat\right|_\Manifold\equiv0$. 
\item The helicity $\Helicity$ in an isolated system evolving
  according to ideal motions is a robust invariant. If
  $\left.\B\cdot\nhat\right|_\Manifold=0$ then
  ${\partial\Helicity}/{\partial t}\equiv0$.
\end{itemize}
\item This expression is only dependent on the lamellar electric field. The
  instantaneous solenoidal (inductive) electric field, which changes the
  normal component of the magnetic field on the boundary, does not contribute
  to helicity transport across the boundary. Thus, \citeauthor{Prior2014}'s
  criticism of \cite{Berger1984a} that when ``the boundary conditions
  $\left.B_n\right|_\Manifold$ are changing in time, ..... the evolution of
  the relative helicity will mix up both real topological changes in $\B$ and
  those simply due to the change of $\BPot$'' \cite[p. 2 in][]{Prior2014}
  does not hold for~(\ref{eqn:RH_Simpliest}).
\item This expression can be unambiguously decomposed into independent
  gauge-invariant expressions for the helicity transport produced by the
  emergence of magnetic field represented by $\ZETAE$, the shearing of field
  magnetic field represented
  $\ZETAT$, and nonideal effects represented by $\ZETANI$, where
  $\ZETA\equiv\ZETAE+\ZETAT+\ZETANI$.
\end{enumerate}
This final conclusion is highly important for studies of the energy
buildup leading to solar eruptions. Given accurate vector magnetograph
data, the expressions derived in \S~\ref{sec:Emerge:Shear} can be
applied to measure the different contributions to the helicity
injection. For the fully general case that includes $\ZETAPot$, the
calculations would be somewhat tedious, but for the special case of a
spherical domain (a coronal domain consisting of the volume between
two spherical shells), then the $\ZETAPot$ drops out, and the
transport reduces to the simple
expression~(\ref{eqn:RH_Spherical}). While this simple expression is
valid only for special domains, the spherical domain should always be
used, if possible, for the solar atmosphere.\par
As discussed in the introduction the concept of helicity is valid
only for a closed system. While the photosphere constitutes a true
boundary to the corona, there are no other physically meaningful
boundaries. Given this fact, the typical assumption is to take as the closing
boundary some spherical surface sufficiently far above the corona so
that the radial field can be approximated as vanishing there. In that
case,~(\ref{eqn:RH_Spherical}) implies that only the photosphere
contributes to the helicity transport. Under this assumption, our
result above,~(\ref{eqn:RH_Spherical}), can be used to determine the
gauge-invariant helicity transport through a spherical photospheric
surface.  \par
Many authors, however, have attempted to measure coronal helicity
transport using box-like rather than spherical domains
\cite[e.g.][]{Chae2001b,Kusano2002a,Nindos2003a,Pevtsov2003,Pariat2005,Demoulin2007a,Vemareddy2012,Liu2014}.
The usual assumption in using such a domain is that there is minimal
helicity transport through the side boundaries in the corona, because
the velocities are small there. The problem with using such a domain
is that the fully general helicity transport expressions, including
calculation of the intrinsically solenoidal part of the reference
electric field, must be used. Given the complexities of the
calculations involved, we defer to a subsequent paper a discussion
and demonstration of the detailed application of our results to finite
Cartesian domains. Another problem with having side walls in the
corona is that even with minimal photospheric velocities, coronal reconnection
may efficiently transport helicity to the walls via the process of
helicity condensation \cite[]{Antiochos2013}.  \par
Our first-principles approach to the surface helicity transport rate
can be connected with the transport discussed by \cite{Mackay2014} in
the context of helicity condensation \cite[]{Antiochos2013}. The rate
of change in the vector potential produced by the electric field
transporting helicity can be written generally for emergence or
shearing in spherical coordinates as
\begin{displaymath}
\left[\frac{\partial \A}{\partial t}\right]_\Helicity=c\,\gradp\ZETA.
\end{displaymath}
This leads to the same form for the evolution of the surface components of
the magnetic field proposed by \cite{Mackay2014},\footnote{Note that
  the difference in notation between \cite{Mackay2014} and the present
  work. The former denotes the cyclonic parameter by
  $\zeta=l^2\,\omega_l/2$ and the present work denotes the lamellar
  potential by $\ZETA$.}
\begin{displaymath}
\left[\frac{\partial \B_\Surf}{\partial t}\right]_\Helicity=\grad\cross\left[\frac{\partial \A}{\partial t}\right]_\Helicity=-c\,\grad\cross\left(\rhat\,\frac{\partial\ZETA}{\partial r}\right),\qquad\left[\frac{\partial B_n}{\partial t}\right]_\Helicity=0.
\end{displaymath}
A solution to this system is
$\v_\Surf=\left(c/B_r\right)\,\rhat\cross\grad\ZETA$ for any
$\ZETA$. If we choose $\zeta$ to be constant on flux surfaces, 
$\v_\Surf=\left(c/B_r\right)\,\rhat\cross\gradp\ZETA\left(B_r\right)$,
then the surface helicity transport can be written
${\partial\HelicityS}/{\partial
  t}\equiv2\,c\,\oint_{\Manifold}dS\,\ZETA\left(B_r\right)\,B_r$ as
purely a function of $B_r$. If we consider a statistical ensemble of
vorticies of scale $l$ and average rotation rate $\omega_l$, the
lamellar potential can be written
$\ZETA\left(B_r\right)=B_r\,l^2\,\omega_l/\left(2\,c\right)$ and the
result ${\partial\HelicityS}/{\partial
  t}\equiv2\,\oint_{\Manifold}dS\,\left(l^2\,\omega_l/2\right)\,B_r^2$
from \cite{Mackay2014} is recovered.\par
The other essential quantity determining solar coronal activity is
magnetic free energy. As discussed in the introduction, energy and
energy transport are generally much more familiar to most researchers
than their helicity counterparts;
however, for completeness, we have derived the expression for free
energy transport across the surface $\Surf$ of a closed volume:
\begin{displaymath}
\frac{dE_\mathrm{F}}{dt}=\frac{c}{4\,\pi}\,\oint_\Surf
dS\,\nhat\cdot\left(\E\cross\Bcl\right).
\end{displaymath}
Note that the free energy transport is manifestly dependent on the
closed field $\Bcl = \B - \BPot$ on the boundary. This result
emphasizes that the field at the boundary must be nonpotential for
free energy to be injected into the corona. Unlike energy flux, there
can be no free energy transport into the volume in the absence of
electric currents within the volume or at the surface $\Surf$.\par
Although the free energy transport expression may seem simpler than
the expressions for helicity transport, it is, in fact, much less
amenable to measurement by observations. The problem is that the free
energy transport depends primarily on the tangential components of the
magnetic field at the boundary. These components are more difficult to
measure than the normal component near the center of the solar disk
where active regions are observed at the highest resolution by present
instruments such as \textit{SDO}/HMI. Hopefully, progress in
instrumentation will result in broader coverage of magnetic and
velocity field measurements over the solar surface with sufficient
accuracy that the expressions derived in this paper can be applied to
yield major new insights into the mechanisms of solar coronal
activity.\par
\acknowledgments
We thank Mark Linton, James Leake, and Alex Glocer for useful and
entertaining discussions. We also thank an anonymous referee for a
number of very helpful comments, which significantly improved the
paper. This work was supported by the NASA Heliophysics Internal
Scientist Funding Model (17-HISFM18-0007) and the NASA Living With a
Star Programs (NNH16ZDA001N and NNH17ZDA001N).
\appendix
\section{General Expression for the Intrinsically Solenoidal Reference
  Electric Field $\epot$ \label{sec:EPot:Revisited}}
Returning to the problem of computing $\epot$ in general geometries,
the Helmholtz decomposition in Appendix~\ref{sec:Helmholtz} ensures
$\epot$ is completely determined when the tangential
components of $\epot$ on $\Surf$ are known; however, this is an
awkward boundary condition for~(\ref{eqn:Faraday}),
(\ref{eqn:eP:Div}), and (\ref{eqn:eP:Neumann}), as $\epot$ is not known
a priori anywhere, and only
$\grad\cross\epot=-c^{-1}\,{\partial \BPot}/{\partial t}$ is known
everywhere.\footnote{The difficulty in applying the Helmholtz
  decomposition here is that~(\ref{eqn:Helmholtz:Differential:f}) is
  written in terms of the free-space Green's function $\GreenF$. All
  the information about the geometry is encoded in the details of the 
  integrals. In contrast, consider the potential magnetic field, where
  the Green's function $\GreenNP$ is solved for an impulse response
  with homogeneous Neumann boundary
  conditions~(\ref{eqn:Green})\--(\ref{eqn:Green:Boundary}). This
  encodes the geometry into the Green's function, permitting direct
  knowledge of the $\phiPot$ everywhere from its normal gradient,
  namely $B_n$, on the bounding surface $\Surf$
  \cite[]{Sakurai1982,Nemenman1999}.}  Thus, we seek to determine the
inverse curl $\grad^{-1}\cross\left(\grad\cross\epot\right)$ knowing
$\partial\BPot/\partial t$ everywhere. In free-space the inverse
curl has been known for roughly 200 years; it is the Biot\---Savart law
\cite[]{Jackson1975}, which plays a fundamental role in fluid mechanics
and electromagnetism by connecting the vorticity, magnetic field, or
current with the fluid velocity, vector potential, or magnetic field
respectively. For a closed volume $\Vol$, the Biot\---Savart operator
\begin{equation}
\epot\left(\x,t\right)=-\frac{1}{4\,\pi\,c}\,\int_\Vol{d^3x'}\,\frac{\partial\BPot\left(\x',t\right)}{\partial t}\cross\frac{\left(\x-\x'\right)}{\left|\x-\x'\right|^3}=-\frac{1}{4\,\pi\,c}\,\grad\cross\int_\Vol{d^3x'}\,\frac{1}{\left|\x-\x'\right|}\frac{\partial\BPot\left(\x',t\right)}{\partial t},\label{eqn:Biot-Savart}
\end{equation}
satisfies Faraday's law~(\ref{eqn:Faraday}) if and only if
$\grad\cdot{\partial\BPot}/{\partial t}=0$ and $\left.{\partial
  \nhat\cdot\BPot}/{\partial t}\right|_\Surf=0$
\cite[]{Cantarella2001}. This is straightforward to see from the application of the curl to the right-hand  side of (\ref{eqn:Biot-Savart}), which leads to
\begin{equation}
\grad\cross\epot\left(\x,t\right)=-c^{-1}\,\frac{\partial\BPot\left(\x',t\right)}{\partial t}-\frac{1}{4\,\pi\,c}\,\grad\left[\int_\Vol{d^3x'}\,\frac{1}{\left|\x-\x'\right|}\,\grad'\cdot\frac{\partial\BPot\left(\x',t\right)}{\partial t}+\oint_\Surf{dS'}\,\frac{1}{\left|\x-\x'\right|}\,\frac{\partial\nhat'\cdot\BPot\left(\x',t\right)}{\partial t}\right],
\end{equation}
with~(\ref{eqn:Laplacian}),(\ref{eqn:Green}), (\ref{eqn:Delta}),
(\ref{eqn:Div_psi_f}), and~(\ref{eqn:Gauss}). The first condition is
satisfied by any magnetic field, but the second condition is
impossible to satisfy for any time-dependent potential
magnetic field.  While a great deal of effort has been expended in the
literature proving the existence and uniqueness of solutions of the
div-curl system represented by~(\ref{eqn:Faraday}),
(\ref{eqn:eP:Div}), and (\ref{eqn:eP:Neumann})
\cite[]{Girault1986,Jiang1994,Amrouche1998,Amrouche2013,Cheng2017},
surprisingly, only recently has a general closed-form solution been
developed for this system in bounded domains of Riemannian three-manifolds
(locally Euclidean three-space). \cite{Enciso2018} constructed an integral
solution to ~(\ref{eqn:Faraday}), (\ref{eqn:eP:Div}), and
(\ref{eqn:eP:Neumann}) of the form
\begin{equation}
\epot\left(\x,t\right)=-c^{-1}\,\int_\Vol{d^3x'}\,\Kernel_\Vol\left(\x,\x'\right)\cdot\frac{\partial \BPot\left(\x',t\right)}{\partial t},\label{eqn:EP:Unique}
\end{equation}
where $\Kernel_\Vol\left(\x,\x'\right)$ is a matrix-valued integral
kernel.  This solution is unique up to a harmonic field
$\grad\cdot\Harmonic=0$, $\grad\cross\Harmonic=0$,
$\left.\nhat\cdot\Harmonic\right|_\Surf=0$ which cannot be supported
in a simply connected volume $\Vol$ (see
footnote~\ref{foot:Simply}). The details of the kernel
$\Kernel_\Vol\left(\x,\x'\right)$ are beyond the scope of this
paper. However, it permits the expression of a formal relationship
between $\epot$ and changes in the normal component of the magnetic
field. \par
The potential reference electric field~(\ref{eqn:EP:Unique}) can be
re-expressed entirely in terms of surface values by noting that a
solution for $\BPot$
in~(\ref{eqn:Potential}),~(\ref{eqn:Laplace})\--(\ref{eqn:Laplacer})
may be constructed as the convolution of a Green's function and the
normal component of $\BPot$ on the surface $\Surf$
\begin{equation}
\phiPot\left(\x,t\right)=\left\langle\phiPot\left(t\right)\right\rangle+\oint_\Surf dS'\,\GreenNP\left(\x,\x'\right)\,\bpot_n\left(\x',t\right),
\label{eqn:Green}
\end{equation}
where $\left\langle\phiPot\left(t\right)\right\rangle$ is arbitrary 
often set to zero or set to the average of $\phiPot$ in the
volume. The Green's function $\GreenNP$ is developed from the Poisson
equation
\begin{subequations}
\begin{equation}
\nabla^2\GreenNP\left(\x,\x'\right)={\nabla'}^2\GreenNP\left(\x,\x'\right)=\delta\left(\x-\x'\right),\qquad\mbox{for
  $\x$ and $\x'$ $\in$ $\Vol$},\label{eqn:Green:Delta}
\end{equation}  
where ${\nabla'}^2$ operates on the primed coordinates with homogeneous
Neumann boundary conditions at $\Surf$,
\begin{equation}
\nhat'\cdot\grad'\GreenNP\left(\x,\x'\right)=-\frac{1}{\Surf}\qquad\mbox{for
  $\x'$ on $\Surf$ and $\x$ within $\Vol$},\label{eqn:Green:Boundary}
\end{equation}
\end{subequations}
and $\Surf$ represents the total surface area bounding $\Vol$. 
\begin{subequations}
A solution to~(\ref{eqn:Laplace})\--(\ref{eqn:Laplace}) exists if only
if the compatibility condition
\begin{equation}
\int_\Vol{d^3x}\,\grad\cdot\grad\phiPot=-\oint_\Surf{dS}\nhat\cdot\grad\phiPot=\oint_\Surf{dS}\,\bpot_n\equiv0,\qquad\mbox{for $\x$ $\in$ $\Vol$},\label{eqn:Laplace:constraint}
\end{equation}  
is satisfied, which follows from the solenoidal properties of the
magnetic field and the boundary condition on the scalar magnetic
potential,
\begin{equation}
-\nhat\cdot\grad\phiPot=\bpot_n\qquad\mbox{for $\x$ $\in$ $\Surf$}.\label{eqn:Pn}
\end{equation}
\end{subequations}
Using~(\ref{eqn:Green}) in~(\ref{eqn:EP:Unique}) permits the formal
expression of $\epot$ entirely in terms of surface values of
${\partial \bpot_n}/{\partial t}$ as a double convolution:
\begin{equation}
\epot\left(\x,t\right)=c^{-1}\,\int_\Vol{d^3x'}\,\Kernel_\Vol\left(\x,\x'\right)\cdot\grad'
\oint_\Surf dS''\,\GreenNP\left(\x',\x''\right)\,\frac{\partial \bpot_n\left(\x'',t\right)}{\partial t}.\label{eqn:EP:Unique:Surface}
\end{equation}

\section{Vector Identities\label{sec:Vector}}
For the convenience of the reader, we include some vector identities
and surface operators used throughout this paper.  \cite{Bladel2007}
contains a fairly comprehensive inventory of vector relations.
\subsection{Triple Products}
\begin{equation}
  \f\cross\left(\g\cross\h\right)=\left(\f\cdot\h\right)\,\g-\left(\f\cdot\g\right)\,\h,\label{eqn:fxgxh} 
\end{equation}
\begin{equation}
\f\cdot\left(\g\cross\h\right)=\g\cdot\left(\h\cross\f\right)=\h\cdot\left(\f\cross\g\right).\label{eqn:f_g_x_h} 
\end{equation}
\subsection{Integration by Parts}
Below, $\XI$, $\f$, and $\g$ are suitably continuously differentiable functions:
\begin{equation}
\grad\left(\f\cdot\g\right)=\f\cross\left(\grad\cross\g\right)+\g\cross\left(\grad\cross\f\right)+\left(\g\cdot\grad\right)\f+\left(\f\cdot\grad\right)\g,\label{eqn:grad_f_g} 
\end{equation}
\begin{equation}
\grad\cdot\left(\XI\,\f\right)=\XI\,\grad\cdot\f+\f\cdot\grad\XI,\label{eqn:Div_psi_f} 
\end{equation}
\begin{equation}
\grad\cross\left(\XI\,\f\right)=\XI\,\grad\cross\f+\grad\XI\cross\f,\label{eqn:Curl_psi_f} 
\end{equation}
\begin{equation}
\grad\cdot\left(\f\cross\g\right)=\g\cdot\left(\grad\cross\f\right)-\f\cdot\left(\grad\cross\g\right).\label{eqn:Div_f_x_g} 
\end{equation}
\begin{equation}
\grad\cross\left(\f\cross\g\right)=\f\,\grad\cdot\g-\g\,\grad\cdot\f+\left(\g\cdot\grad\right)\f-\left(\f\cdot\grad\right)\g.\label{eqn:Curl_f_x_g} 
\end{equation}
\subsection{Laplacian}
The vector Laplacian is
\begin{equation}
\nabla^2\f=\grad\left(\grad\cdot\f\right)-\grad\cross\left(\grad\cross\f\right).\label{eqn:Laplacian} 
\end{equation}
The curl of this equation leads to
\begin{equation}
\grad\cross\left(\nabla^2\f\right)=-\grad\cross\left[\grad\cross\left(\grad\cross\f\right)\right].\label{eqn:Curl_Laplacian} 
\end{equation}
\section{Dupin Surface Coordinates, Surface Vectors, and Operators\label{sec:Surface}}
\begin{figure}
\centerline{\includegraphics[width=4in,viewport=58 384 483 730,clip=]{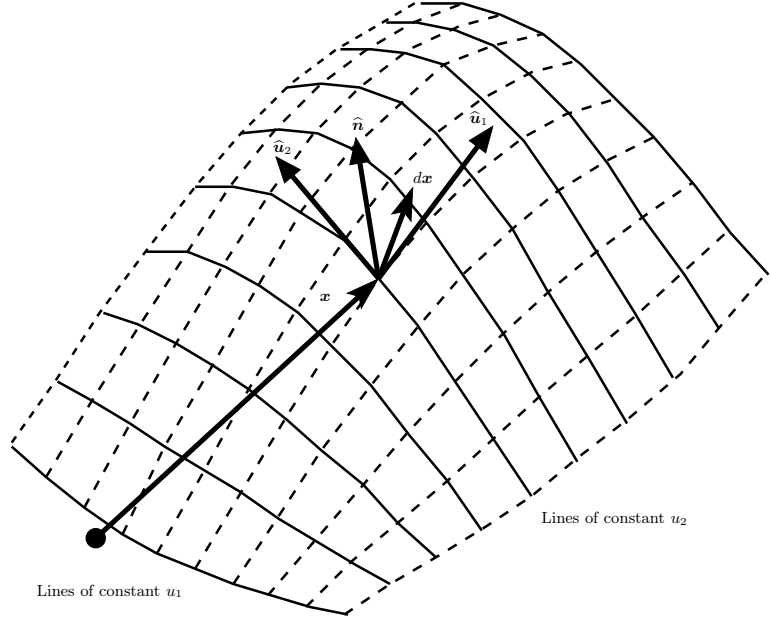}}
\caption{The Dupin surface coordinate system. Here,
  $\nhat=\uhat_1\cross\uhat_2$ and the total differential of the position
 vector $\x$ from a point on the surface to a neighboring point in space
  $\x+d\x$ is
  $d\x=h_1\,du_1\,\uhat_1+h_2\,du_2\,\uhat_2+dn\,\nhat$. After  \cite{Tai1992}.\label{fig:Dupin}}
\end{figure}
Key to performing vector calculus on a surface $\Manifold$ is Dupin
surface coordinates. A brief overview is provided below for convenience, and
the reader is referred to seminal texts by \cite{Weatherburn1955},
\cite{Tai1992}, and \cite{Bladel2007}. Below, the
expressions from \cite{Weatherburn1955} and \cite{Bladel2007} are used, but
\cite{Tai1992} is an excellent reference for connecting the different
definitions and occasionally ambiguous notations employed for surface
operators. For clarity, the surface operators used in this paper will
be explicitly defined and connected with their familiar volumetric
counterparts.  As \cite{Tai1991} note,  several familiar
coordinate systems have natural surfaces corresponding to the Dupin
system: Cartesian $\left(x,y,z\right)$ with $\nhat=\xhat$, $\yhat$, or
$\zhat$; spherical $\left(r,\vartheta,\varphi\right)$ with
$\nhat=\rhat$; and cylindrical $\left(\rho,\theta,z\right)$ with
$\nhat=\rhohat$ or $\nhat=\zhat$. In contrast, the toroidal system
does not have a natural Dupin surface. Chapter 2 in \cite{Tai1992}
provides an example of deriving the Dupin surface for a conical
section in cylindrical coordinates.\par
\subsection{Dupin Surface Coordinates}
A diagram of the Dupin surface coordinate system is shown in
Figure~\ref{fig:Dupin}. A point $\x$ on a regular, but open or closed,
surface may be labeled with the coordinates $u_1$ and $u_2$ with unit
vectors $\left(\uhat_1,\uhat_2,\nhat\right)$, where $\uhat_1$ and
$\uhat_2$ are tangent to the surface and $\nhat=\uhat_1\cross\uhat_2$
is normal to the surface.  The coordinates $u_1$ and $u_2$ can be
chosen concomitant with the principal directions, which are orthogonal
$\uhat_1\cdot\uhat_2=0$, and the coordinate $n$ denotes the normal
distance measured linearly from the surface. Along the principal
directions $\uhat_1$ and $\uhat_2$, the normals $\nhat$ at contiguous
points intersect the normal at point $\x$. The distance from point
$\x$ to this point of intersection is a principal radius of curvature
at $\x$ correspondingly denoted $R_1$ and $R_2$ for $\uhat_1$ and
$\uhat_2$ respectively. The total differential of the position vector
$\x$ from a point on the surface to a neighboring point in space
$\x+d\x$ is $d\x=h_1\,du_1\,\uhat_1+h_2\,du_2\,\uhat_2+dn\,\nhat$
where
\begin{subequations}
\begin{equation}
\frac{\partial \x}{\partial u_1}=h_1\,\uhat_1,\qquad  
\frac{\partial \x}{\partial u_2}=h_2\,\uhat_2,\qquad
\frac{\partial \x}{\partial n}=h_n\,\nhat,\label{eqn:Manifold}
\end{equation}
with
\begin{equation}
h_n\equiv1,\label{eqn:hn}
\end{equation}
\end{subequations}
where $h_1$, $h_2$, and $h_n$ are the scale factors that form the
metric tensor $h_i=\sqrt{\Metric_{ii}}$, and $i=1,2,n$ with
$\Metric_{ij}=0$ for $i\neq{j}$. The continuity $\Continuity^k$ of the
surface implies that $\Metric_{ii}$ is $\Continuity^{k-1}$ \cite[see
  p. 296 in][]{Morrey1966}. From~(\ref{eqn:Manifold}), a
$\Continuity^1$ surface is sufficient to define a proper normal
$\nhat$ and the tangent vectors $\uhat_1$ and $\uhat_2$.  The scale
factors are not completely independent as \cite[see p. 26
  in][]{Morse1953a}\footnote{For pedagogical purposes, the
  scale factor $h_n$ is carried through, understanding that~(\ref{eqn:hn})
  holds.}
\begin{subequations}
\begin{align}
  \frac{\partial\uhat_1}{\partial u_1}=&-\frac{\uhat_2}{h_2}\,
  \frac{\partial h_1}{\partial u_2}-\frac{\nhat}{h_n}\,\frac{\partial
    h_1}{\partial n}, &
 \qquad\frac{\partial\uhat_1}{\partial
   u_2}=&\frac{\uhat_2}{h_1}\,\frac{\partial h_2}{\partial u_1}, &
  \qquad\frac{\partial\uhat_1}{\partial n}=&\frac{\nhat}{h_1}\,\frac{\partial
    h_n}{\partial u_1}=0,\label{eqn:top_row}\\
  \frac{\partial\uhat_2}{\partial u_2}=&-\frac{\uhat_1}{h_1}\,
  \frac{\partial h_2}{\partial u_1}-\frac{\nhat}{h_n}\,\frac{\partial
    h_2}{\partial n}, &
 \qquad\frac{\partial\uhat_2}{\partial
   u_1}=&\frac{\uhat_1}{h_2}\,\frac{\partial h_1}{\partial u_2}, &
  \qquad\frac{\partial\uhat_2}{\partial n}=&\frac{\nhat}{h_2}\,\frac{\partial
    h_n}{\partial u_2}=0,\label{eqn:middle_row}\\
  \frac{\partial\nhat}{\partial n}=&-\frac{\uhat_1}{h_1}\,
  \frac{\partial h_n}{\partial u_1}-\frac{\uhat_2}{h_2}\,\frac{\partial
    h_n}{\partial u_2}=0,&
 \qquad\frac{\partial\nhat}{\partial
   u_1}=&\frac{\uhat_1}{h_n}\,\frac{\partial h_1}{\partial n},&
  \qquad\frac{\partial\nhat}{\partial u_2}=&\frac{\uhat_2}{h_n}\,\frac{\partial
    h_2}{\partial n}.\label{eqn:bottom_row}
\end{align}
\end{subequations}
A $\ContinuityHodgeMin$ surface $\Manifold$ is sufficient
for~(\ref{eqn:top_row})\--(\ref{eqn:bottom_row}) to be well defined,
which in turn are sufficient for a well-defined surface divergence,
surface curl, and Laplace\---Beltrami operator (surface Laplacian) on a
scalar described below. However, the vector fields in physical
problems are not only defined on the surface but also in some
neighborhood of it. For example, the potential field
constraint~(\ref{eqn:n_Curl_Curl_EPot}) requires that
$\grad\cross\grad\cross\EPot=0$ be well defined on the surface
$\Surf$. A surface with $\ContinuityCurrentMin$ continuity almost
everywhere is sufficient to satisfy this physical
requirement.\footnote{\cite{Schulz2016} notes on p. 44 that a
  $\ContinuityCurrentMin$ surface is usually sufficient in physics,
  but mathematicians assume $\Continuity^\infty$ so that they can
  avoid the ``boring'' details of the surface.} However, a
$\ContinuityCurrentMin$ surface implies additional relationships for
the $h_i$'s. Differentiating the first and last equations
in~(\ref{eqn:top_row}) by $\partial/\partial n$ and $\partial/\partial
u_1$, respectively,  produces
\begin{subequations}
\begin{align}
 \frac{\partial^2\uhat_1}{\partial n\partial u_1}=&-\uhat_2\,\left(\frac{\partial h_1}{\partial u_2}\,\frac{\partial}{\partial n}\frac{1}{h_2}+\frac{1}{h_2}\,
  \frac{\partial^2 h_1}{\partial u_2\,\partial n}\right)-\frac{\nhat}{h_n}\,\frac{\partial^2
    h_1}{\partial n^2},\\
\frac{\partial^2\uhat_1}{\partial u_1\partial n}=&\frac{\partial
  h_n}{\partial u_1}\frac{\partial}{\partial
  u_1}\left(\frac{\nhat}{h_1}\right)+\frac{\nhat}{h_1}\,\frac{\partial^2
  h_n}{\partial u_1^2}=0.
\end{align}  
Equating these leads to
\begin{equation}
\frac{\partial^2 h_1}{\partial u_2\partial n}=\frac{1}{h_2}\,\frac{\partial
  h_1}{\partial u_2}\,\frac{\partial h_2}{\partial n}\qquad\mbox{and}\qquad\frac{\partial^2
    h_1}{\partial n^2}=0.\label{eqn:h_u2_n}
\end{equation}
\end{subequations}
Cross-differentiating the first two equations in~(\ref{eqn:top_row}) and
substituting the appropriate terms from~(\ref{eqn:middle_row})
and~(\ref{eqn:bottom_row}) produce
\begin{subequations}
\begin{align}
\frac{\partial^2\uhat_1}{\partial u_2\partial u_1}=&\frac{\uhat_1}{h_2}\,\frac{\partial h_1}{\partial u_2}\,\frac{1}{h_1}\,
\frac{\partial h_2}{\partial u_1}
-\uhat_2\,\left(\frac{1}{h_2}\,
    \frac{\partial^2 h_1}{\partial u_2^2}
-\frac{1}{h_2}\,\frac{\partial h_1}{\partial u_2}\,\frac{1}{h_2}\,\frac{\partial h_2}{\partial u_2}
+\frac{1}{h_n}\,\frac{\partial
  h_1}{\partial n}\,\frac{1}{h_n}\,\frac{\partial
    h_2}{\partial n}\right)
+\frac{\nhat}{h_n}\,\left(\frac{1}{h_2}\,\frac{\partial h_1}{\partial u_2}\,\frac{\partial
    h_2}{\partial n}
-\frac{\partial^2
    h_1}{\partial u_2\partial n}\right),\\
 \frac{\partial^2\uhat_1}{\partial u_1\partial
   u_2}=&\frac{\uhat_1}{h_1}\,\frac{\partial h_2}{\partial
   u_1}\,\frac{1}{h_2}\,\frac{\partial h_1}{\partial u_2}
+\uhat_2\,\left(\frac{1}{h_1}\,\frac{\partial^2 h_2}{\partial u_1^2}-\frac{1}{h_1}\,\frac{\partial h_2}{\partial
     u_1}\,\frac{1}{h_1}\,\frac{\partial h_1}{\partial
   u_1}\right).
\end{align}
Equating the $\uhat_2$ component leads to a new relationship
\begin{equation}
\frac{1}{h_1}\,\frac{\partial^2 h_2}{\partial u_1^2}+\frac{1}{h_2}\,
    \frac{\partial^2 h_1}{\partial u_2^2}
-\frac{1}{h_1}\,\frac{\partial h_2}{\partial
     u_1}\,\frac{1}{h_1}\,\frac{\partial h_1}{\partial
  u_1}
-\frac{1}{h_2}\,\frac{\partial h_1}{\partial u_2}\,\frac{1}{h_2}\,\frac{\partial h_2}{\partial u_2}
+\frac{1}{h_n}\,\frac{\partial
  h_1}{\partial n}\,\frac{1}{h_n}\,\frac{\partial
    h_2}{\partial n}
=0.\label{eqn:curl_curl}
\end{equation}
\end{subequations}
Equations~(\ref{eqn:top_row}\---\ref{eqn:bottom_row}),~(\ref{eqn:h_u2_n}),
and~(\ref{eqn:curl_curl}) ensure that~(\ref{eqn:Curl_Laplacian}) is
satisfied when $\f=\psi\,\nhat$.\par
\subsection{Surface Vectors and Operators}
A surface $\Manifold$ of at least $\Continuity^1$ is sufficient to define
the differential operators and relationships below.  A surface vector
may be decomposed into a normal vector and a tangent vector denoted
with a subscripts ``$n$'' and ``$\Manifold$,'' respectively,
\begin{equation}
\f=\f_n+\f_\tang=\nhat\,\left(\nhat\cdot\f\right)-\nhat\cross\left(\nhat\cross\f\right).\label{eqn:fS} 
\end{equation}
The familiar gradient operator written in Dupin surface coordinates is
\begin{subequations}
\begin{equation}
\grad=\frac{\uhat_1}{h_1}\,\frac{\partial}{\partial u_1}+\frac{\uhat_2}{h_2}\,\frac{\partial}{\partial u_2}+ \nhat\,\frac{\partial}{\partial n}.
\end{equation}
The surface gradient operator is defined as the projection of the gradient
operator on the surface $\Manifold$
\begin{equation}
\gradp=\frac{\uhat_1}{h_1}\,\frac{\partial}{\partial u_1}+\frac{\uhat_2}{h_2}\,\frac{\partial}{\partial u_2}.
\end{equation}
Symbolically, the relationship between the surface gradient operator and the
familiar gradient operator is determined from~(\ref{eqn:fS}) as
\begin{equation}
\gradp\XI\equiv\left(\nhat\cross\grad\XI\right)\cross\nhat=\grad\XI-\nhat\nhat\cdot\grad\XI.\label{eqn:GradS_psi}  
\end{equation}  
\end{subequations}\par
A surface $\Manifold$ of at least $\ContinuityHodgeMin$ is sufficient
to define the differential operators and relationships below.  The
``first curvature'' \cite[see p. 78 in][]{Weatherburn1955} is the sum
of the principal curvatures
\begin{equation}
\Jfirst\equiv\frac{1}{R_1}+\frac{1}{R_2}=-\frac{\partial\log\left(h_1\,h_2\right)}{\partial n}.\label{eqn:JFirst}
\end{equation}
The familiar divergence operator in Dupin surface coordinates is
\begin{subequations}
\begin{equation}
\grad\cdot\f=\frac{1}{h_1\,h_2}\,\left[\frac{\partial\left(h_2\,f_1\right)}{\partial
    u_1}+\frac{\partial\left(h_1\,f_2\right)}{\partial u_2}\right]-\Jfirst\,f_n+\frac{\partial f_n}{\partial n}.  
\end{equation}
The surface divergence is defined
\begin{equation}
\gradp\cdot\f=\frac{1}{h_1\,h_2}\,\left[\frac{\partial\left(h_2\,f_1\right)}{\partial
    u_1}+\frac{\partial\left(h_1\,f_2\right)}{\partial u_2}\right]-\Jfirst\,f_n.
\end{equation}
Symbolically, the relationship between the surface divergence operator and the
familiar divergence operator is then
\begin{equation}
\gradp\cdot\f\equiv\grad\cdot\f-\nhat\cdot\frac{\partial \f}{\partial
  n}=\gradp\cdot\f_\tang-\Jfirst\,f_n,\label{eqn:DivS_f} 
\end{equation}  
Note that when the vector $\f$ is tangent to the surface $\Surf$ with $f_n=0$, ${\partial f_n}/{\partial n}=0$, then
\begin{equation}
\grad\cdot\f_\tang=\gradp\cdot\f_\tang.\label{eqn:Div:fS}
\end{equation}
\end{subequations}
The surface divergence satisfies a  relationship formally equivalent
to~(\ref{eqn:Div_psi_f})
\begin{equation}
\gradp\cdot\left(\XI\,\f\right)=\f\cdot\gradp\XI+\XI\,\gradp\cdot\f.\label{eqn:DivS_psi_f} 
\end{equation}\par
 The familiar curl operator in Dupin surface coordinates takes the form
\begin{subequations}
\begin{equation}
  \grad\cross\f=\frac{1}{h_1\,h_2}\,\left\lbrace
  h_1\,\left[\frac{\partial f_n}{\partial u_2}-\frac{\partial\left(h_2 f_2\right)}{\partial n}\right]\,\uhat_1
  +h_2\,\left[\frac{\partial \left(h_1\,f_1\right)}{\partial
      n}-\frac{\partial f_n}{\partial u_1}\right]\,\uhat_2
  +\left[\frac{\partial \left(h_2\,f_2\right)}{\partial
      u_1}-\frac{\partial \left(h_1\,f_1\right)}{\partial u_2}\right]\,\nhat
\right\rbrace.\label{eqn:Curl}
\end{equation}
The surface curl is defined
\begin{equation}
  \gradp\cross\f=\frac{1}{h_1\,h_2}\,\left\lbrace
  \left[h_1\,\frac{\partial f_n}{\partial u_2}-h_1 f_2\frac{\partial h_2}{\partial n}\right]\,\uhat_1
  -\left[h_2\,\frac{\partial f_n}{\partial u_1}-h_2 f_1\frac{\partial h_1}{\partial n}\right]\,\uhat_2
  +\left[\frac{\partial \left(h_2\,f_2\right)}{\partial
      u_1}-\frac{\partial \left(h_1\,f_1\right)}{\partial u_2}\right]\,\nhat
\right\rbrace.\label{eqn:Weatherburn:CurlS}     
\end{equation}
The normal component of the curl and surface curl are
equivalent,
\begin{equation}
\nhat\cdot\grad\cross\f=\nhat\cdot\gradp\cross\f,\label{eqn:n_Curl=n_CurlS}
\end{equation} 
but the relationship between the tangential components is not immediately
apparent from~(\ref{eqn:Curl}) and~(\ref{eqn:Weatherburn:CurlS}). However, symbolically, the familiar curl and the surface curl are related by
\begin{equation}
\grad\cross\f=\gradp\cross\f+\nhat\cross\frac{\partial \f}{\partial n}.  
\end{equation}
\end{subequations}\par
The normal $\nhat$ satisfies
\begin{subequations}
\begin{align}
\gradp\cdot\nhat=&-\Jfirst,\label{eqn:DivS_n}\\ 
\grad\cdot\nhat=&-\Jfirst,\label{eqn:DIV_n}\\
\gradp\cross\nhat=&0,\label{eqn:CurlS_n}\\ 
\grad\cross\nhat=&0,\label{eqn:Curl_n}\\
\nhat\cdot\grad\XI\equiv&\frac{\partial\XI}{\partial n},\label{eqn:normal_derivative}\\ 
\nhat\cdot\frac{\partial\nhat}{\partial n}=&0.\label{eqn:Weatherburn}
\end{align}  
\end{subequations}
The surface divergence satisfies a relationship formally equivalent
to~(\ref{eqn:Div_f_x_g}),
\begin{subequations}
\begin{equation}
\gradp\cdot\left(\f\cross\g\right)=\g\cdot\gradp\cross\f-\f\cdot\gradp\cross\g, 
\end{equation}
which for $\f=\nhat$ with~(\ref{eqn:CurlS_n}) becomes
\begin{equation}
\gradp\cdot\left(\nhat\cross\g\right)=-\nhat\cdot\gradp\cross\g,\label{eqn:DivS_n_x_f}
\end{equation}
\end{subequations}
The surface curl satisfies a relationship formally equivalent
to~(\ref{eqn:Curl_psi_f}),
\begin{equation}
\gradp\cross\left(\XI\,\f\right)=\XI\,\gradp\cross\f+\gradp\XI\cross\f.\label{eqn:CurlS_psi_f} 
\end{equation}
The tangent derivative operator may be written as \cite[see p. 712 in][]{Wu2007}
\begin{equation}
\nhat\cross\grad=\nhat\cross\gradp.\label{eqn:Tangent:Der}  
\end{equation}
For any tangent vectors $\f_\tang$ and $\g_\tang$ satisfying
$\nhat\cdot\f_\tang=0$ and $\g_\tang=\nhat\cross\f_\tang$, \cite[see
  p. 712 in][]{Wu2007},
\begin{equation}
\gradp\cdot\f_\tang=\left(\nhat\cross\grad\right)\cdot\g_\tang=\nhat\cdot\left(\grad\cross\g_\tang\right)=\nhat\cdot\left(\gradp\cross\g_\tang\right).\label{eqn:Wu}   
\end{equation}  
The familiar Laplacian
operator in Dupin surface coordinates takes the form
\begin{subequations}
\begin{equation}
\grad\cdot\grad\XI=\frac{1}{h_1\,h_2}\,\sum_{j=1,2}\frac{\partial}{\partial u_j}\,\left(\frac{h_1\,h_2}{h_j^2}\,\frac{\partial\XI}{\partial u_j}\right)-\Jfirst\,\frac{\partial\XI}{\partial n}+\frac{\partial^2\XI}{\partial n^2}
\end{equation}
The surface Laplacian or Laplace\---Beltrami operator may be expressed as
\begin{equation}
\gradp\cdot\gradp\XI=\Lapp\XI=\frac{1}{h_1\,h_2}\,\sum_{j=1,2}\frac{\partial}{\partial u_j}\,\left(\frac{h_1\,h_2}{h_j^2}\,\frac{\partial\XI}{\partial u_j}\right).
\end{equation}
Symbolically, these are related by
\begin{equation}
\Lapp\XI=\nabla^2\XI
+\Jfirst\,\frac{\partial\XI}{\partial
  n}-\frac{\partial^2\XI}{\partial
  n^2}.\label{eqn:DivS_GradS_psi}
\end{equation}
\end{subequations}
Note that~(\ref{eqn:Wu}) with $\f_\tang=\gradp\XI$ and
$\g_\tang=\nhat\cross\f_\tang$ implies that
\begin{equation}
\nhat\cdot\grad\cross\left(\nhat\cross\gradp\XI\right)=\nhat\cdot\gradp\cross\left(\nhat\cross\gradp\XI\right)=\Lapp\XI.\label{eqn:n_Curl_n_x_GradS_psi}
\end{equation}
Unlike the volumetric operator which satisfies $\grad\cross\grad\XI=0$,
only the normal component vanishes identically for the surface operators
\cite[see p. 233 in][]{Weatherburn1955}
\begin{subequations}
\begin{equation}
\nhat\cdot\left(\gradp\cross\gradp\XI\right)=0,\label{eqn:n_CurlS_GradS_psi}
\end{equation}
but with~(\ref{eqn:n_Curl=n_CurlS}), there is also
\begin{equation}
\nhat\cdot\left(\grad\cross\gradp\XI\right)=0.\label{eqn:n_Curl_GradS_psi}
\end{equation}
Using this with~(\ref{eqn:DivS_n_x_f}) implies
\begin{equation}
\gradp\cdot\left(\nhat\cross\gradp\XI\right)=0.\label{eqn:DivS_n_x_GradS_psi:Surf}
\end{equation}
\end{subequations}
More specifically, $\gradp\cross\gradp\XI=0$ only for surfaces with zero
Gaussian curvature \cite[see p. 122 in][]{Ludu2012}.
\section{Integral Relationships\label{sec:Integral}}
\cite{Arapura2016} provides a introductory treatment of the
Gauss\---Ostrogradsky and Stokes' theorems.  Below, $\f$ is continuously
differentiable $\left(\ContinuouslyDifferentiable\right)$, $\Surf$
is a closed $\ContinuityStokes$ surface bounding $\Vol$, and $\nhat$ is
the inwardly directed normal. The Gauss\---Ostrogradsky theorems are
\begin{subequations}
\begin{align}
\int_{\Vol}d^3x\,\grad\cdot\f=&-\oint_{\Surf}dS\,\nhat\cdot\f,\label{eqn:Gauss}\\
\int_{\Vol}d^3x\,\grad\cross\f=&-\oint_{\Surf}dS\,\nhat\cross\f.\label{eqn:Ostrogradsky}
\end{align}
\end{subequations}
An inwardly directed normal to the coronal volume corresponds to
radially outward vector in the photosphere, and so our convention is to
measure positive flux into the coronal volume.\par
Below, $\Cont$ is the $\ContinuityStokes$ contour bounding the
$\ContinuityStokes$ open surface $\Surf$ with the line element $d\ELL$. The
normal $\nhat$ is oriented according to the right-hand rule in relation to the
contour $\Cont$ with $d\ELL\cdot\nhat=0$ locally because $d\ELL$ lies in the
surface $\Surf$. Stokes' theorems for an open surface take the form
\begin{subequations}
\begin{align}
\int_\Surf{dS}\,\nhat\cdot\left(\grad\cross\f\right)=&\oint_\Cont{d\ELL}\cdot\f,\label{eqn:Stokes}\\
\int_\Surf{dS}\,\left(\nhat\cross\grad\XI\right)=&\oint_\Cont{d\ELL}\,\XI.\label{eqn:Stokes2}
\end{align}
\end{subequations}
\begin{subequations}
and for a closed surface,
\begin{align}
\oint_\Surf{dS}\,\nhat\cdot\left(\grad\cross\f\right)=&0,\label{eqn:StokesClosed}\\
\oint_\Surf{dS}\,\left(\nhat\cross\grad\XI\right)=&0.\label{eqn:StokesClosed2} 
\end{align}
\end{subequations}
\par
\begin{figure}
  \centerline{\includegraphics[width=3in,viewport=61 556 226 731]{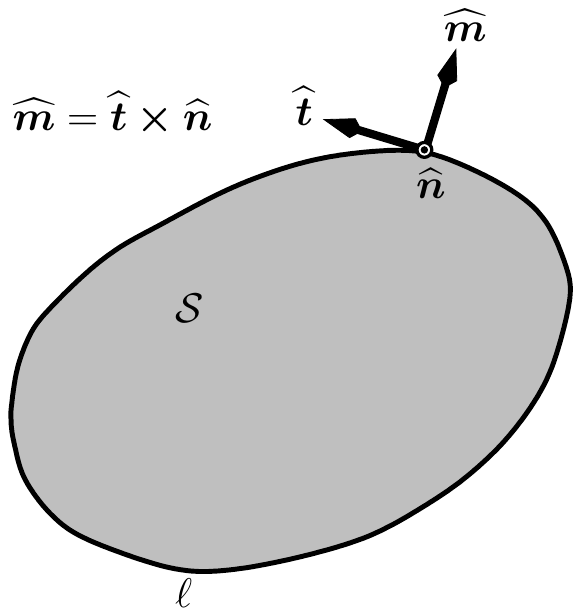}}
  \caption{Unit vectors on an open surface $\Surf$. After \cite{Bladel2007}.\label{fig:Bladel}}
\end{figure}
Below, the unit vector $\um$ is in the tangent plane and perpendicular to
contour $\Cont$; it points outward from the $\ContinuityStokes$ surface
$\Manifold$ enclosed by contour $\Cont$. The unit vector $\ut$ is tangent to
contour $\Cont$, $\nhat$ is normal to the surface at the edge, and
$\ut=\nhat\cross\um$ as shown in Figure~\ref{fig:Bladel}. The surface
divergence theorem is
\begin{subequations}
\begin{equation}
\int_\Manifold{dS}\,\gradp\cdot\f+\int_\Manifold{dS}\Jfirst\,\left(\f\cdot\nhat\right)=\int_\Cont\,d\eLL\,\f\cdot\um.\label{eqn:SDT}  
\end{equation}
The first integral vanishes for a closed surface if $\f$ is everywhere tangent to
the surface~$\Manifold$,
\begin{equation}
\oint_\Manifold{dS}\,\gradp\cdot\f_\tang=0.\label{eqn:SDT:closed} 
\end{equation}
\end{subequations}  
The integrals~(\ref{eqn:Gauss}\---\ref{eqn:SDT:closed}) have been
generalized to rough surfaces and even fractal surfaces that have no
proper unit normal vector $\nhat$ within the framework of differential
forms \cite[]{Harrison1999}. Mathematicians have noticed that one side
of these integrals can be used to define the other side under more
general conditions. Indeed, \cite{Harrison1993} notes that
\cite{Whitney1957} used the left-hand side of~(\ref{eqn:Stokes}) to
define the right-hand side for a rough boundary. Generally, the
smoother $\f$ is, the rougher $\Surf$ or $\Contour$ can be.
Furthermore, these integrals~(\ref{eqn:Gauss}\---\ref{eqn:SDT:closed})
can be extended to the piecewise $\Continuity^1$ surfaces $\Surf$ and
piecewise $\Continuity^1$ contours $\Contour$ by considering
$\Continuity^1$ surface patches $\Surf_i$ that comprise $\Surf$ and
dividing the corresponding bounding contours $\Contour_i$ into
$\Continuity^1$ segments.\par
\section{Vector Decompositions\label{sec:Helmholtz}}
Fundamental to the new formula~(\ref{eqn:RH_Simpliest}) for helicity
transport across a surface $\Surf$ is determining $\epot$ uniquely in
the enclosed volume $\Vol$ and on the enclosing surface $\Surf$. To
accomplish this, two major theorems of vector calculus are employed: the
Helmholtz decomposition in a finite volume $\Vol$ and Helmholtz\---Hodge
decomposition on a surface $\Surf$. \cite{Zhou2006} provides a recent
rigorous treatment of the former and \cite{Bladel1993b} discusses the
latter. \cite{Bladel1958} is a comprehensive reference that describes
the splitting of the Helmholtz decomposition in a volume for various
combinations of of boundary conditions on $\F$ \cite[see the table on
  p. 22 of][]{Bladel1958}.  \cite{Cantarella2002} provides a
comprehensive introduction to the connection between topology and the
Helmholtz\---Hodge decompositions for divergence-free vectors in
multiply connected volumes.  These theorems are stated below and
the reader is referred to the cited references for the proofs.
\subsection{Helmholtz Decomposition in a Volume\label{sec:Helmholtz:Volume}}
\citeauthor{Zhou2006} (\citeyear{Zhou2006}, p. 95) states the Helmholtz decomposition theorem as \cite[see also \cite{Gui2007},][pp. 52\---54]{Morse1953a}:
\begin{theorem}
Any finite, integrable, and continuously differentiable vector
function $\F\left(\x\right)$ given in a simply connected volume $\Vol$
enclosed by $\Surf$ can be completely and uniquely decomposed into a sum of an
irrotational part and a solenoidal part. The two parts are independent.
Mathematically, it is the identity:
\begin{subequations}
\begin{equation}
\F\left(\x\right)=-\overbrace{\grad\XI}^\mathrm{Irr}+\overbrace{\grad\cross\f}^\mathrm{Sol};\quad\grad\cdot\f=0\quad\in\quad\Vol,\label{eqn:Helmholtz:Differential}
\end{equation}
where
\begin{align}
\f\left(\x\right)=&\int_\Vol{d^3x'}\,\GreenF\left(\x,\x'\right)\,\grad'\cross\F\left(\x'\right)+\oint_\Surf{dS'}\,\GreenF\left(\x,\x'\right)\,\nhat\left(\x'\right)\cross\F\left(\x'\right),\label{eqn:Helmholtz:Differential:f}\\
\XI\left(\x\right)=&\int_\Vol{d^3x'}\,\GreenF\left(\x,\x'\right)\,\grad'\cdot\F\left(\x'\right)-\oint_\Surf{dS'}\,\GreenF\left(\x,\x'\right)\,\nhat\left(\x'\right)\cdot\F\left(\x'\right).\label{eqn:Helmholtz:Differential:XI}
\end{align}
where $\GreenF$ is the free-space Green's function
\begin{equation}
\GreenF\left(\x,\x'\right)=-\frac{1}{4\,\pi}\,\frac{1}{\left|\x-\x'\right|}.\label{eqn:GreenF}  
\end{equation}
which satisfies
\begin{equation}
\nabla\GreenF\left(\x,\x'\right)=\delta\left(\x-\x'\right).\label{eqn:Delta}
\end{equation}
\end{subequations}
\end{theorem}
The exact statement in \cite{Zhou2006} is more general than what is stated
above, being applicable to piecewise continuous functions in a
multiply connected volume, which is beyond the scope of this paper.  Strictly
speaking,~(\ref{eqn:Helmholtz:Differential})\--(\ref{eqn:GreenF}) applies to a
single boundary surface, but disconnected bounding surfaces, such as a volume
between spherical shells $R_<<r<R_>$, can be incorporated directly with the
appropriate auxiliary surfaces. Furthermore, a multiply connected region can
be transformed into a simply connected region by $\genus$ cuts and
$b_1=2\,\genus$ auxiliary surfaces, where $b_1$ is the first Betti number and
$\genus$ is the genus of the bounding surface $\Surf$ which is equivalent to
the number of holes \cite[]{Yoshida1998,Boulmezaoud1999}. Each cut generates
two new surfaces in the multiply connected volume, denoted $\surf$ bracketing
the cut. The fluxes at each of these surfaces must then be specified, namely
$\int_{\surf_i}{dS}\,\nhat_i\cdot\F$, where $i=1\ldots{b_1}$ and $\nhat_i$ is
the normal to the surface $\surf_i$, which in the convention of this paper
points into the volume $\Vol$. Essentially, the closed harmonic field lines in
the original multiply connected volume become open field lines in a new simply
connected volume $\Vol$. By combining the two terms
in~(\ref{eqn:Helmholtz:Differential:f}) with (\ref{eqn:Ostrogradsky})
and~(\ref{eqn:Curl_psi_f}) and
$\grad'\GreenF\left(\x,\x'\right)=-\grad\GreenF\left(\x,\x'\right)$ it becomes
clear that the condition $\grad\cdot\f=0$ asserted
in~(\ref{eqn:Helmholtz:Differential}) holds:
\begin{equation}
\f\left(\x\right)=\grad\cross\int_{\Vol}d^3x'\,\GreenF\left(\x,\x'\right)\,\F\left(\x'\right).
\end{equation}\par
\citeauthor{Zhou2006} (\citeyear{Zhou2006} p. 97), states the general
uniqueness theorem as:
\begin{theorem}
A vector function $\F\left(\x\right)$ in $\Vol$ bounded by the surface $\Surf$
can be uniquely determined by its divergence, curl and boundary values (both
normal and tangential components) over the boundary $\Surf$, i.e., the
solution to the system
\begin{subequations}
  \begin{align}
    \grad\cross\F=&\boldsymbol{s}_\mathrm{curl},\\
  \grad\cdot\F=&s_\mathrm{div},\\
  \left.\F\right|_\Surf=&\left.\F_0\right|_\Surf,
  \end{align}
\end{subequations}
is unique.
\end{theorem}
Here $\F_0=\left.\F\right|_\Surf$ is the vector boundary condition on $\F$, and
$\boldsymbol{s}_\mathrm{curl}$ and $s_\mathrm{div}$ are the sources of the curl and
divergence in $\Vol$, respectively.
In ``Corollary 2'' of the general uniqueness theorem of a vector function in
\citeauthor{Zhou2006} (\citeyear{Zhou2006}, p. 99) the situation for $\epot$ is stated concisely:
\begin{corollary}
An intrinsically solenoidal vector function $\F\left(\x\right)$ can be
uniquely determined by its curl and tangential components over the boundary
$\Surf$ (because the normal components are zero). That is, the solution to the
mathematical problem
\begin{subequations}
  \begin{align}
    \grad\cross\F=&\boldsymbol{s}_\mathrm{curl},\\
  \left.\nhat\cross\F\right|_\Surf=&\left.\nhat\cross\F_0\right|_\Surf,\\
  \grad\cdot\F=&0,\label{sec:Corollary:DivF}\\
  \left.\nhat\cdot\F\right|_\Surf=&0,\label{sec:Corollary:Fn}
  \end{align}
\end{subequations}
is unique.  
\end{corollary}
When (\ref{sec:Corollary:DivF}) and~(\ref{sec:Corollary:Fn}) hold, this
corollary has the following consequences for $\f\left(\x\right)$ and
$\XI\left(\x\right)$
in~(\ref{eqn:Helmholtz:Differential})\---(\ref{eqn:Helmholtz:Differential:XI})
\begin{subequations}
\begin{align}
\grad\cross\f=&\F,\label{eqn:Curl_F}\\
\left.\nhat\cross\left(\grad\cross\f\right)\right|_\Surf=&\left.\nhat\cross\F_0\right|_\Surf,\\
\XI=&\XI_0\quad\mbox{(meaningless constant)},\\
\left.\nhat\cdot\grad\cross\f\right|_\Surf=&0.\label{eqn:Neumann:F}
\end{align}  
\end{subequations}
This representation can be achieved, more intuitively, albeit less
rigorously, by noting that a goal of representing a vector
$\F\left(\x\right)$ in $\Vol$ by the Helmholtz
decomposition~(\ref{eqn:Helmholtz:Differential}) is to decompose it
into linear independent orthogonal representations, which requires
that
\begin{equation}
\int_\Vol{d^3x}\,\grad\cross\f\cdot\grad\XI=0.
\end{equation}
Integrating by parts with~(\ref{eqn:Div_f_x_g})
or~(\ref{eqn:Div_psi_f}) and using~(\ref{eqn:Gauss})
and~(\ref{eqn:f_g_x_h}) produce
\begin{subequations}
\begin{align} 
 \int_\Vol{d^3x}\,\grad\cdot\left(\f\cross\grad\XI\right)=&-\oint_\Surf{dS}\,\f\cdot\left(\grad\XI\cross\nhat\right)=-\oint_\Surf{dS}\,\grad\XI\cdot\left(\nhat\cross\f\right)=0,\\
  \int_\Vol{d^3x}\,\grad\cdot\left(\XI\,\grad\cross\f\right)=&-\oint_\Surf{dS}\,\nhat\cdot\left(\XI\,\grad\cross\f\right)=0.
\end{align}
\end{subequations}
Noting~(\ref{eqn:StokesClosed}) for the last relationship implies that any of the boundary conditions
\begin{subequations}
\begin{align}
\left.\XI\right|_\Surf=&\mbox{constant},\label{eqn:Dirichlet}\\
\left.\nhat\cross\f\right|_\Surf=&0,\label{eqn:Homogeneous}\\
\left.\nhat\cdot\left(\grad\cross\f\right)\right|_\Surf=&0.\label{eqn:Neumann}
\end{align}
\end{subequations}
are sufficient for orthogonality. Note that~(\ref{eqn:Homogeneous})
automatically implies (\ref{eqn:Neumann}).  All conditions holding
simultaneously are sufficient for orthogonality but are  not necessary
\cite[]{Denaro2003}. The first boundary condition corresponds to
homogeneous Dirichlet conditions $\left.\XI\right|_\Surf=\XI_0$, and
the second and third boundary conditions correspond to homogeneous
Neumann conditions $\left.\partial\XI/\partial
n\right|_\Surf=-\nhat\cdot\F$.\par
For the intrinsically solenoidal reference electric field $\F=\epot$
discussed in~\S~\ref{sec:lamellar}, with $\grad\cdot\F=0$ and
$\left.\nhat\cdot\F\right|_\Surf=0$, either choice leads to the same
result for $\XI$. The function $\XI$ is the solution to
\begin{subequations}
\begin{equation}
\nabla^2\XI=0\quad\in\Vol,
\end{equation}  
with 
\begin{equation}
\left.\XI\right|_\Surf=\XI_0,
\end{equation}  
or 
\begin{equation}
\left.\nhat\cdot\left(\grad\cross\f\right)\right|_\Surf=0\quad\mbox{leading to}\quad\left.\nhat\cdot\grad\XI\right|_\Surf=\left.\nhat\cdot\F\right|_\Surf=0.
\end{equation}  
\end{subequations}
Consequently, either choice has the solution
$\XI=\XI_0=\mbox{constant}$
and~(\ref{eqn:Curl_F})\--(\ref{eqn:Neumann:F}). In other words,
$\epot$ is unique, but note that the complete reference electric
field $\EPot=\epot+\grad\GaugeP$ does admit the gradient of a scalar
potential $\GaugeP$, which satisfies Dirichlet boundary conditions on
$\Surf$.
\subsection{Helmholtz\---Hodge Decomposition on a Surface\label{sec:Helmholtz:Surface}}
The development of the Helmholtz\---Hodge decomposition on a surface is more
modern \cite[]{Hodge1952} and has several integral forms
\cite[]{Scharstein1991,Bladel1993a,Backus1996,Imbert-Gerard2016,Kustepeli2017}. \cite{Reusken2019}
provides a comprehensive exposition of the the Helmholtz\---Hodge
Decomposition on a $\ContinuityHodgeMin$ surface, which is sufficient for our
analysis
\begin{theorem}
Any finite, square integrable, vector function $\f\left(\x\right)$ on
a $\ContinuityHodgeMin$ surface $\Surf$ may be represented by a normal
component $\f_n$ and a tangent vector $\f_\Surf$
using~(\ref{eqn:fS}). The tangent vector $\f_\Surf$ can be further
uniquely decomposed into a solenoidal component (with no divergence)
and lamellar component (with no normal component of the
\underline{surface} curl)\footnote{In some literature,
  $\nhat\cdot\gradp\cross\f_\Surf$, despite being a scalar, is denoted
  ``the surface curl'' of $\f_\Surf$ \cite[]{Scharstein1991}.} and a
harmonic field $\Harmonic_\Surf$ 
\begin{equation}
\f=\overbrace{\tau\,\nhat}^{\f_n}+\overbrace{\underbrace{\nhat\cross\gradp\phi}_{\mathrm{solenoidal}}+\underbrace{\gradp\psi}_{\mathrm{\!\!lamellar\!\!}}+\Harmonic_\Surf}^{\f_\Surf}.\label{eqn:Helmholtz:Decomposition}  
\end{equation}
The scalars $\tau$, $\psi$, and
$\phi$ may be determined from
\begin{subequations}
\begin{equation}
\nhat\cdot\f=f_n=\tau,\label{eqn:Helmholtz:normal}
\end{equation}
\begin{equation}
\Lapp\psi=\gradp\cdot\left[\left(\nhat\cross\f_\Surf\right)\cross\nhat\right],  \label{eqn:Helmholtz:lamellar}
\end{equation}
and
\begin{equation}
\Lapp\phi=\nhat\cdot\left(\gradp\cross\f_\Surf\right).\label{eqn:Helmholtz:solenoidal}
\end{equation}
\end{subequations}
The harmonic term can be determined by
\begin{equation}
\Harmonic_\Surf\equiv\f_\Surf-\nhat\cross\gradp\phi-\gradp\psi.
\end{equation}
\end{theorem}
The second term in~(\ref{eqn:Helmholtz:Decomposition}) is purely
solenoidal as $\grad\cdot\left(\nhat\cross\gradp\phi\right)=0$. While
the third term is irrotational with respect to the normal component of
the surface curl $\nhat\cdot\gradp\cross\gradp\psi=0$, it is not
necessarily irrotational in three dimensions as discussed at the end
of Appendix~\ref{sec:Surface}. The harmonic term is both surface divergence
free $\gradp\cdot\Harmonic_\Surf=0$ and surface curl free
$\nhat\cdot\left(\gradp\cross\Harmonic_\Surf\right)=0$. For a simply
connected $\Continuity^k$ surface with $k\ge2$, the harmonic term must
be zero \cite[see Lemma 4.3 in][]{Reusken2019}.  \par
The operator
$\nhat\cdot\left(\grad\cross\f\right)$ can be written in this
representation with~(\ref{eqn:Curl_psi_f}) and~(\ref{eqn:Wu}) as
\begin{equation}
\nhat\cdot\left(\grad\cross\f\right)=\nhat\cdot\left[\grad\cross\gradp\psi+\grad\cross\left(\nhat\cross\gradp\phi\right)+\tau\,\grad\cross\nhat-\nhat\cross\gradp\tau\right],\label{eqn:null_space_i}
\end{equation}
and using~(\ref{eqn:n_Curl_n_x_GradS_psi}),~(\ref{eqn:n_Curl_GradS_psi}) and~(\ref{eqn:Curl_n})
\begin{equation}
\nhat\cdot\left(\grad\cross\f\right)=\Lapp\phi.  
\end{equation}
The null space of this operator on $\Surf$ is
\begin{equation}
\f_\Surf=\tau\,\nhat+\gradp\psi+\Harmonic_\Surf.\label{eqn:null_space_f}
\end{equation}\par
Here, $\tau$ is just as smooth as $\f$, and $\psi$ and $\phi$ are somewhat
smoother. The normal component is orthogonal to the surface component
$\f_n\cdot\f_\Surf=0$ in a point wise sense, and the lamellar, solenoidal, and
harmonic components in are mutually orthogonal in an average sense over a closed
surface:
\begin{subequations}
\begin{equation}
\oint_\Surf{dS}\,\nhat\cross\gradp\phi\cdot\gradp\psi=\oint_\Surf{dS}\,\gradp\cdot\left[\nhat\cross\left(\psi\grad\phi\right)\right]=0,\label{eqn:Hodge:Orthogonality}  
\end{equation}
where integration by parts~(\ref{eqn:Div_psi_f})
has been used and~(\ref{eqn:SDT:closed})
has been invoked. Similarly,  
\begin{equation}
\oint_\Surf{dS}\,\nhat\cross\gradp\phi\cdot\Harmonic_\Surf=\oint_\Surf{dS}\,\gradp\cdot\left(\Harmonic_\Surf\cross\phi\,\nhat\right)-\oint_\Surf{dS}\,\phi\,\nhat\cdot\gradp\cross\Harmonic_\Surf=0,\label{eqn:Hodge:Orthogonality:Harmonic1}  
\end{equation}
where integration by parts~(\ref{eqn:Div_f_x_g}) has been used
and~(\ref{eqn:SDT:closed}) has been invoked, and
\begin{equation}
\oint_\Surf{dS}\,\gradp\psi\cdot\Harmonic_\Surf=\oint_\Surf{dS}\,\gradp\cdot\left(\psi\,\Harmonic_\Surf\right)-\oint_\Surf{dS}\,\psi\,\gradp\cdot\Harmonic_\Surf=0,\label{eqn:Hodge:Orthogonality:Harmonic2}  
\end{equation}
\end{subequations}
where integration by parts~(\ref{eqn:Div_psi_f})
has been used and~(\ref{eqn:SDT:closed})
has been invoked.  

\vspace{5mm}
\facilities{NASA/GSFC}

\bibliography{bibliography,extra}

\begin{thebibliography}{}
\expandafter\ifx\csname natexlab\endcsname\relax\def\natexlab#1{#1}\fi
\providecommand{\url}[1]{\href{#1}{#1}}
\providecommand{\dodoi}[1]{doi:~\href{http://doi.org/#1}{\nolinkurl{#1}}}
\providecommand{\doeprint}[1]{\href{http://ascl.net/#1}{\nolinkurl{http://ascl.net/#1}}}
\providecommand{\doarXiv}[1]{\href{https://arxiv.org/abs/#1}{\nolinkurl{https://arxiv.org/abs/#1}}}

\bibitem[{Amrouche {et~al.}(1998)Amrouche, Bernardi, Dauge, \&
  Girault}]{Amrouche1998}
Amrouche, C., Bernardi, C., Dauge, M., \& Girault, V. 1998, Mathematical
  Methods in the Applied Sciences, 21, 823,
  \dodoi{10.1002/(SICI)1099-1476(199806)21:9<823::AID-MMA976>3.0.CO;2-B}

\bibitem[{Amrouche \& Seloula(2013)}]{Amrouche2013}
Amrouche, C., \& Seloula, N. E.~H. 2013, Mathematical Models and Methods in
  Applied Sciences, 23, 37, \dodoi{10.1142/S0218202512500455}

\bibitem[{{Antiochos}(2013)}]{Antiochos2013}
{Antiochos}, S.~K. 2013, \apj, 772, 72, \dodoi{10.1088/0004-637X/772/1/72}

\bibitem[{Arapura(2016)}]{Arapura2016}
Arapura, D. 2016, Introduction to differential forms,
  \url{https://www.math.purdue.edu/~dvb/preprints/diffforms.pdf}

\bibitem[{{Aulanier} {et~al.}(2013){Aulanier}, {D{\'e}moulin}, {Schrijver},
  {Janvier}, {Pariat}, \& {Schmieder}}]{Aulanier2013}
{Aulanier}, G., {D{\'e}moulin}, P., {Schrijver}, C.~J., {et~al.} 2013, \aap,
  549, A66, \dodoi{10.1051/0004-6361/201220406}

\bibitem[{Backus(1986)}]{Backus1986}
Backus, G. 1986, Reviews of Geophysics, 24, 75, \dodoi{10.1029/RG024i001p00075}

\bibitem[{{Backus} {et~al.}(1996){Backus}, {Parker}, \&
  {Constable}}]{Backus1996}
{Backus}, G., {Parker}, R., \& {Constable}, C. 1996, {Foundations of
  Geomagnetism} (Cambridge University Press), 370

\bibitem[{Berger({1984})}]{Berger1984b}
Berger, M.~A. {1984}, {Geophysical and Astrophysical Fluid Dynamics}, {30}, 79,
  \dodoi{{10.1080/03091928408210078}}

\bibitem[{{Berger}(1997)}]{Berger1997}
{Berger}, M.~A. 1997, \jgr, 102, 2637, \dodoi{10.1029/96JA01896}

\bibitem[{Berger(1999)}]{Berger1999}
Berger, M.~A. 1999, Plasma Physics and Controlled Fusion, 41, B167

\bibitem[{{Berger} \& {Field}(1984)}]{Berger1984a}
{Berger}, M.~A., \& {Field}, G.~B. 1984, Journal of Fluid Mechanics, 147, 133

\bibitem[{{Berger} \& {Ruzmaikin}(2000)}]{Berger2000}
{Berger}, M.~A., \& {Ruzmaikin}, A. 2000, \jgr, 105, 10481,
  \dodoi{10.1029/1999JA900392}

\bibitem[{Bhatia {et~al.}(2013)Bhatia, Norgard, Pascucci, \&
  Bremer}]{Bhatia2013}
Bhatia, H., Norgard, G., Pascucci, V., \& Bremer, P.~T. 2013, IEEE Transactions
  on Visualization and Computer Graphics, 19, 1386,
  \dodoi{10.1109/TVCG.2012.316}

\bibitem[{Boulmezaoud {et~al.}(1999)Boulmezaoud, Maday, \&
  Amari}]{Boulmezaoud1999}
Boulmezaoud, T.-Z., Maday, Y., \& Amari, T. 1999, ESAIM: Mathematical Modelling
  and Numerical Analysis - Mod\'elisation Math\'ematique et Analyse
  Num\'erique, 33, 359

\bibitem[{Brouwer(1912)}]{Brouwer1912}
Brouwer, L. 1912, Mathematische Annalen, 71, 97

\bibitem[{Butt {et~al.}(1976)Butt, Newton, \& Verhage}]{Butt1976}
Butt, E., Newton, A., \& Verhage, A. 1976, in Pulsed High Beta Plasmas, ed.
  D.~Evans (Pergamon), 419 -- 423.
\newblock
  \url{https://www.sciencedirect.com/science/article/pii/B978008020941850069X}

\bibitem[{Cantarella {et~al.}(2001)Cantarella, DeTurck, \&
  Gluck}]{Cantarella2001}
Cantarella, J., DeTurck, D., \& Gluck, H. 2001, Journal of Mathematical
  Physics, 42, 876, \dodoi{10.1063/1.1329659}

\bibitem[{Cantarella {et~al.}(2002)Cantarella, DeTurck, \&
  Gluck}]{Cantarella2002}
---. 2002, The American Mathematical Monthly, 109, 409,
  \dodoi{10.1080/00029890.2002.11919870}

\bibitem[{{Chae}(2001)}]{Chae2001b}
{Chae}, J. 2001, \apjl, 560, L95

\bibitem[{Chandra {et~al.}(2010)Chandra, Pariat, Schmieder, Mandrini, \&
  Uddin}]{Chandra2010}
Chandra, R., Pariat, E., Schmieder, B., Mandrini, C.~H., \& Uddin, W. 2010,
  Solar Physics, 261, 127, \dodoi{10.1007/s11207-009-9470-2}

\bibitem[{Cheng \& Shkoller(2017)}]{Cheng2017}
Cheng, C. H.~A., \& Shkoller, S. 2017, Journal of Mathematical Fluid Mechanics,
  19, 375, \dodoi{10.1007/s00021-016-0289-y}

\bibitem[{Clegg {et~al.}(2000)Clegg, Browning, Laurence, Bromage, \&
  Stredulinsky}]{Clegg2000b}
Clegg, J.~R., Browning, P.~K., Laurence, P., Bromage, B. J.~I., \&
  Stredulinsky, E. 2000, Journal of Mathematical Physics, 41, 6783,
  \dodoi{10.1063/1.1287923}

\bibitem[{Crick(1976)}]{Crick1976}
Crick, F.~H. 1976, Proceedings of the National Academy of Sciences, 73, 2639

\bibitem[{C\u{a}lug\u{a}reanu(1959)}]{Calugareanu1959}
C\u{a}lug\u{a}reanu, G. 1959, Math. Pures Appl., 4, 5

\bibitem[{{D{\'e}moulin}(2007)}]{Demoulin2007a}
{D{\'e}moulin}, P. 2007, Advances in Space Research, 39, 1674,
  \dodoi{10.1016/j.asr.2006.12.037}

\bibitem[{Dennis \& Hannay(2005)}]{Dennis2005}
Dennis, M., \& Hannay, J. 2005, Proceedings of the Royal Society A:
  Mathematical, Physical and Engineering Sciences, 461 (2062), 3245 ,
  \dodoi{10.1098/rspa.2005.1527}

\bibitem[{Enciso {et~al.}(2018)Enciso, Ángeles García-Ferrero, \&
  Peralta-Salas}]{Enciso2018}
Enciso, A., Ángeles García-Ferrero, M., \& Peralta-Salas, D. 2018, Journal de
  Math\'{e}matiques Pures et Appliqu\'{e}es, 119, 85 ,
  \dodoi{https://doi.org/10.1016/j.matpur.2017.11.004}

\bibitem[{Esparza-L\'{o}pez {et~al.}(2016)Esparza-L\'{o}pez, Ley-Koo, \&
  Rendve\'{o}n}]{Esparza-Lopez2016}
Esparza-L\'{o}pez, C., Ley-Koo, E., \& Rendve\'{o}n, P. 2016, {Revista mexicana
  de f\~A\-sica E}, 62, 40

\bibitem[{Feynman {et~al.}(1989)Feynman, Leighton, \& Sands}]{Feynman1989}
Feynman, R., Leighton, R., \& Sands, M. 1989, The Feynman Lectures on Physics:
  Commemorative Issue, Advanced book program (Addison-Wesley).
\newblock \url{https://books.google.com/books?id=IROUjgEACAAJ}

\bibitem[{Finn \& Antonsen(1985)}]{Finn1985}
Finn, J., \& Antonsen, T. 1985, Comments Plasma Phys. Controlled Fusion, 9, 111

\bibitem[{{Forbes}(2000)}]{Forbes2000}
{Forbes}, T.~G. 2000, \jgr, 105, 23153, \dodoi{10.1029/2000JA000005}

\bibitem[{Fuller(1971)}]{Fuller1971}
Fuller, F.~B. 1971, Proceedings of the National Academy of Sciences, 68, 815

\bibitem[{Gauss(1867)}]{GaussWerkeV}
Gauss, C.~F. 1867, in Werke, ed. C.~Sch\"{a}fe, Vol.~5 (Leipzig, Berlin:
  K\"{o}nigliche Gesellschaft der Wissenschaften zu G\"{o}ttingen)

\bibitem[{Girault \& Raviart(1986)}]{Girault1986}
Girault, V., \& Raviart, P. 1986, Finite Element Methods for Navier-Stokes
  Equations: Theory and Algorithms, Springer Series in Computational
  Mathematics (Springer-Verlag Berlin Heidelberg).
\newblock \url{https://books.google.com/books?id=8C7vCAAAQBAJ}

\bibitem[{Gui \& Dou(2007)}]{Gui2007}
Gui, Y.~F., \& Dou, W.-B. 2007, Progress In Electromagnetics Research, 69, 287

\bibitem[{Halmos(1974)}]{Halmos1974}
Halmos, P.~R. 1974, Measure theory / [by] Paul R. Halmos (Springer-Verlag New
  York), xi, 304 p. ;.
\newblock \url{http://www.loc.gov/catdir/enhancements/fy0814/74010690-t.html}

\bibitem[{Harrison(1993)}]{Harrison1993}
Harrison, J. 1993, Bulletin of the American Mathematical Society, 29, 235

\bibitem[{Harrison(1999)}]{Harrison1999}
---. 1999, Journal of Physics A: Mathematical and General, 32, 5317,
  \dodoi{10.1088/0305-4470/32/28/310}

\bibitem[{Hodge(1959)}]{Hodge1952}
Hodge, W. 1959, The theory and applications of harmonic integrals, reprinted
  2nd edn. (Cambridge: University Press).
\newblock \url{https://books.google.com/books?id=P0jvAAAAMAAJ}

\bibitem[{{Imbert-Gerard} \& {Greengard}(2016)}]{Imbert-Gerard2016}
{Imbert-Gerard}, L.-M., \& {Greengard}, L. 2016, ArXiv e-prints.
\newblock \doarXiv{1608.04436}

\bibitem[{Jackson(1975)}]{Jackson1975}
Jackson, J.~D. 1975, Classical {E}lectrodynamics, 2nd edn. (New York: John
  Wiley \& Sons)

\bibitem[{{Jackson} \& {Okun}(2001)}]{Jackson2001}
{Jackson}, J.~D., \& {Okun}, L.~B. 2001, Reviews of Modern Physics, 73, 663,
  \dodoi{10.1103/RevModPhys.73.663}

\bibitem[{{Jiang} {et~al.}(1994){Jiang}, {Loh}, \& {Povinelli}}]{Jiang1994}
{Jiang}, B.-N., {Loh}, C.~Y., \& {Povinelli}, L.~A. 1994, NASA STI/Recon
  Technical Report N, 94

\bibitem[{Karpen {et~al.}(2012)Karpen, Antiochos, \& DeVore}]{Karpen2012}
Karpen, J.~T., Antiochos, S.~K., \& DeVore, C.~R. 2012, The Astrophysical
  Journal, 760, 81

\bibitem[{Kemmer(1977)}]{Kemmer1977}
Kemmer, N. 1977, Vector Analysis: A Physicist's Guide to the Mathematics of
  Fields in Three Dimensions (Cambridge University Press),
  \dodoi{10.1017/CBO9780511569524}

\bibitem[{Klimchuk(2006)}]{Klimchuk2006}
Klimchuk, J.~A. 2006, Solar Physics, 234, 41, \dodoi{10.1007/s11207-006-0055-z}

\bibitem[{{Knizhnik} {et~al.}(2017){Knizhnik}, {Antiochos}, {DeVore}, \&
  {Wyper}}]{Knizhnik2017}
{Knizhnik}, K.~J., {Antiochos}, S.~K., {DeVore}, C.~R., \& {Wyper}, P.~F. 2017,
  \apjl, 851, L17, \dodoi{10.3847/2041-8213/aa9e0a}

\bibitem[{{Kusano} {et~al.}(2002){Kusano}, {Maeshiro}, {Yokoyama}, \&
  {Sakurai}}]{Kusano2002a}
{Kusano}, K., {Maeshiro}, T., {Yokoyama}, T., \& {Sakurai}, T. 2002, \apj, 577,
  501

\bibitem[{Kusano {et~al.}(2003)Kusano, Maeshiro, Yokoyama, \&
  Sakurai}]{Kusano2003}
Kusano, K., Maeshiro, T., Yokoyama, T., \& Sakurai, T. 2003, Advances in Space
  Research, 32, 1917 , \dodoi{https://doi.org/10.1016/S0273-1177(03)90626-0}

\bibitem[{{Kusano} {et~al.}(1995){Kusano}, {Suzuki}, \&
  {Nishikawa}}]{Kusano1995}
{Kusano}, K., {Suzuki}, Y., \& {Nishikawa}, K. 1995, \apj, 441, 942,
  \dodoi{10.1086/175413}

\bibitem[{Kustepeli(2016)}]{Kustepeli2017}
Kustepeli, A. 2016, Electromagnetics, 36, 135,
  \dodoi{10.1080/02726343.2016.1149755}

\bibitem[{{Liu} {et~al.}(2014){Liu}, {Hoeksema}, {Bobra}, {Hayashi}, {Schuck},
  \& {Sun}}]{Liu2014}
{Liu}, Y., {Hoeksema}, J.~T., {Bobra}, M., {et~al.} 2014, \apj, 785, 13,
  \dodoi{10.1088/0004-637X/785/1/13}

\bibitem[{Ludu(2012)}]{Ludu2012}
Ludu, A. 2012, Nonlinear Waves and Solitons on Contours and Closed Surfaces,
  Springer Series in Synergetics (NY: Springer Berlin Heidelberg).
\newblock \url{https://books.google.com/books?id=e7eAkiAehUUC}

\bibitem[{Mackay {et~al.}(2014)Mackay, DeVore, \& Antiochos}]{Mackay2014}
Mackay, D.~H., DeVore, C.~R., \& Antiochos, S.~K. 2014, The Astrophysical
  Journal, 784, 164, \dodoi{10.1088/0004-637x/784/2/164}

\bibitem[{{MacNeice} {et~al.}(2004){MacNeice}, {Antiochos}, {Phillips},
  {Spicer}, {DeVore}, \& {Olson}}]{MacNeice2004}
{MacNeice}, P., {Antiochos}, S.~K., {Phillips}, A., {et~al.} 2004, \apj, 614,
  1028

\bibitem[{Maria~Denaro(2003)}]{Denaro2003}
Maria~Denaro, F. 2003, International Journal for Numerical Methods in Fluids,
  43, 43, \dodoi{10.1002/fld.598}

\bibitem[{Moffatt(1969)}]{Moffatt1969}
Moffatt, H.~K. 1969, Journal of Fluid Mechanics, 35, 117–129,
  \dodoi{10.1017/S0022112069000991}

\bibitem[{Moffatt(2014)}]{Moffatt2014}
---. 2014, Proceedings of the National Academy of Sciences, 111, 3663,
  \dodoi{10.1073/pnas.1400277111}

\bibitem[{Moffatt \& Ricca(1992)}]{Moffatt1992}
Moffatt, H.~K., \& Ricca, R.~L. 1992, Proceedings of the Royal Society of
  London A: Mathematical, Physical and Engineering Sciences, 439, 411,
  \dodoi{10.1098/rspa.1992.0159}

\bibitem[{Morrey(1966)}]{Morrey1966}
Morrey, C. 1966, Multiple Integrals in the Calculus of Variations, Classics in
  Mathematics (Berlin Heidelberg: Springer-Verlag), 506,
  \dodoi{10.1007/978-3-540-69952-1}.
\newblock \url{https://www.springer.com/us/book/9783540699156}

\bibitem[{Morse(1953)}]{Morse1953a}
Morse, P.~M. 1953, International Series in Pure and Applied Physics, Vol.~1,
  Methods in {T}heoretical {P}hysics (New York: McGraw-Hill Pubublishing Co.)

\bibitem[{{Nemenman} \& {Silbergleit}(1999)}]{Nemenman1999}
{Nemenman}, I.~M., \& {Silbergleit}, A.~S. 1999, Journal of Applied Physics,
  86, 614, \dodoi{10.1063/1.370775}

\bibitem[{{Nindos} {et~al.}(2003){Nindos}, {Zhang}, \& {Zhang}}]{Nindos2003a}
{Nindos}, A., {Zhang}, J., \& {Zhang}, H. 2003, \apj, 594, 1033

\bibitem[{{O'Neil}(2018)}]{ONeil2018}
{O'Neil}, M. 2018, Adv. Comput Math,
  \dodoi{https://doi.org/10.1007/s10444-018-9587-7}

\bibitem[{{Pariat} {et~al.}(2005){Pariat}, {D{\'e}moulin}, \&
  {Berger}}]{Pariat2005}
{Pariat}, E., {D{\'e}moulin}, P., \& {Berger}, M.~A. 2005, \aap, 439, 1191,
  \dodoi{10.1051/0004-6361:20052663}

\bibitem[{{Pariat} {et~al.}(2007){Pariat}, {D{\'e}moulin}, \&
  {Nindos}}]{Pariat2007}
{Pariat}, E., {D{\'e}moulin}, P., \& {Nindos}, A. 2007, Advances in Space
  Research, 39, 1706, \dodoi{10.1016/j.asr.2007.02.047}

\bibitem[{{Parker}(1988)}]{Parker1988}
{Parker}, E.~N. 1988, \apj, 330, 474, \dodoi{10.1086/166485}

\bibitem[{{Pevtsov} {et~al.}(2003){Pevtsov}, {Balasubramaniam}, \&
  {Rogers}}]{Pevtsov2003}
{Pevtsov}, A.~A., {Balasubramaniam}, K.~S., \& {Rogers}, J.~W. 2003, \apj, 595,
  500, \dodoi{10.1086/377339}

\bibitem[{Pohl(1968)}]{Pohl1968}
Pohl, W.~F. 1968, Journal of Mathematics and Mechanics, 17, 975

\bibitem[{{Prior} \& {Yeates}(2014)}]{Prior2014}
{Prior}, C., \& {Yeates}, A.~R. 2014, \apj, 787, 100,
  \dodoi{10.1088/0004-637X/787/2/100}

\bibitem[{Reusken(Accepted for publication 2018)}]{Reusken2019}
Reusken, A. Accepted for publication 2018, IMA Journal of Numerical Analysis,
  ??, ??, \dodoi{10.1093/imanum/dry062}

\bibitem[{Ricca(2002)}]{Ricca2002}
Ricca, R.~L. 2002, in Tubes, Sheets and Singularities in Fluid Dynamics, ed.
  K.~Bajer \& H.~K. Moffatt (Dordrecht: Springer Netherlands), 139--144

\bibitem[{Ricca \& Nipoti(2011)}]{Ricca2011}
Ricca, R.~L., \& Nipoti, B. 2011, Journal of Knot Theory and Its Ramifications,
  20, 1325, \dodoi{10.1142/S0218216511009261}

\bibitem[{{Romano, P.} {et~al.}(2011){Romano, P.}, {Pariat, E.}, {Sicari, M.},
  \& {Zuccarello, F.}}]{Romano2011}
{Romano, P.}, {Pariat, E.}, {Sicari, M.}, \& {Zuccarello, F.} 2011, A\&A, 525,
  A13, \dodoi{10.1051/0004-6361/201014437}

\bibitem[{{Sakurai}(1982)}]{Sakurai1982}
{Sakurai}, T. 1982, \solphys, 76, 301, \dodoi{10.1007/BF00170988}

\bibitem[{Scharstein(1991)}]{Scharstein1991}
Scharstein, R.~W. 1991, in The Twenty-Third Southeastern Symposium on System
  Theory, IEEE (Los Alamitos, CA: IEEE Computer Society Press), 424--426

\bibitem[{Schulz \& Schulz(2016)}]{Schulz2016}
Schulz, A.~E., \& Schulz, W.~C. 2016, A Practical Introduction to Differential
  Forms (Flagstaff, Vienna, Cosmopolis: Transgalactic Publishing Co.)

\bibitem[{Shubin \& Andersson(2001)}]{Shubin2001}
Shubin, M., \& Andersson, S. 2001, Pseudodifferential Operators and Spectral
  Theory, Pseudodifferential Operators and Spectral Theory (Berlin: Springer
  Verlag), \dodoi{10.1007/978-3-642-56579-3}.
\newblock \url{https://books.google.com/books?id=p4xENSPhhw0C}

\bibitem[{Tai(1992)}]{Tai1992}
Tai, C. 1992, Generalized vector and dyadic analysis: applied mathematics in
  field theory, IEEE/OUP series on electromagnetic wave theory (Institute of
  Electrical and Electronics Engineers).
\newblock \url{https://books.google.com/books?id=HPtQAAAAMAAJ}

\bibitem[{Tai \& Fang(1991)}]{Tai1991}
Tai, C.~., \& Fang, N. 1991, IEEE Transactions on Education, 34, 167,
  \dodoi{10.1109/13.81596}

\bibitem[{{Taylor}(1974)}]{Taylor1974}
{Taylor}, J.~B. 1974, Physical Review Letters, 33, 1139,
  \dodoi{10.1103/PhysRevLett.33.1139}

\bibitem[{Thomson(1868)}]{Thomson1868}
Thomson, S. 1868, Transactions of the Royal Society of Edinburgh, 25,
  217–260, \dodoi{10.1017/S0080456800028179}

\bibitem[{Van~Bladel(1958)}]{Bladel1958}
Van~Bladel, J. 1958, {On Helmholtz's theorem in finite regions}, Tech. Rep.
  MURA-440, Midwestern Univ. Res. Ass., Madison, WI

\bibitem[{{Van Bladel}(1993{\natexlab{a}})}]{Bladel1993a}
{Van Bladel}, J. 1993{\natexlab{a}}, AEU-Archiv Fur Elektronik Und
  Ubertragungstechnik-International Journal Of Electronics And Communications,
  47, 131

\bibitem[{{Van Bladel}(1993{\natexlab{b}})}]{Bladel1993b}
---. 1993{\natexlab{b}}, Electromagnetics, 13, 95,
  \dodoi{10.1080/02726349308908332}

\bibitem[{{Van Bladel}(2007)}]{Bladel2007}
---. 2007, Electromagnetic Fields, 2nd edn., IEEE Press Series on
  Electromagnetic Wave Theory (Hoboken, NJ: Wiley - IEEE Press).
\newblock \url{http://books.google.com/books/about/Electromagnetic_Fields.html}

\bibitem[{Vemareddy(2015)}]{Vemareddy2015}
Vemareddy, P. 2015, The Astrophysical Journal, 806, 245

\bibitem[{{Vemareddy} {et~al.}(2012){Vemareddy}, {Ambastha}, {Maurya}, \&
  {Chae}}]{Vemareddy2012}
{Vemareddy}, P., {Ambastha}, A., {Maurya}, R.~A., \& {Chae}, J. 2012, \apj,
  761, 86, \dodoi{10.1088/0004-637X/761/2/86}

\bibitem[{Watson \& Craig(2001)}]{Watson2001}
Watson, P.~G., \& Craig, I. J.~D. 2001, Journal of Geophysical Research: Space
  Physics, 106, 15735, \dodoi{10.1029/2000JA000418}

\bibitem[{Weatherburn(1955)}]{Weatherburn1955}
Weatherburn, C. 1955, Differential Geometry of Three Dimensions, Vol.~1,
  Differential geometry of three dimensions (NY: Cambridge University Press).
\newblock \url{https://books.google.com/books?id=z0fvAAAAMAAJ}

\bibitem[{Weyl(1919)}]{Weyl1919}
Weyl, H. 1919, Annalen der Physik, 364, 101, \dodoi{10.1002/andp.19193641002}

\bibitem[{White(1969)}]{White1969}
White, J.~H. 1969, American Journal of Mathematics, 91, 693

\bibitem[{Whitney(1957)}]{Whitney1957}
Whitney, H. 1957, Geometric Integration Theory, Princeton mathematical series
  (Princeton University Press).
\newblock \url{https://books.google.com/books?id=-IcEtQEACAAJ}

\bibitem[{{Woltjer}(1958)}]{Woltjer1958}
{Woltjer}, L. 1958, Proceedings of the National Academy of Science, 44, 489,
  \dodoi{10.1073/pnas.44.6.489}

\bibitem[{Wu {et~al.}(2007)Wu, Ma, \& Zhou}]{Wu2007}
Wu, J., Ma, H., \& Zhou, M. 2007, Vorticity and Vortex Dynamics, Lecture notes
  in mathematics (Springer Berlin Heidelberg).
\newblock \url{https://books.google.com/books?id=P5yNCu44PiwC}

\bibitem[{{Wyper} {et~al.}(2017){Wyper}, {Antiochos}, \& {DeVore}}]{Wyper2017}
{Wyper}, P.~F., {Antiochos}, S.~K., \& {DeVore}, C.~R. 2017, \nat, 544, 452,
  \dodoi{10.1038/nature22050}

\bibitem[{{Yamamoto} \& {Sakurai}(2009)}]{Yamamoto2009}
{Yamamoto}, T.~T., \& {Sakurai}, T. 2009, \apj, 698, 928,
  \dodoi{10.1088/0004-637X/698/1/928}

\bibitem[{Yoshida(1998)}]{Yoshida1998}
Yoshida, Z. 1998, Nonlinear physics of twisted magnetic field lines, Tech.
  rep., Japan

\bibitem[{Zhou(2006)}]{Zhou2006}
Zhou, X. 2006, Progress In Electromagnetics Research, 65, 93,
  \dodoi{10.2528/PIER06081202}

\end{thebibliography}



\end{document}